%% file: thesis.tex
\documentclass{ucbthesis}
\usepackage[utf8]{inputenc}
\usepackage[T1]{fontenc}
\usepackage{newunicodechar}
\newunicodechar{̀}{\`{}}

\usepackage{biblatex}
\usepackage{rotating} 
\usepackage{amsmath}
\usepackage{amssymb}

\usepackage{mathtools}
\usepackage{cancel}

\usepackage{graphicx}
\usepackage{dcolumn}
\usepackage{bm}
\usepackage{xr}
\usepackage{cleveref}
\usepackage{soul} 

\usepackage{graphicx}
\usepackage{dcolumn}
\usepackage{bm}
\usepackage{appendix}

\AtBeginEnvironment{subappendices}{%
\chapter*{Chapter appendix}
\counterwithin{figure}{section}
\counterwithin{table}{section}
}

\usepackage{comment}

\usepackage{graphicx}
\usepackage{dcolumn}
\usepackage{bm}
\usepackage{algorithm}
\usepackage{algpseudocode}

\algdef{SE}[REPEATN]{RepeatN}{End}[1]{\algorithmicrepeat\ #1 \textbf{times}}{\algorithmicend}

\usepackage{capt-of}

\usepackage{soul} 
\usepackage{xcolor} 

\usepackage[normalem]{ulem} 

\usepackage{chngcntr}

\newcommand{\tee}{{\mathrm{t}}}
\newcommand{\tf}{{\mathrm{\tau}}}
\newcommand{\A}{{\mathrm{A}}}
\newcommand{\B}{{\mathrm{B}}}
\newcommand{\C}{{\mathrm{C}}}
\newcommand{\D}{{\mathrm{D}}}
\newcommand{\U}{{\mathrm{U}}}
\newcommand{\W}{{\mathrm{W}}}
\newcommand{\bolddx}{{\Delta {x}}}

\newcommand{\boldx}{{x}} 
\newcommand{\boldX}{{X}} 
\newcommand{\boldzero}{{0}} 

\newcommand{\boldv}{{v}} 
\newcommand{\boldb}{{b}} 

\newcommand{\rhoeq}{\rho^\mathrm{eq}}

\newcommand{\boldnabla}{\nabla}

\newcommand{\boldxi}{\xi} 

\newcommand{\dee}{{\mathrm{d}}}

\newcommand{\dX}{{\mathrm{d}X}}
\newcommand{\dt}{{\,\mathrm{d}t}}
\newcommand{\dx}{{\,\mathrm{d}x}}

\newcommand{\dP}{{\mathrm{d}P}}
\newcommand{\dXtilde}{{\mathrm{d}\tilde{X}}}
\newcommand{\dtau}{{\,\mathrm{d}\tau}}

\newcommand{\ds}{{\,\mathrm{d}s}}

\newcommand{\kBT}{{k_\mathrm{B}T}}

\newcommand{\kB}{{k_\mathrm{B}}}

\newtheorem{theorem}{Jibberish}

\bibliography{BIBLIOGRAPHY}

\begin{document}

\sloppy


\title{Transporting Baguettes With Minimal Action: \\  The Geometry of Optimal Nonequilibrium Processes in Stochastic Thermodynamics}
\author{Adrianne Zhong}
\degreesemester{Summer}
\degreeyear{2025}
\degree{Doctor of Philosophy}
\chair{Professor Michael R. DeWeese}
\othermembers{Professor Jonathan S. Wurtele \\
  Professor Ahmad Omar}
\numberofmembers{3}
\field{Physics}


\maketitle

\copyrightpage

\include{abstract}

\begin{frontmatter}

\begin{dedication}
\null\vfil
\begin{center}
To my cherished friends and family.\\\vspace{12pt}
\end{center}
\vfil\null
\end{dedication}


\tableofcontents
\clearpage
\listoffigures

\begin{acknowledgements}

There are a few individuals I would like to thank. But first, I acknowledge the support from the NDSEG fellowship as well as the COCOSYS grant, for funding in part my PhD. Having this financial support has greatly facilitated the completion of my degree. 

I would like to thank my qualifying exam committee members: Ahmad Omar, Hernan Garcia, Jonathan Wurtele, and Michael DeWeese; as well as my qualifying exam committee: Ahmad Omar, Michael DeWeese, and Jonathan Wurtele. Thank you for taking the time out of your very busy schedules to give me very helpful feedback on my research. Along with aforementioned members of my PhD committees, there are a number of other professors who have also been instrumental in my development as a scientist: Andy Charman, Bruno Olshausen, Fritz Sommer, Michael Lindsey, Na Ji, and Samantha Lewis. I appreciate the time and patience you placed in me, and through our discussions I have learned so much. 

I give my nod to the ``Non-Eq Stat-Mech Reading Group'' that took place in 2022-2023---what motley bunch we were! As regular members, Adam Frim, Ben Kuznets-Speck, Cory Hargus, Jorge Rosa-Raíces, Nivedina Sarma, Songela Chen, and I would meet up every week to discuss whatever topic or paper in nonequilibrium statistical mechanics we found interesting. Although much of the time it felt like feeling in the dark, I had so much fun! Having this regular reading group truly sparked my interest in nonequilibrium statistical mechanics. Thank you, Adam, Ben, Cory, Jorge, Nividena, and Songela! 

To the folks at the Redwood Center of Theoretical Neuroscience, thank you for providing me such a welcoming academic home. Although in the end my PhD research turned out not to be in theoretical neuroscience, I have had such a lovely experience getting to know you and work in the same space as you. I will miss our many conversations and coffee-runs to the building next door. 

To the rotating cast of DeWeese lab members, thank you for your support and keeping me on my toes! I wish the current members the best of luck in navigating the rest of your PhDs; please never feel afraid to reach out if you are in need of any advice! Additionally, to Sam D'Ambrosia and Hana Mir: May nonequilibrium statistical mechanics continue to live on in your hearts!

To my PhD advisor Michael DeWeese, it is crazy to believe that we have known each other for more than decade. You inspire me as one of the most creative individuals I know, always chock-full of ideas. Thank you so much for encouraging me when I did not believe in myself, and advocating for me in so many different ways over all these years. I look forward to seeing you in Chicago! 

I would like to thank the psychotherapist I have had for the past four-and-a-half years.\footnote{For confidentiality reasons, I do not name this therapist.} I would not be the person I am today without all the support I have received from them. I highly recommend to any student and scientist the professional services of a psychotherapist, to help navigate the turbulent waters that are unavoidably encountered in a career in academia and the sciences---especially in this day and age.

I would like to thank the following individuals for their friendship: Adam Frim, Alexis Plos Miller, Aly Lidayan, Andrea Feher, Ashley Song, Ashwin Singh, Ben Kuznets-Speck, Carlos Sierra, Caryn Tran, Chad Harper, Chenling Xu, Chris Branner-Augmon, Chris Kymn, Chrissy Scardina, Christina Stogsdill, Collin Hoover, Cory Hargus, Daniel Goldberg, Danielle Shi, Diana Martinez, Eduardo Sandoval, Elizabeth Dresselhaus, Evie Pai, Finn Wurtz, Gabriella Licata, Galen Chuang, Holly Flores, Hunter Akins, Irian d'Andrea, Jeske Dioquino, Jessica Quiroga, Joseph Fierro, Kailin Zhaung, Kate McDowell, Keyanna Beavers, Madeline Monroy, Malcolm Science Lazarow, Mar Thomas, Micah Lykken, Nann Tsehay, Nekeia Rideout, Nicole Wong, Nivedina Sarma, Q. Zheng, Rachaël Longuépée, Roy Tu, S. Davis, Stefan Divic, Storee Moss, Victor Mendes, and Yoyo Li. There would not have been nearly as much music and joy without your presence in my life; nor is it possible for me to imagine getting to where I am today without your love and support. For that I am deeply appreciative. 

Unfortunately, the past six years have not gone by 
without certain losses: 
\begin{itemize}[leftmargin=*]
  \item[] To my friend Ariadne Belsito, I wish you did not have to endure all the hardship in your life just for being who you are. I miss the days of reverse-engineering hardware with you. I wish you could know how dearly you are missed by so many.
  \item[] To my graduate-level statistical mechanics professor Phill Geissler, all the time and care you have put into pedagogy, to be able to explain things with such rich clarity, has inspired me to become a much better teacher. I really wish we could have worked together before you so suddenly left.
  \item[] To my undergraduate advisor Joel Fajans, thank you for believing in me and looking out for me throughout all of these years. You taught me just how much fun it is to do physics, and how through cleverness (among other things) we can learn things about the universe around us. I am so glad to have had the chance to finish this one last plasma physics project with you before you left,\footnote{I am extraordinarily grateful to Jonathan Wurtele for helping see this last project \cite{zhong2024shot} to completion.} and I will always carry with me the kindness and inspiration you have given me.
\end{itemize}

To my best friends Michael Alan Chang and Cora Leigh Moss, thank you for being a beacon of light. You were by my side during both the wonderful times and the darker moments in my life. It has been beautiful to witness each others' growth over the past six years; your presence in my life has changed me for the better, inspiring me to be much more caring and empathetic human being. From the bottom of my heart, I thank you for all your love and care you have shared.

Last but not least, I would like to thank my family: Lefan, Rongxiang, Emily, and Suanna. Your continued love and support mean the world to me. 



\end{acknowledgements}

\end{frontmatter}

\pagestyle{headings}


\chapter{Introduction}
What is the \emph{laziest way possible} to do something, subject to the fundamental laws of physics?\newline\newline
%
Before attempting to answer that question, here is a pair of paradoxes:

\begin{itemize}
   \item How does one reconcile the fundamental time-reversibility of classical physics with the obvious time-irreversibility we observe in living organisms?\footnote{This is a specific case of what is known as Loschmidt's paradox \cite{Loschmidt}. Having the planets orbit clockwise instead of counter-clockwise around the sun is just as valid of a solution to Newton's laws of motion, while on the other hand, \cite{england2013statistical} points out that we virtually \textit{never} observe time-reversed cellular division, e.g., two daughter bacteria fusing back into a single mother cell with two copies of DNA becoming ``un-replicated'' back into a single copy under reversed meiosis. Nor do we ever really observe rabbits running backwards, newborn chicks crawling back into eggshells that later unhatch, or cows sucking in their ``moo''s through having the sound pressure wave propagating from the outside world to the back into their mouths.}   
   \item The Second Law of Thermodynamics states that entropy (disorder) should \textit{always} increase---how is it that biological organisms seem to be violating it, through assembling and maintaining a high degree of complex bodily structure throughout their lifespans?\footnote{This is the central problem tackled in \cite{schrodinger1992life}, which I highly recommend reading!}
\end{itemize}
Resolutions to both paradoxes lie within the realm of nonequilibrium thermodynamics, where familiar expressions of standard (equilibrium) thermodynamics, such as ``$PV = N \kBT$'' and ``$\mathrm{Prob}(\mathrm{microstate} \ x) \propto \exp \big( -{\mathrm{Energy}(\mathrm{microstate} \ x)} / {\kBT} \big) $'', are no longer valid.\footnote{Actually, I think that as an undergrad at UC Berkeley, the equilibrium thermodynamics course, \textit{Physics 112: Introduction to Statistical and Thermal Physics}, was simultaneously the most tedious and least satisfying physics course I took. It was only through studying nonequilibrium statistical physics that such a seemingly random assortment of formulas in equilibrium thermodynamics began to make sense.} The breaking of time-reversibility turns out to be a simple consequence within nonequilibrium thermodynamics, and surprisingly, it \textit{can} be derived from the time-reversible laws of classical physics when you do not know the exact state, i.e., positions and velocities, of the environment surrounding a large system of interest, e.g., every one of the $\sim 10^{23}$ air molecules in a piston, or the locations and velocities of the (mostly) water molecules surrounding a protein in a cell. See Appendix A for a derivation of irreversible stochastic dynamics of a system (e.g., a protein) interacting with a large, not fully knowable environment (e.g., the bath of water molecules surrounding it), from reversible Hamiltonian dynamics for the composite system: the system \emph{and} environment together.

As far as the second paradox is concerned, it is important to not forget about the existence of the sun, which is very gently but constantly baking the Earth at temperature $\sim 6000$ Kelvin\footnote{This is $\sim 10000^\circ$ Fahrenheit!} with a whopping $\sim 1000$ Watts per square meter on a cloudless day. One must never forget that the Second Law of Thermodynamics is a statement about \textit{closed} systems---systems that do not exchange energy or matter with an ``outside world''---whereas the global ecosystem that is the Earth is an \textit{open} system being constantly bombarded by sunlight. Nonequilibrium thermodynamics deals with such cases, and through it one can show that while the entropy of biological life can be sustained to remain low, the Second Law of Thermodynamics still holds when considering the entropy of the Earth \textit{and} Sun together, which is ever-increasing.

Of course, there is reason to believe that down to the molecular level, biological systems should not be wasteful with expending energy, which through thermodynamics is related to management of entropy. We know however, that the molecular environment within a living cell is noisy, and can only be adequately addressed through nonequilibrium statistical physics. Recapitulating the first sentence, the central problem we tackle in this thesis is: What are the fundamental limitations placed by the laws of thermodynamics on the energy expenditure needed to carry out a given task in nonequilibrium environments? 

In this thesis we explore the connections between the optimal control of nonequilibrium thermodynamic systems, and (as is suggested by the thesis title) the optimal transportation problem. The conclusion is that the former is fundamentally geometric.

\section{Problem setup}

\begin{figure*}[t]
    \centering
    \includegraphics[width=.7\linewidth]{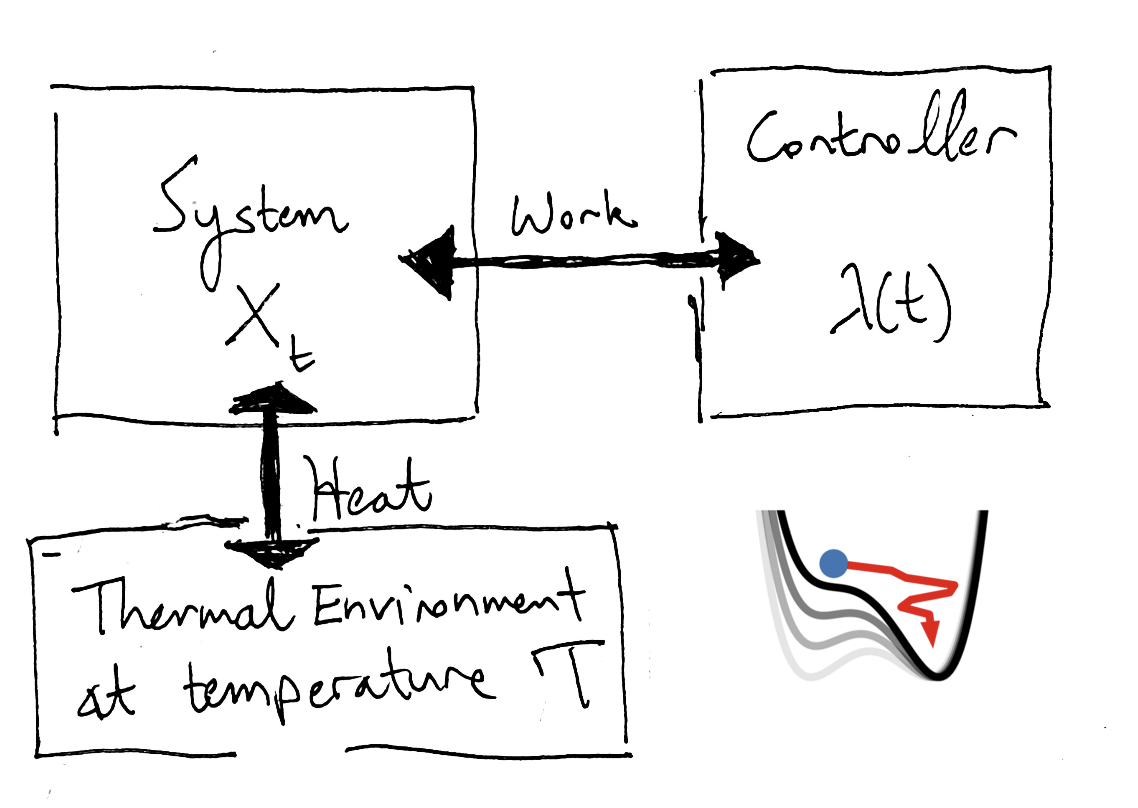}
    \caption[Cartoon schematic of a thermodynamic system.]{A system with (stochastic) state $X_t \in \mathbb{R}^d$ is connected to an environment at temperature $T$, and may be manipulated by a controller whose control state is indicated by $\lambda(t)$. The dark bidirectional arrows indicate the exchange of energy, which we identify as the work (between the system and controller) and heat (between the system and environment). In the bottom right figure, the blue dot indicates the state of the system $X_t \in \mathbb{R}$, and $\lambda(t)$ modulates a tilt in the energy function that biases the state towards the right. The specific trajectory that the state undergoes is stochastic due to the interaction with the heat bath.}
    \label{fig:cartoon}
\end{figure*}

In this section we sketch the setup of this ``thermodynamically optimal protocols'' problem, which is accompanied by the cartoon Fig.~\ref{fig:cartoon}. We consider a system described with state $X_t \in \mathbb{R}^d$ at time $t$ that is connected to an environment (i.e., ``heat bath'') at temperature $T$, and may be externally manipulated by a controller whose controls are indicated by $\lambda(t) \in \mathcal{M}$. (We call $\mathcal{M}$ the \textbf{control manifold}.) Specifically, $\lambda \in \mathcal{M}$ parameterizes an energy function for the states of the system $U_\lambda(x)$, with manipulations of $\lambda(t)$ perturbing the energy function. With these interactions, energy can flow both between the system and environment which is the thermodynamic heat, and also between the system and controller which is the thermodynamic work. 

Due to the interactions between the system and the thermal environment, the dynamics for the system state $X_t$ are necessarily stochastic.\footnote{It is for this reason we distinguish the notation for the time-dependence of the state with the subscript $\boldX_t$, vs. for the controls we notate $\lambda(t)$.} Nonetheless, it is possible to talk about the statistical or \textbf{thermodynamic state} of the system which is described by the \textbf{probability density} of the system $\rho_t$ at time $t$, which is defined as $\rho_t(x) = \mathrm{Prob}(X_t = x)$. Throughout this thesis we shall often refer to the probability density $\rho_t(x)$ as the thermodynamic state of the system. 

The optimal protocols problem is: what are the optimal controls $\lambda^*(t)$ that transform a thermodynamic state in a finite amount of time, that uses the least amount of work?


\subsection{Note}

In the following, the material in Sections 1.2-6 is largely adapted from the review \cite{seifert2012stochastic}, although the particular proof given for the claim in section 1.6 is my own. The material in Section 1.7 is adapted from \cite{sivak2012thermodynamic}, and the material in Section 1.8 is adapted from \cite{aurell2011optimal} and \cite{villani2021topics}. We summarize the original contributions made in this thesis in Section 1.9.

\section{The Langevin equation}

The specific dynamics we concern ourselves with is the \textbf{overdamped Langevin}\footnote{This is pronounced, for English speakers, as ``lawn-je-VON''.} \textbf{Equation}, which is a stochastic differential equation (SDE)
\begin{equation}
    \frac{\dee \boldX_t }{\dt} = \underbrace{- \mu \nabla U_{\lambda(t)}(\boldX_t)}_\text{interaction with controller} + \underbrace{\sqrt{2 \mu \kBT} \, \boldxi_t}_\text{interaction with environment} .  \label{eq:langevin-eq}
\end{equation}
Here, $\mu$ is the motility (units are $\text{velocity}/\text{force}$), and $U_\lambda(\boldx)$ is an energy function that is parameterized by the controls $\lambda \in \mathcal{M}$ in an $m$-dimensional manifold, with $\boldnabla$ denoting the spatial gradient with regards to $\boldx$. In the second term (arising from interactions with the heat bath), $\kB$ is Boltzmann's constant,  $T$ is the \emph{fixed} temperature of the environment connected to the system,\footnote{It turns out by considering changes in the temperature adds a considerable amount of subtlety to the discussion---there is already enough mathematical richness in the isothermal case.} and $\boldxi_t$ is a standard delta-correlated Gaussian white noise with statistics $\langle \boldxi_t\rangle = \boldzero$ and $\langle  \boldxi_t \, \boldxi_{t'}^\mathsf{T} \rangle = \boldsymbol{I}_d \, \delta(t - t')$.


Initial conditions for the Langevin Equation are specified by the initial thermodynamic state of the system $\rho_i(\boldx) \in \mathcal{P}(\mathbb{R}^d)$, which gives the probability where a trajectory starts at time $t = 0$:
\begin{equation}
    \boldX_0 \sim \rho_i(\cdot). 
\end{equation}
(A physical derivation for the Langevin equation, (i.e., from the deterministic and reversible Hamiltonian equations of motion for $Q_t = (X_t, Y_t)$ the system variables $X_t$ \emph{and} the environment degrees of freedom $Y_t$) may be found in Appendix \ref{chapter:physical-derivation-langevin}.)

\begin{figure*}[t]
    \centering
    \includegraphics[width=1.\linewidth]{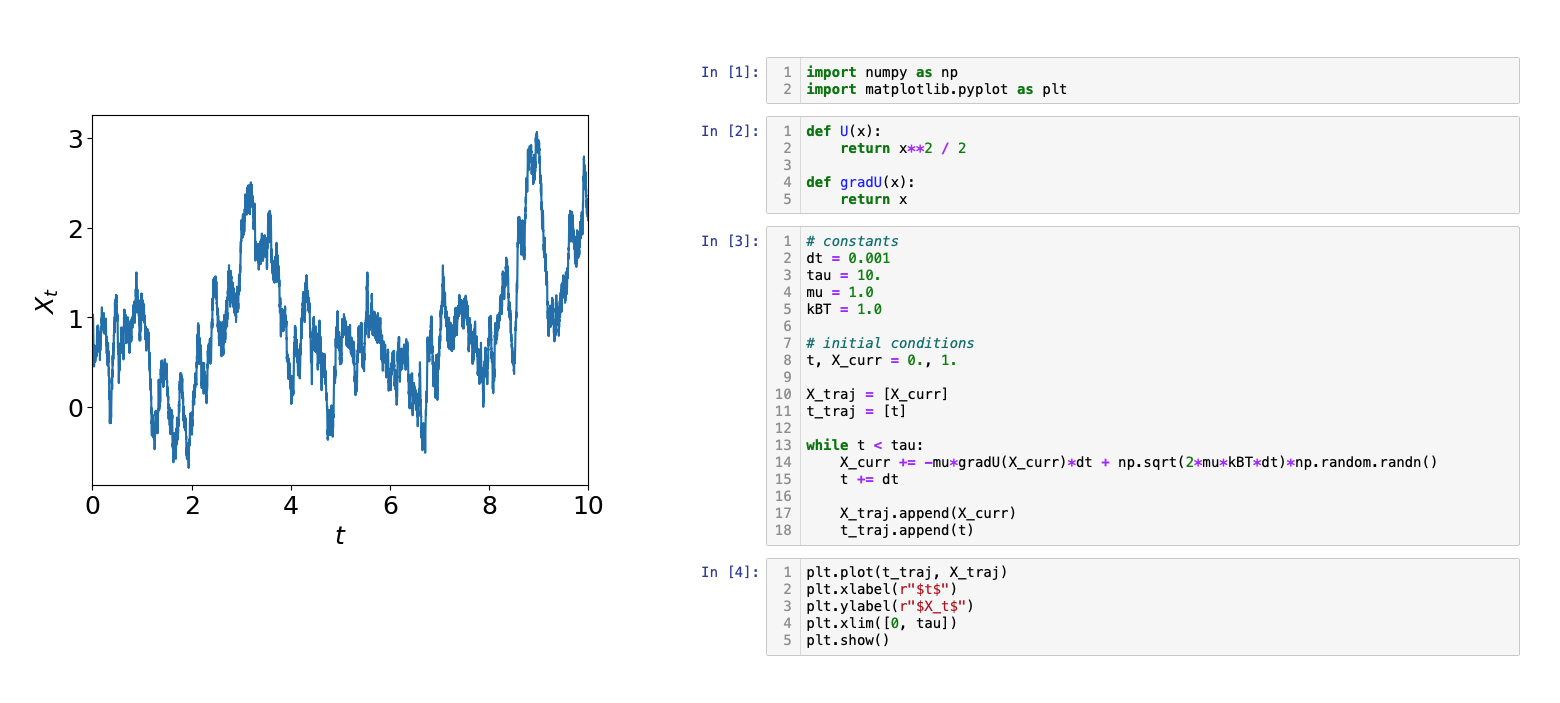}
    \caption[An example Langevin trajectory and Python code.]{An example solution trajectory for the Langevin stochastic differential equation (Eq.~\eqref{eq:langevin-eq}), and the code in Python that generated it. Here, $\mu = \kBT = 1$, and the potential energy function is a quadratic function $U_\lambda(x) = x^2 / 2$. The initial condition is deterministic $X_0 = 1$, which is to say that the initial distribution is a Dirac delta function $\rho_i(x) = \delta(x - 1)$.}
    \label{fig:Langevin-traj}
\end{figure*}

Unlike ``normal'' differential equations (i.e., ones that are deterministic), the Langevin equation yields stochastic trajectories as solutions. (An example solution trajectory is given in Fig.~\ref{fig:Langevin-traj} with $d = 1$ for the quadratic, $\lambda$-independent potential energy $U_\lambda(x) = x^2 / 2$.) Of course, some trajectories are more probable than others. We can consider the set of all possible stochastic trajectories and the probability of realizing each one as the ``statistical ensemble'' of the Langevin equation. 

\section{The Fokker-Planck equation}

At any given time $t$, we can define the probability density $\rho_t(x)$, which is the probability that the stochastic Langevin trajectory is in state $\boldx$ at time $t$. The probability density is normalized so that $\int \rho_t(x) \dx = 1$. 

While the Langevin equation (Eq.~\eqref{eq:langevin-eq}) yields stochastic trajectories, the time evolution for the thermodynamic state trajectory (i.e., the time-dependent configuration space probability density) $\rho_{[0,\tau]}(\boldx)$ is deterministic, given by the Fokker-Planck partial differential equation (PDE) \cite{risken1996fokker}: 
\begin{equation}
    \frac{\partial \rho_t}{\partial t} = \mathcal{L}_{\lambda(t)}[\rho_t] = \boldnabla \cdot \{\rho_t\boldnabla ( \mu U_{\lambda(t)} +  \mu\kBT \ln \rho_t) \}. \label{eq:FP-eq}
\end{equation}
It may be readily seen that the Fokker-Planck operator $\mathcal{L}_\lambda$ is a linear operator, as expanding the seconding term yields $\mathcal{L}_{\lambda}[\rho] = \mu \boldnabla \cdot \{\rho \boldnabla ( \mu U_{\lambda})\} + \mu \kBT \nabla^2 \rho$ which is linear in $\rho$. 

The Fokker-Planck equation (Eq.~\eqref{eq:FP-eq}) may be seen also as a continuity equation, in that an ensemble of trajectories undergoing the deterministic dynamics $\dot{\boldX}_t = \boldb_t (\boldX_t)$, where $\boldb_t(\boldx)$ is a (possibly time-dependent) vector field, has its ensemble density evolve as $\partial_t \rho_t = -\boldnabla \cdot (\rho_t \boldb_t)$.\footnote{In this case of deterministic dynamics, all sources of stochasticity come from the initial condition $\rho_0(\boldx)$.} In Eq.~\eqref{eq:FP-eq}, the two terms in the sum correspond exactly to the two terms in Eq.~\eqref{eq:langevin-eq}. In particular, writing out the Fokker-Planck equation in this form [Eq.~\eqref{eq:FP-eq}] underlines the entropic underpinnings of diffusion: spatial gradients of the log probability density serve as an additional ``potential force'', on equal footing with gradients of the potential $\boldnabla U_\lambda$. In other words, Gaussian white noise $\boldxi_t$ leads to a statistical tendency for trajectories to flow from regions of high (log-)probability to low (log-)probability. 

If $\lambda(t) = \lambda$ were held constant, no matter what the initial thermodynamic state of the system $\rho_0$ is, the system ``relaxes'' to equilibrium, which is given by the \textbf{Boltzmann distribution}
\begin{equation}
    \lim_{t \rightarrow \infty } \rho_t(\boldx) = \rhoeq_{\lambda}(\boldx) := \frac{e^{-U_\lambda(\boldx) / \kBT }}{Z(\lambda) } = e^{ -\{ U_\lambda(\boldx) - F^\mathrm{eq}(\lambda) \} / \kBT }, \label{eq:relaxation}
\end{equation}
where
\begin{equation}
    Z(\lambda) = \int e^{-U_\lambda(\boldx')/\kBT}\dx'
\end{equation}
is the partition function that ensures the probability density normalization $\int \rhoeq_\lambda(\boldx') \dx' = 1$, and 
\begin{equation} \label{eq:free-energy-definition}
    F^\mathrm{eq}(\lambda) := - \kBT \ln Z(\lambda)
\end{equation} 
is known as the \textbf{equilibrium free energy}.\footnote{A proof for Eq.~\eqref{eq:relaxation} comes from considering the nonequilibrium free energy functional $F^\mathrm{neq}_\lambda$ (Eq.~\eqref{eq:F-neq}), which can be shown to be a Lyapunov function for the Fokker-Planck dynamics Eq.~\eqref{eq:FP-eq} with fixed $\lambda$: $\frac{\dee}{\dt} {F}^\mathrm{neq}[\rho_t; \lambda_\mathrm{fixed}] = -\mu \int \rho_t | \boldnabla (U_{\lambda_\mathrm{fixed}} + \kBT \ln \rho_t )|^2 \dx \leq 0$ with uniquely $\dot{F}^\mathrm{neq} = 0$ if and only if $\rho_t = \rhoeq_{\lambda_\mathrm{fixed}}$. Actually, it is possible to show that $\dot{F}^\mathrm{neq} = \dot{S}^\mathrm{tot}$ as defined in Eq.
~\eqref{eq:total-entropy}, which I leave as an exercise for the reader.
} 

The equilibrium energy and equilibrium Gibbs entropy may also be defined
\begin{equation} \label{eq:eq-energy-eq-entropy}
    E^\mathrm{eq}(\lambda) = \int U_{\lambda}(\boldx) \rhoeq_\lambda(\boldx) \dx , \quad\quad\mathrm{and}\quad\quad S^\mathrm{eq}(\lambda) = -\kB \int \rhoeq_\lambda(\boldx) \ln \rhoeq_\lambda(\boldx) \dx , 
\end{equation}
and can be shown to be consistent with the Helmholtz Free Energy definition, 
\begin{equation} \label{eq:helmholtz-free-energy}
    F^\mathrm{eq} = E^\mathrm{eq} - T S^\mathrm{eq},
\end{equation}
which is, in in the author's opinion, quite satisfying.

\subsection{Extending to nonequilibrium}

What it means for a system obeying Langevin dynamics (Eq.~\eqref{eq:langevin-eq}) to be out of equilibrium is that the current thermodynamic state deviates from the equilibrium thermodynamic state $\rho_t(\boldx) \neq \rhoeq_{\lambda(t)}(\boldx)$. This occurs when, e.g., the specified initial thermodynamic state is not in equilibrium ($\rho_0 \neq \rho_{\lambda(0)}$), and/or the protocol $\lambda(t)$ changes quickly.\footnote{Note that this is an issue one also faces in quantum mechanics, i.e., when a Hamiltonian is changed too quickly, the quantum system no longer remains in the ground state.}

For a specified potential energy function $U_\lambda(x)$, it is straightfoward to generalize the energy, Gibbs entropy, and free energy to the nonequilibrium setting for arbitrary thermodynamic states $\rho(x)$ :
\begin{gather} \label{eq:E-S-neq}
  E^\mathrm{neq}[\rho; \lambda] = \int U_\lambda(\boldx) \rho(\boldx) \dx, \quad S^\mathrm{neq}[\rho; \lambda] = -\kB \int \rho(\boldx) \ln \rho(\boldx) \dx, 
\end{gather}
and
\begin{equation} \label{eq:F-neq}
    F^\mathrm{neq}[\rho; \lambda] = E^\mathrm{neq} - T S^\mathrm{neq} = \int \rho(\boldx) \{U_\lambda(\boldx) + \kBT \ln \rho(\boldx) \} \dx.
\end{equation}
The equilibrium definitions are readily recovered as $E^\mathrm{eq}(\lambda) = E^\mathrm{neq}[\rhoeq_\lambda, \lambda]$, etc. 

We note that the nonequilibrium free energy is lower-bounded by the equilibrium energy, in that for any $\rho(x)$:
\begin{equation}
    F^\mathrm{neq} [\rho; \lambda] \geq F^\mathrm{eq}(\lambda),
\end{equation}
with equality if and only if $\rho = \rhoeq_{\lambda}$. It is straightforward to show\footnote{perhaps, as an exercise to the reader!} that the difference between nonequilibrium and equilibrium free energies is
\begin{equation} \label{eq:free-energy-D_KL}
    F^\mathrm{neq}[\rho; \lambda] - F^\mathrm{eq}(\lambda)=  \kBT D_\mathrm{KL}(\rho | \rhoeq_\lambda) \geq 0,
\end{equation}
%
where 
\begin{equation} 
    D_\mathrm{KL}(\rho_A | \rho_B ) = \int \rho_A(x) \ln \frac{\rho_A(x)}{\rho_B(x)} \dx
\end{equation}
is the Kullbeck-Leibler divergence from information theory, which is strictly non-negative. 

\section{Stochastic thermodynamics}

Stochastic thermodynamics is a framework for nonequilibrium statistical mechanics that is consistent in both macroscopic and equilibrium (quasistatic) limits. For Langevin dynamics it was first established by \cite{sekimoto1998langevin}. See \cite{seifert2012stochastic} for an excellent review.  

For the overdamped Langevin Equation (Eq.~\eqref{eq:langevin-eq}), the energy of a stochastic trajectory $\boldX_{[0,\tau]}$ at time $t$ is given by
\begin{equation}
    E_\mathrm{traj}(t) = U_{\lambda(t)} \big( \boldX_t\big), 
\end{equation}
and thus has the total time derivative via the chain rule
\begin{equation}
    \dot{E}_\mathrm{traj}  = 
    {\dot{\lambda}^\mu \frac{\partial U_{\lambda}}{\partial \lambda^\mu}
    }
    + { \dot{\boldX}_t \circ \boldnabla U_{\lambda}.
    }
\end{equation}
We can interpret the two terms 
\begin{equation}
    \dot{W}_\mathrm{traj}(t) =  \dot{\lambda}^\mu(t) \frac{\partial U_{\lambda}}{\partial \lambda^\mu} \quad \quad \mathrm{and} \quad \quad \dot{Q}_\mathrm{traj}(t) =  \dot{\boldX}_t \circ \boldnabla U_{\lambda},  \quad
\end{equation}
as the energy added to the system from modulating the potential energy through changing $\lambda$, and energy coming in from the heat bath---this is the first law of thermodynamics!\footnote{Due to the properties of the Gaussian white noise term in the Langevin equation, we must specify the dot product $\circ$ as a Stratonovich product. I.e., in an evenly-spaced time discretization for the trajectory $\boldX_{[0,\tau]} \rightarrow \{ \boldX_0, \boldX_1, ..., \boldX_N\}$ with timestep $\Delta t = \tau / N$, the latter term $\dot{Q}_\mathrm{traj}$ would be $\big( (\boldX_{n+1} - \boldX_n) /\Delta t \big) \cdot \boldnabla U_{\lambda_{n+1/2}}\big( \boldX_{n+1/2} \big)$, evaluated at the midpoint $\lambda_{n+1/2}$ and $\boldX_{n+1/2} =  (\boldX_{n+1} + \boldX_n) / 2$.} (Figure 1.1 in \cite{crooks1999excursions} illustrates this quite well for a discrete-state system.) Because trajectories are stochastic, depending on both the initial condition $\boldX_0 \sim \rho_i(\cdot)$ and the instantiation of the noise $\boldxi_{[0,\tau]}$, the total work $W_\mathrm{traj}[\boldX_{[0,\tau]}] = \int_0^\tau \dot{W}_\mathrm{traj}(t) \,\dt $ and total heat $Q_\mathrm{traj}[\boldX_{[0,\tau]}] = \int_0^\tau \dot{Q}_\mathrm{traj}(t) \,  \dt$ are also stochastic quantities.\footnote{Quite surprisingly, the \emph{entropy} of a trajectory $S_\mathrm{traj}(t)$ may also be defined as $S_\mathrm{traj}(t) = - k_\mathrm{B} \ln \rho_t \big( \boldX_t \big)$ where $\rho_t$ is the ensemble probability density, and is thermodynamically consistent, in that the expectation of the entropy over an ensemble's trajectories $S(t) = \langle S_\mathrm{traj}(t) \rangle = -k_\mathrm{B} \int  \rho_t(\boldx) \ln \rho_t(\boldx) \dx$, which is the Gibbs entropy of the probability distribution $\rho_t(\cdot)$! This was first established by \cite{seifert2005entropy}.}

While the work and heat added to each trajectory is stochastic, the ensemble average energy, heat, and work ($E(t) := \langle E_\mathrm{traj}(t) \rangle$, $\dot{W}(t) := \langle \dot{W}_\mathrm{traj}(t) \rangle$, and $\dot{Q}(t) := \langle \dot{Q}_\mathrm{traj}(t) \rangle$ ) are determined by the thermodynamic state trajectory $\rho_{[0,\tau]}$:

\begin{equation}
  \begin{aligned} \label{eq:ensemble-energy-work-heat}
   E(t) &= \int U_{\lambda(t)} (\boldx)  \,\rho_t(\boldx) \dx \\ 
   \dot{W}(t) &= \dot{\lambda}^\mu \int  \frac{ \partial U_{\lambda(t)} (\boldx) }{\partial \lambda^\mu} \rho_t(\boldx) \dx \\ 
   \dot{Q}(t)  &= \int U_{\lambda(t)} (\boldx) \, \frac{\partial \rho_t(\boldx)}{\partial t} \dx  \\ 
   \end{aligned}
\end{equation}
and, notably, are \emph{consistent} with macroscopic equilibrium thermodynamics, in the following sense: if the potential energy $U_{\lambda(t)}(\boldx)$ is modulated sufficiently slowly as to be quasistatic (meaning $\rho_t(\boldx) \approx \rhoeq_{\lambda(t)}(\boldx)$), then 
\begin{equation}
    E(t) \approx  E^\mathrm{eq}\big( \lambda(t) \big) := \int U_{\lambda(t)}(\boldx) \,\rhoeq_{\lambda(t)} (\boldx) \dx. 
\end{equation}
It may be seen that partial derivatives of the free energy satisfy
\begin{align}
    \frac{\partial F^\mathrm{eq}}{\partial \lambda^\mu} &= - \kBT \frac{\partial\ln Z(\lambda) }{\partial \lambda^\mu}  = - \kBT \frac{1}{Z(\lambda)} \frac{\partial}{\partial \lambda^\mu} \int  e^{-U_\lambda(\boldx)/\kBT}\dx \nonumber \\ 
    &=  - \kBT \frac{1}{Z(\lambda)} \int  e^{-U_\lambda(\boldx)/\kBT}  \bigg[-\frac{1}{\kBT} \frac{\partial U_\lambda(x)}{\partial \lambda^\mu} \bigg] \dx  \nonumber \\ 
    &= \int \frac{\partial U_\lambda(x)}{\partial \lambda^\mu} \frac{ e^{-U_\lambda(\boldx)/\kBT}}{Z(\lambda)} \dx  \nonumber \\ 
    &=  \int \frac{\partial U_\lambda(x)}{\partial \lambda^\mu} \rhoeq_{\lambda}(x) \dx, \label{eq:F-eq-partial-derivatives}
\end{align}
and so
\begin{equation}
    \dot{F}^\mathrm{eq} = \dot{W}^\mathrm{eq} :=  \dot{\lambda}^\mu \int  \frac{ \partial U_{\lambda(t)} (\boldx) }{\partial \lambda^\mu} \rhoeq_{\lambda(t)}(\boldx) \dx .
\end{equation}
%
Thus, in the quasistatic limit $\rho_t \approx \rhoeq_{\lambda(t)}$, wherein $\lambda(t)$ is changed sufficiently slowly, the work (Eq.~\eqref{eq:ensemble-energy-work-heat}) satisfies 
\begin{equation}
    W \approx \Delta W^\mathrm{eq} = \Delta F^\mathrm{eq}. 
\end{equation}
The heat may be shown to obey the classical isothermic thermodynamic relationship through the first law of thermodynamics $\Delta E^\mathrm{eq} = \Delta W^\mathrm{eq} + \Delta Q^\mathrm{eq} = (\Delta E^\mathrm{eq} - T\Delta S^\mathrm{eq}) + \Delta Q^\mathrm{eq}$ (using the definition of the free energy and that the temperature $T$ is fixed), and so
\begin{equation}
     Q \approx \Delta Q^\mathrm{eq} =  T \Delta S^\mathrm{eq} ,
\end{equation}
where $S^\mathrm{eq}(\lambda)$ is the Gibbs entropy (Eq.~\eqref{eq:eq-energy-eq-entropy}).

It is in this way that Langevin dynamics serves as a fantastic ``under-the-hood'' example of nonequilibrium thermodynamics, and becomes consistent with equilibrium thermodynamics in the quasistatic limit. For the author, thinking about the ensemble of stochastic trajectories in this limit has provided a great deal of insight to equilibrium thermodynamics!

\section{Optimal nonequilibrium protocols}
We are now ready to state the general problem tackled in this PhD thesis: that of optimal nonequilibrium protocols.  

Supposing we insist that the controls start at some specified $\lambda(0) = \lambda_i$, and at some terminal time $t = \tau$ they must be fixed to $\lambda(\tau) = \lambda_f$. What is the \emph{optimal protocol} $\lambda^*(t)|_{t\in[0,\tau]}$ that minimizes the total ensemble expected input work 
\begin{equation}
  W [\lambda(t)]= \int_0^\tau \dot{\lambda}^\mu \bigg\langle \frac{\partial U_\lambda}{\partial \lambda^\mu} \bigg\rangle \,\dt \, = \int_0^\tau \dot{\lambda}^\mu  \bigg\{\int  \frac{ \partial U_{\lambda(t)} (\boldx) }{\partial \lambda^\mu} \rho_t(\boldx) \dx   \bigg\} \dt  ? \label{eq:cost-function}
\end{equation}
The brackets $\langle \cdot \rangle$ denote an ensemble average over the statistical trajectory ensemble, which in this case could be written as an integral over thermodynamic state $\rho_t(x)$. The system is initialized to be in equilibrium $\rho_0 =\rho^\mathrm{eq}_{\lambda_i}$, but does \emph{not} need to end in equilibrium $\rho_\tau \neq \rhoeq_{\lambda_f}$.

\begin{figure*}[t]
    \centering
    \includegraphics[width=.5\linewidth]{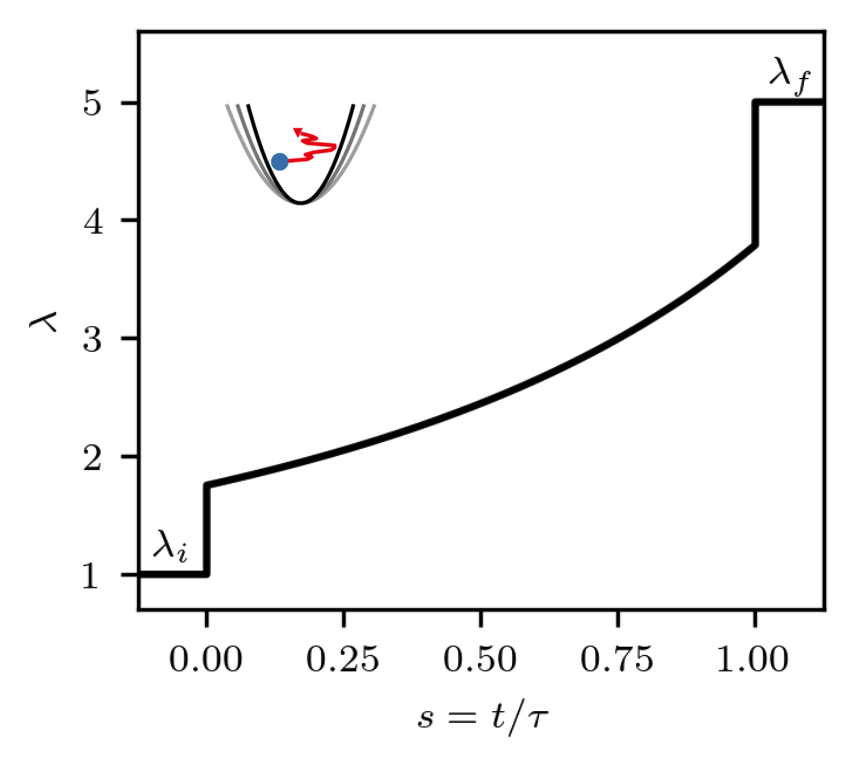}
    \caption[Optimal protocols for a harmonic oscillator have discontinuities!]{The optimal protocol $\lambda^*(t)$ that minimizes the work Eq.~\eqref{eq:cost-function} for the variable-stiffness harmonic oscillator $U_\lambda(x) = \lambda x^2 / 2$; $\lambda_i = 1$; $\lambda_f = 5$, $\mu = \kBT = 1$, and $\tau = 0.2$. There are jumps at $t = 0$ and $t = \tau$; \emph{who ordered that?}}
    \label{fig:jumps}
\end{figure*}

What makes this a challenging problem is that the thermodynamic state of system $\rho_t$ is dependent on the entire history of the protocol up to that point in time $\lambda(t')|_{t' \in [0, t]}$. Furthermore, it has been known that optimal protocols have discontinuities at $t= 0$ and $t = \tau$! For instance, see Fig.~\ref{fig:jumps} for the $d = 1$ harmonic oscillator $U_\lambda(x) = \lambda x^2/ 2$ as discovered in \cite{schmiedl2007optimal} in 2007! This strange feature has turned out to be quite ubiquitous~\cite{schmiedl2007optimal, 
then2008computing, bonancca2018minimal, naze2022optimal, blaber2021steps, zhong2022limited, whitelam2023train, rolandi2023optimal, engel2023optimal, esposito2010finite}.

This is the specific problem we tackle in this thesis. 

\section{Excess work}

In this section, we introduce the so-called ``excess work''
\begin{equation}
    W^\mathrm{ex} = W - W^\mathrm{eq} = W - \big[ F^\mathrm{eq}(\lambda_f) - F^\mathrm{eq}(\lambda_i) \big],
\end{equation}
which is an non-negative quantity that approaches zero for quasistatic processes, as \newline\newline
\textbf{Claim:} it follows from the Second Law of Thermodynamics that the work from a finite-time protocol $W \geq \Delta F^\mathrm{eq} = F^\mathrm{eq}(\lambda_f) - F^\mathrm{eq}(\lambda_i)$, so $W^\mathrm{ex} \geq 0$.  
%
%

This is the mathematical statement that it always takes more work to do something in a finite amount of time than if you do it quasistatically, that is to say, to dilly-dally and take (literally!) forever. \newline\newline
\textbf{Proof:} The Second Law of Thermodynamics states that the total entropy of the composite system-and-environment must increase over time 
\begin{equation}
    \Delta S^\mathrm{tot} = \Delta S^\mathrm{sys} + \Delta S^\mathrm{env} \geq 0. \label{eq:total-entropy}
\end{equation}
The entropy in the system is given by the nonequilibrium Gibbs entropy $S^\mathrm{sys}(t) = S^\mathrm{neq}[\rho_t, \lambda(t)]$ defined in Eq.~\eqref{eq:E-S-neq}. On the other hand, for Langevin dynamics it is assumed that the environment is so big that it is always thermalized at temperature $T$, and so its change in entropy is given by the flow of heat out from the system into the environment 
\begin{equation}
    T \Delta S^\mathrm{env} = Q^\mathrm{sys \to env} = -  Q^\mathrm{env \to sys}. \nonumber 
\end{equation}

By the first law of thermodynamics $\Delta E = W + Q^\mathrm{env \rightarrow sys}$, the work may be rewritten as  
\begin{equation}
  W = - Q^\mathrm{env \rightarrow sys} + \Delta E =T \Delta S^\mathrm{env} +  E(\tau) - E(0) . \nonumber
\end{equation}
We can add and subtract $T\Delta S^\mathrm{sys} = T\{S(\tau) - S(0)\}$ the left hand side to get 
\begin{align}
 W &=   T \big( \Delta S^\mathrm{sys} + \Delta S^\mathrm{env}\big) + \bigg\{ E(\tau) - T S(\tau)  - \big[ E(0) - T S(0) \big] \bigg\} \nonumber \\
 &= T \Delta S^\mathrm{tot} + \bigg\{ F^\mathrm{neq}[\rho_\tau , \lambda_f] -  F^\mathrm{neq}[\rho_0 , \lambda_i] \bigg\} . \nonumber
\end{align}
We are given that the system starts in equilibrium $\rho_0 = \rhoeq_{\lambda_i}$, so the initial free energy is $F^\mathrm{neq}(0) = F^\mathrm{eq}(\lambda_i)$ defined in Eq.~\eqref{eq:eq-energy-eq-entropy}. Using Eq.~\eqref{eq:free-energy-D_KL}, we obtain the final expression 
\begin{equation}
    W = \kBT D_\mathrm{KL}(\rho_\tau | \rhoeq_{\lambda_f}) + F^\mathrm{eq}(\lambda_f) - F^\mathrm{eq}(\lambda_i) + T \Delta S^\mathrm{tot}, \nonumber
\end{equation}
or after some rearranging, 
\begin{equation}
    W - \Delta F^\mathrm{eq} = \underbrace{T \Delta S^\mathrm{tot}}_{\geq 0} +  \underbrace{\kBT D_\mathrm{KL}(\rho_\tau | \rhoeq_{\lambda_f})}_{\geq 0}.\nonumber
\end{equation}
Thus concludes our proof! We shall show throughout the thesis that the total increase in entropy is a fundamentally geometric quantity in terms of the thermodynamic state $\Delta S^\mathrm{tot} = \int_0^\tau \| \dot{\rho}_t\|^2_\mathrm{thermo} \dt$ ...

\section{Thermodynamic geometry}

In 2012, the scientists at UC Berkeley, David Sivak and Gavin Crooks showed through a linear-response argument that, for a large class of thermodynamic systems (of which overdamped Langevin dynamics is just one case), if the the protocol duration is sufficiently long (in comparison with a characteristic thermalization time $\tau \gg \tau_\mathrm{thermal}$), then the excess work may be expressed as  
\begin{equation}
    W^\mathrm{ex}[\lambda(t)] \approx \int_0^\tau |\dot\lambda|^2_\mathrm{friction} \,\dt = \int_0^\tau \dot{\lambda}^\mu \dot{\lambda}^\nu g_{\mu \nu} \big( \lambda(t) \big) \dt, \label{eq:Wex-geometry}
\end{equation}
which becomes exact in the limit $\tau \rightarrow \infty$.\footnote{A more precise statement is $W^\mathrm{ex} = \int_0^\tau |\dot\lambda|^2_\mathrm{friction} \,\dt + o(\tau^{-2})$.} Here, $g_{\mu \nu}(\lambda)$ is known as the \textbf{friction tensor}, as it expresses how much ``excess friction'' is experienced to push the controls from $\lambda^\mu \rightarrow \lambda^\mu + \delta \lambda^\mu$ in this slow, large-$\tau$ limit. It may be calculated in terms of a time-correlation
\begin{equation}
    g_{\mu \nu}(\lambda) = \kBT \int_0^\infty \bigg\langle \frac{\partial \ln \rhoeq_{\lambda}(X_{t'})}{\partial \lambda^\mu}  \,  \frac{\partial \ln \rhoeq_{\lambda}(X_{0})}{\partial \lambda^\mu}  \bigg\rangle^\mathrm{eq}_\lambda \dt',  \label{eq:g-mu-nu-time-correlation}
\end{equation}
where the brackets $\langle \cdot \rangle^\mathrm{eq}_\lambda$ represent an equilibrium average\footnote{which is to say, to run the Langevin dynamics $\dot{X}_t = {- \mu \nabla U_{\lambda}(\boldX_t)} + {\sqrt{2 \mu \kBT} \, \boldxi_t}$ with $X_0 \sim \rhoeq_\lambda(\cdot)$ with the controls $\lambda$ \emph{fixed}.}. Note, due to the definition of $\rhoeq_\lambda$ (Eq.~\eqref{eq:relaxation}), we have 
\begin{equation}
    \frac{\partial \ln \rhoeq_{\lambda}(x)}{ \partial \lambda^\mu} = -\frac{1}{\kBT} \bigg( \frac{\partial U_\lambda(x)}{\partial \lambda^\mu} - \frac{\partial F(\lambda)}{\partial \lambda^\nu}\bigg) = -\frac{1}{\kBT} \bigg( \frac{\partial U_\lambda(x)}{\partial \lambda^\mu} - \bigg\langle\frac{\partial U_\lambda}{\partial \lambda^\mu}\bigg\rangle^\mathrm{eq}_\lambda \bigg),
\end{equation}
which is straightforward to evaluate.

\begin{figure*}[t]
    \centering
    \includegraphics[width=0.4\linewidth]{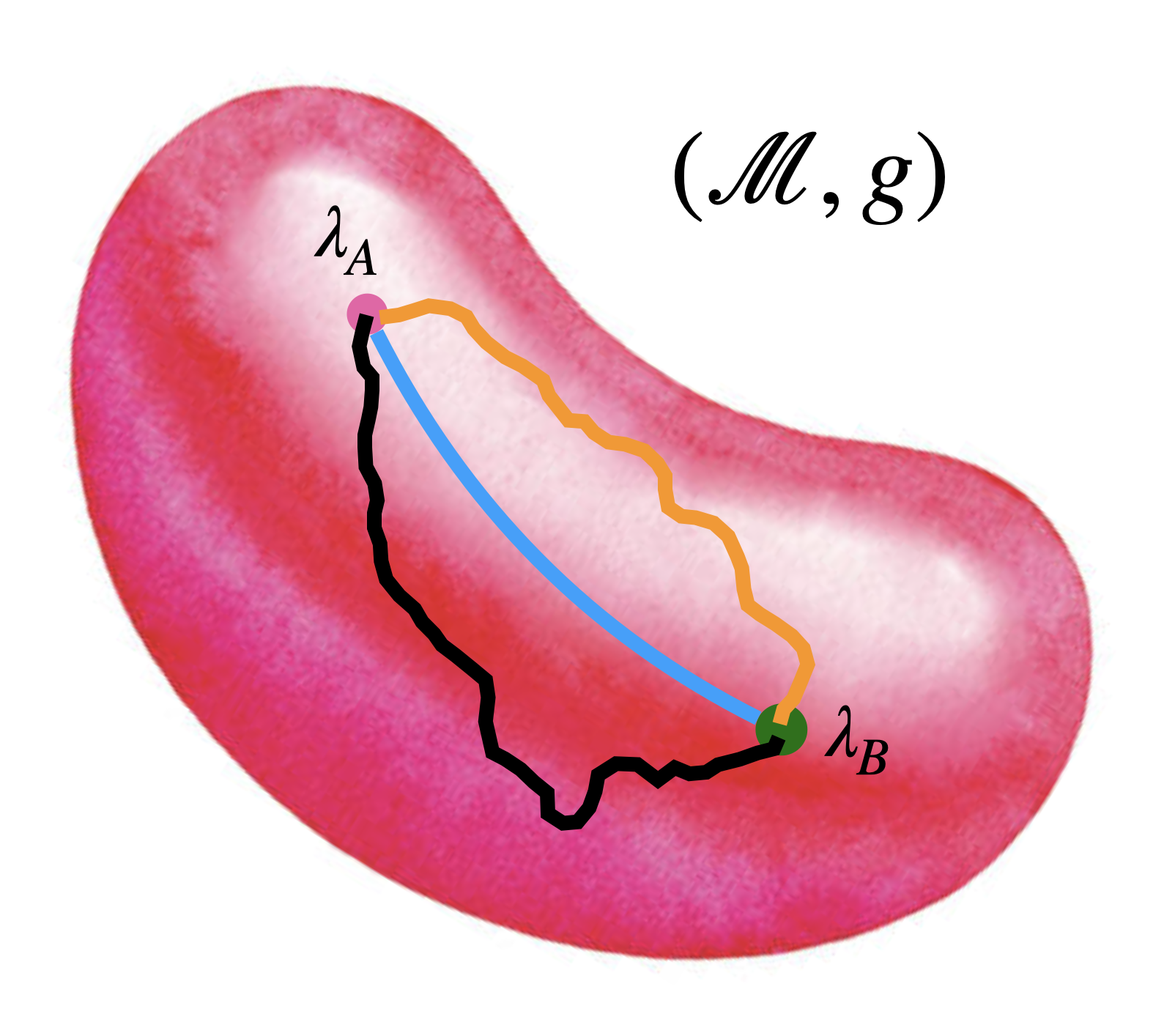}
    \caption[Distances on a Riemannian manifold.]{The thermodynamic distance between two points $\lambda_A, \lambda_B \in \mathcal{M}$ is given by the shortest-length path connecting $\lambda_A$ to $\lambda_B$. The metric tensor $g$ allows the measurement of local lengths $|\delta \lambda|^2_{\lambda} = \delta \lambda^\mu \delta \lambda^\nu g_{\mu \nu}(\lambda)$ for small $\delta \lambda \ll 1$, and is needed to define in the first place the path length $\ell[\lambda(s)] = \int_0^1 \sqrt{|\dot{\lambda}(s)|^2} \ds = \int_0^1 \sqrt{\dot{\lambda}^\mu \dot{\lambda}^\nu g_{\mu \nu}} \ds $. The shortest-length path (here the blue path!) is called the geodesic between $\lambda_A$ and $\lambda_B$. }
    \label{fig:geodesic}
\end{figure*}

What is so intriguing about this result, is that first $g_{\mu \nu}(\lambda) $ can be numerically calculated with Eq.~\eqref{eq:g-mu-nu-time-correlation}, and second that the expression Eq.~\eqref{eq:Wex-geometry} is a \emph{geometric} expression in the following way: the friction tensor $g_{\mu \nu}$, which is demonstrably positive, can be interpreted as a \textbf{Riemannian metric tensor}, which transforms the control space $\lambda \in \mathcal{M}$ into an \textbf{Riemannian manifold} $(\mathcal{M}, g)$; see Fig.~\ref{fig:geodesic}. This allows for the definition of a ``thermodynamic length'' \cite{crooks2007measuring} for a path $\lambda(t) \in \mathcal{M}$ given by 
\begin{equation}
    \ell[\lambda(t)]= \int_0^\tau \sqrt{|\dot{\lambda}(t)|^2_\mathrm{friction}} \, \dt = \int_0^\tau \sqrt{\dot{\lambda}^\mu \dot{\lambda}^\nu g_{\mu \nu} \big( \lambda(t) \big)} \, \dt
\end{equation}
($\sqrt{|\dot{\lambda}(t)|^2_\mathrm{friction}}$ is the ``local velocity''), as well as a (squared) ``thermodynamic distances'' between two points $\lambda_A, \lambda_B \in \mathcal{M}$ 
\begin{align}
    \mathcal{T}(\lambda_A, \lambda_B) = \min_{ \lambda(s)|_{s\in[0,1]}} \bigg\{  \ell[\lambda(s)] \  \bigg| \ \lambda(0)= \lambda_A, \lambda(1) = \lambda_B \bigg\}. 
\end{align}
The path of shortest length $\lambda^*(s)$, i.e., the argmin of the above expression, is known as the \text{geodesic} between $\lambda_A$ and $\lambda_B$.

Furthermore, the (squared) thermodynamic distance may be expressed equivalently as
\begin{align}
    \mathcal{T}^2(\lambda_A, \lambda_B) = \min_{\lambda(s)|_{s\in[0,1]}} \bigg\{ \int_0^1 |\dot{\lambda}(s)|^2_\mathrm{friction} \ds  \  \bigg| \ \lambda(0)= \lambda_A, \lambda(1) = \lambda_B \bigg\}. \label{eq:argmin-thermo-geo}
\end{align}
A particularly nice thing about the second expression is that the argmin (i.e., the shortest possible path $\lambda(s)|_{s\in[0,1]}$ connection $\lambda(0) = \lambda_A$ to $\lambda(1) = \lambda_B$) is the \textbf{constant speed geodesic}.  

What is so remarkable about Sivak and Crooks' result may be seen through comparing their result Eq.~\eqref{eq:Wex-geometry} with Eq.~\eqref{eq:argmin-thermo-geo}. From their expression, the optimal protocol in the large-$\tau$ limit \emph{is} the constant speed geodesic between $\lambda_A$ and $\lambda_B$ on the control manifold $\mathcal{M}$ with $g$ as its the metric tensor. For long protocol durations, observed optimal protocols $\lambda^*(t)$ are indeed geodesics \cite{zhong2022limited}! This gives a foothold into computing optimal protocols, as solving for geodesics on the finite-dimensional control manifold $\mathcal{M}$, instead of considering the full problem on nonequilibrium thermodynamic states. 

However, as elegant as this result is---relating nonequilibrium thermodynamic control to the geometry of curved surfaces---the geometric expression is still approximate for large-$\tau$. 
%
%
%
%
This raises the following questions: is there an exact geometric description for this nonequilibrium control problem? And, how can you possibly explain the jumps found in analytically and numerically obtained optimal protocols, as in Fig.~\ref{fig:jumps} !?

\section{The geometry of optimal transport}

\begin{figure*}[t]
    \centering
    \includegraphics[width=0.5\linewidth]{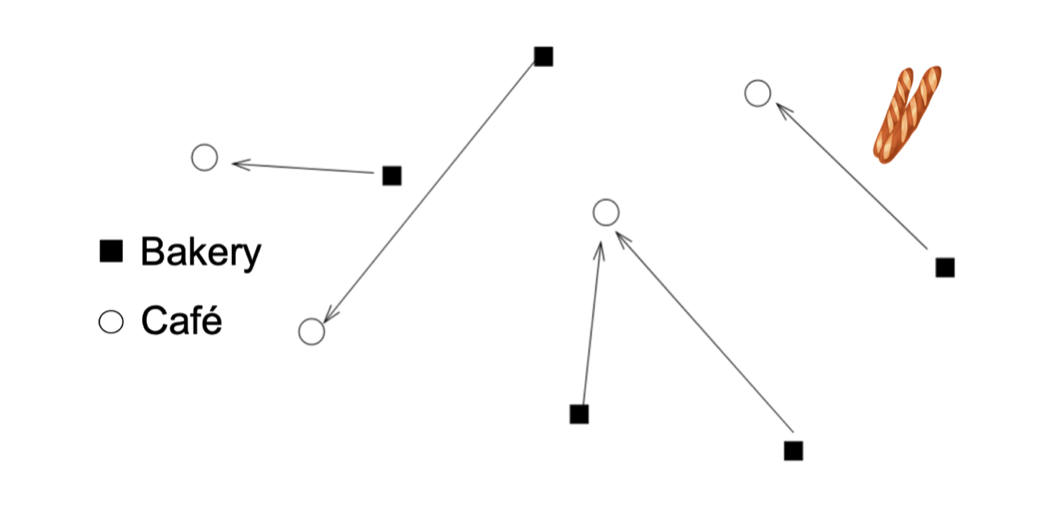}
    \caption[Cartoon of the optimal transport problem.]{The optimal transport problem may be illustrated with: given that there are bakeries producing baguettes and cafes that sell baguettes, what is the transportation map sending the baguettes from bakeries to cafes, subject to the specific supply and demand constraints, that requires the least amount of distance traveled by the baguette? This figure was adapted from \cite{villani2009optimal}.}
    \label{fig:OT-schematic}
\end{figure*}

The last piece of the puzzle in this introduction is optimal transport theory, which is a field of applied mathematics studying the following optimization problem: given two distributions $\rho_A(x)$ and $\rho_B(x)$ with $\int \rho_A(x) \dx = \int \rho_B(x) \dx$, what is the transportation map between thee two distributions that requires the least amount of integrated squared distance traveled? Fig.~\ref{fig:OT-schematic} illustrates this optimization problem in a fun way with the transportation of baguettes.\footnote{In this fun example, $\rho_A$ would correspond to the ``supply distribution'' in the bakeries, while $\rho_B$ would correspond to the ``demand distribution'' in the cafes.} From this optimal transport problem, the \textbf{Wasserstein metric distance} could be defined between distributions $ \mathcal{W}(\rho_A, \rho_B)$ whose value is the minimum total transported distance that is required. I highly recommend \cite{villani2021topics} as a great resource to learn optimal transport.\footnote{Thank you Malcolm Lazarow for introducing me to Cédric Villani!}

Surprisingly, Jean-David Benamou and Yann Brenier discovered in 2000 \cite{benamou2000computational} that the Wasserstein metric distance is also a geodesic distance, in the following sense: Think about moving a pile of sand from one configuration described by the density function $\rho_A(x)$ to another described by $\rho_B(x)$. Now considering the movie of the moving sandpile $\rho_t(x)$, if it happens to be that the motion of each individual grain of sand (indexed by $i$) can be described as the the gradient of a scalar field $\dot{X}^{(i)}_t = -\nabla \Phi_t (X^{(i)}_t)$,\footnote{This implies the continuity equation as the dynamics for the distribution $\partial_t \rho_t = \nabla \cdot (\rho_t \nabla \Phi_t)$.} then the total instantaneous squared motion is
\begin{equation} 
 \| \dot{\rho}_t\|^2_\mathrm{transport} =  \sum_{\mathrm{sand} \ i} \big|\dot{X}^{(i)}_t \big|^2.
\end{equation}
Benamou and Brenier discovered that the squared Wasserstein distance may also be expressed as:
\begin{equation}
    \mathcal{W}^2 (\rho_A, \rho_B) = \min_{\rho_s|_{s \in [0, 1]}} \bigg\{ \int_0^1 \| \dot{\rho}_s \|^2_\mathrm{transport} \, \ds \ \bigg| \  \rho_0 = \rho_A, \rho_1 = \rho_B \bigg\}. 
\end{equation}
This has the very intuitive interpretation, that by considering $\rho_A$ and $\rho_B$ as two separate piles of baguettes, over all possible movies of traveling baguettes, the Wasserstein distance is given by the one movie in which the total distance traveled by all the transported baguettes is minimized.

This gives a geometric structure to the space of probability distributions $\rho \in \mathcal{P}(\mathbb{R}^d)$, as giving a metric distance between pairs of probability distributions, as well as allowing to characterizing the ``velocity'' and ``distance traveled'' of a time-dependent probability distribution, in a relatively intuitive way. 

\subsection{Relationship to optimal protocols}

In 2011, Aurell et al \cite{aurell2011optimal} showed that for a nonequilibrium Langevin equation the total entropy production (Eq.~\eqref{eq:total-entropy}) can be expressed in terms of thermodynamic state $\rho_t(x)$ that solves the Fokker-Planck equation Eq.~\eqref{eq:FP-eq} as
\begin{equation}
    \Delta S^\mathrm{tot} = \frac{1}{ \mu \kBT } \int_0^\tau \|  \dot{\rho}_t \|^2_\mathrm{transport} \dt.
\end{equation}
Unlike the Sivak and Crooks result which states that excess work ($W^\mathrm{ex} \approx \int_0^\tau |\dot{\lambda}|^2_\mathrm{friction} \dt$) is \emph{approximately} geometric for the control parameter trajectory $\lambda(t)|_{t \in [0, \tau]}$ under the friction tensor geometry on the control manifold $\lambda \in \mathcal{M}$, Aurell et al found that the total entropy production is \emph{exactly} geometric for the time-dependent thermodynamic state probability distribution $\rho_t|_{t \in [0, \tau]}$ in the transport geometry on the space of all probability distributions defined on configuration space $\rho (x) \in \mathcal{P}(\mathbb{R}^d)$. 

This is a strikingly elegant result: indeed there \emph{is} an exact geometric description for the entropy produced in optimal protocols! However, it does not give a good prescription for how to obtain an optimal protocol $\lambda(t)$, in that the optimal transport optimization over probability distributions $\rho_A(x)$ and $\rho_B(x)$ runs into the curse of dimensionality ...

\section{Main results and thesis outline} 

All in all, this thesis attempts to address the question that opened this introduction: ``What is the \emph{laziest way possible} to do something, subject to the fundamental laws of physics?'', through exploring the optimal protocols problem for the overdamped Langevin equation, and discovering the surprising fundamental geometric structure behind optimal nonequilibrium processes. 

The following three chapters of the thesis are closely adapted from the author's published works:\footnote{Unfortunately, due to time constraints I cannot guarantee consistent notation between chapters. However, the notation should be consistent within a chapter.}

\begin{itemize}
    \item[2.] \textbf{Optimal Protocols Arbitrarily Far from Equilibrium: Pontryagin's Principle on the Fokker-Planck Equation.} Here we develop an algorithm for numerically calculating globally optimal protocols, based on optimal control theory of the thermodynamic state. The big surprise is that there exist optimal protocols that are \emph{non-monotonic}! The algorithm simulates the Fokker-Planck equation, which has the shortcoming becoming computationally intractable for high dimensional systems due to the curse of dimensionality.
    \item[3.] \textbf{Efficient Free Energy Estimation from the Time-Asymmetric Work.} This chapter explores a specific application of the optimal protocols problem, namely estimating the equilibrium free energy difference $\Delta F^\mathrm{eq}$ between two different potential energy functions $U_A(x)$ and $U_B(x)$ through nonequilibrium protocols. We introduce the notion of counterdiabatic driving: what should the controlled protocol $\lambda(t)$ be so that the thermodynamic state follows a desired trajectory $\rho_t = \rho_\mathrm{desired}$? We show that with counterdiabatic driving, a modified definition of the work we call the ``time-asymmetric work'' satisfies certain properties that allows for a significantly lower variance estimate for $\Delta F^\mathrm{eq}$. We propose a efficient, numerically tractable adaptive protocol optimization algorithm that does not suffer from the curse of dimensionality, based on importance sampling the Langevin equation under different protocols. 
    \item[4.] \textbf{Equivalence between Thermodynamic Geometry and Optimal Transport.} This chapter presents the main result of this thesis: we demonstrate that the Sivak and Crooks friction tensor-based thermodynamic geometry is equivalent to optimal transport geometry through the equality $|\dot{\lambda}|^2_\mathrm{friction} = \| \dot{\rho}^\mathrm{eq}_{\lambda(t)} \|^2_\mathrm{transport}$. From it, we find that optimal protocols are of geodesic-counterdiabatic form: $\lambda^*(t) = \gamma(t) + \eta(t)$, where $\gamma(t)$ is a friction tensor geodesic, and $\eta(t)$ gives a counterdiabatic forcing that is needed to ensure that the thermodynamic state follows the geodesic $\rho_t = \rhoeq_{\gamma(t)}$. In this way, optimization may be done purely on $\mathcal{M}$, as opposed to the (infinite-dimensional) space of all possible thermodynamic states. This explains the discontinuities in  Fig.~\ref{fig:jumps}! 
\end{itemize}
Finally, we end the thesis with

\begin{itemize}
    \item[5.] \textbf{Conclusions.} Here we give a brief conclusion, and sketch out work that is \emph{almost} done but has not been able to make its way into this thesis. 
\end{itemize}
Also included in this thesis are the Appendices: 

\begin{itemize}
    \item[A.] \textbf{Physical Derivation of the Langevin Equation.} Here we derive the overdamped Langevin equation (Eq.~\eqref{eq:langevin-eq}) for a system, from reversible Hamiltonian dynamics of the system coupled to a large bath of harmonic oscillators. From the unknown initial conditions of the harmonic oscillator bath, irreversibility arises! 
    \item[B.] \textbf{Counterdiabatic Forcing for Arbitrary Geometries.} Here we characterize counterdiabatic forcing when the configuration space of the thermodynamic system is \emph{not} Euclidean space. We use the Hodge decomposition from differential geometry to characterize counterdiabatic vector fields for thermodynamic systems defined on other configuration spaces that are non-Euclidean Riemannian manifolds, and illustrate our results with two $2$-dimensional toy examples: a sphere and torus.
\end{itemize}
To thee, dear reader, I wish a happy reading!

\newcommand{\boldrho}{{\boldsymbol{\rho}}}
\newcommand{\boldU}{{\boldsymbol{U}}}
\newcommand{\boldpi}{{\boldsymbol{\pi}}}
\newcommand{\boldpsi}{{\boldsymbol{\psi}}}
\newcommand{\curlyL}{{\mathcal{L}}}
\newcommand{\curlyW}{{\mathcal{W}}}

\chapter{Optimal Protocols Arbitrarily Far from Equilibrium: Pontryagin's Principle on the Fokker-Planck Equation}

Recent studies have explored finite-time dissipation-minimizing protocols for stochastic thermodynamic systems driven arbitrarily far from equilibrium, when granted full external control to drive the system. However, in both simulation and experimental contexts, systems often may only be controlled with a limited set of degrees of freedom. Here, going beyond slow- and fast-driving approximations employed in previous studies, we obtain exact finite-time optimal protocols for this unexplored limited-control setting. By working with deterministic Fokker-Planck probability density time evolution, we can frame the work-minimizing protocol problem in the standard form of an optimal control theory problem. We demonstrate that finding the exact optimal protocol is equivalent to solving a system of Hamiltonian partial differential equations, which in many cases admit efficiently calculatable numerical solutions. Within this framework, we reproduce analytical results for the optimal control of harmonic potentials, and numerically devise novel optimal protocols for two anharmonic examples: varying the stiffness of a quartic potential, and linearly biasing a double-well potential. We confirm that these optimal protocols outperform other protocols produced through previous methods, in some cases by a substantial amount. We find that for the linearly biased double-well problem, the mean position under the optimal protocol travels at a near-constant velocity. Surprisingly, for a certain timescale and barrier height regime, the optimal protocol is also non-monotonic in time. 



\subsection{Note}

The work in this chapter is based off of: Adrianne Zhong and Michael R DeWeese. ``Limited-control optimal protocols arbitrarily far from equilibrium''. In: \emph{Physical Review E} 106.4 (2022), p. 044135.'' This project was born out of a curiosity behind a deeper structure for optimal nonequilibrium control protocols beyond the Riemanniann geometry obtained via a linear response approximation (pioneered by \cite{sivak2012thermodynamic}), and many ways was the first research project that I saw myself to complete.

If there is anything I would change in retrospect, and of course this is with three years more of knowledge, is that it was completely unnecessary to absorb the boundary conditions into the integral Eq.~\eqref{eq:work-integral}, but could be kept as $W[\lambda(t)] = (\boldU_f^T \boldrho(\tau) - \boldU_{i}^T \boldrho_{\mathrm{eq},i} ) - \int_0^{\tau} \boldU_\lambda^T \mathcal{L}_\lambda \boldrho \, \dt $, with instead the terminal condition for the conjugate momentum $\boldpi(\tau) = - \boldU_f$. (In fact, by adding the additional term to get $W[\lambda(t)] = (\boldU_f + \ln \boldrho(\tau))^T  \boldrho(\tau) - ( \boldU_{i} + \ln \boldrho_{\mathrm{eq},i})^T \boldrho_{\mathrm{eq},i}  - \int_0^{\tau} (\boldU_\lambda + \ln \boldrho )^T \mathcal{L}_\lambda \boldrho \, \dt$, with the logarithm taken elementwise, the integral can be shown to be fully geometric, as the path divergence $\int_0^{\tau} (\boldU_\lambda + \ln \boldrho )^T \mathcal{L}_\lambda \boldrho \, \dt = \int_0^\tau \| \dot{\boldrho} \|^2_\mathcal{W_1} \,  \dt$ under a discrete-state Wasserstein metric \cite{van2022thermodynamic}!)

Of course, this is all in retrospect---in the end, I am proud to finally have had seen a research project to completion.

\section{Chapter introduction}
There has been much recent progress in the study of non-equilibrium stochastic thermodynamics \cite{seifert2012stochastic, jarzynski2013nonequilibrium, ciliberto2017experiments}. In particular, optimal finite-time protocols have been derived for a variety of systems, with applications to
finite-time free-energy difference estimation \cite{schmiedl2007optimal, blaber2020skewed, gomez2008optimal} engineering optimal bit erasure \cite{proesmans2020optimal, zulkowski2014optimal}, and the design of optimal nanoscale heat engines \cite{blickle2012realization, frim2022optimal, frim2022geometric}. 

For finite-time dissipation-minimizing protocols, there are two related optimization problems that are typically studied: designing protocols that transition between two specified distributions within finite time that minimize entropy production \cite{aurell2011optimal, dechant2019thermodynamic, nakazato2021geometrical, chen2019stochastic}, and designing protocols that minimize the amount work needed to shift between two different potential energy landscapes within finite time \cite{schmiedl2007optimal, sivak2012thermodynamic}. For the first problem, methods have been devised to fully control probability density evolution arbitrarily far from equilibrium  \cite{frim2021engineered, ilker2021counterdiabatic, martinez2016engineered}, establishing deep ties with optimal transport theory \cite{aurell2011optimal, villani2009optimal, dechant2019thermodynamic} and culminating in the derivation of an absolute geometric lower bound for finite-time entropy production in terms of the $L^2$-Wasserstein distance \cite{dechant2019thermodynamic, nakazato2021geometrical, chen2019stochastic, chennakesavalu2022unifying}. Crucially, however, full control over the potential energy is needed to satisfy arbitrarily specified initial and terminal conditions for this problem. 

Here, we consider the second problem for the case in which there is only limited, finite-dimensional control of the potential. Only for the simplest case of a Brownian particle in a harmonic potential has the fully non-equilibrium optimal protocol been analytically solved and studied \cite{schmiedl2007optimal, aurell2011optimal, then2008computing, plata2019optimal}. For arbitrary potentials, limited control optimal protocol approximations for the slow near-equilibrium $\tau \gg 1$ \cite{sivak2012thermodynamic, sivakcrooksbarriercrossing, zulkowski2012geometry, rotskoff2015optimal, lucero2019optimal, deffner2020thermodynamic, abiuso2022thermodynamics} and the fast $\tau \ll 1$ \cite{blaber2021steps} regimes have been derived, but these approximations generally are optimal only within the specified limits. Very recently, gradient methods have been devised to calculate fully non-equilibrium optimal protocols through sampling many stochastic trajectories \cite{automaticdifferentiation, yan2022learning, das2022direct}. 

In this work, we show that optimal control theory is a principled and powerful framework to derive exact optimal protocols for limited-control potentials arbitrarily far from equilibrium. Optimal control theory (OCT), having roots in Lagrange's calculus of variations, is a well-studied field of applied mathematics that deals with finding controls of a dynamical system that optimize a specified objective function, with numerous applications to science and engineering \cite{liberzon2011calculus, lenhart2007optimal}, including experimental physics \cite{bechhoefer2021control}. By working directly with the probability density undergoing deterministic Fokker-Planck dynamics (as opposed to individual stochastic trajectories), and rewriting the objective function using the first law of thermodynamics, we show that the problem of finding optimal protocols can be recast in the standard OCT problem form. We may then apply Pontryagin’s maximum principle, one of OCT's foundational theorems, to yield Hamiltonian partial differential equations whose solutions directly give optimal protocols. We note that the optimal control of fields and stochastic systems has been previously studied within applied mathematics and engineering literature \cite{bakshi2020open, annunziato2013fokker, fattorini1999infinite, palmer2011hamiltonian, evans2021spatio, theodorou2012stochastic, fleig2017optimal, annunziato2010optimal, popescu2010existence, chernyak2013stochastic}, but to our knowledge it has never been used to derive exact optimal work-minimizing protocols in stochastic thermodynamics. 

An outline of this paper is as follows. First, we use OCT to derive Hamiltonian partial differential equations whose solutions give optimal protocols for the cases of Markov jump processes over discrete states and Langevin dynamics over continuous configuration space. We then solve these equations analytically for harmonic potential control to reproduce known optimal protocols. Finally, we describe and use a computationally efficient algorithm to numerically calculate optimal protocols for two anharmonic examples: controlling the stiffness of a quartic trap, and linearly biasing a quartic double-well potential. We demonstrate the superiority in performance of these optimal protocols compared to the protocols derived through approximation methods. We discover that for the linearly biased double-well problem, the mean position travels with near-constant velocity under the optimal protocol, and that certain optimal protocols have a remarkably counter-intuitive property --- the control parameter is non-monotonic in time within a certain time and barrier height parameter regime. Finally, we discuss our findings and the implications of our work for the study of non-equilibrium stochastic thermodynamics. 

\section{\label{sec:setup}Discrete state derivation}

We start by considering a continuous-time Markov jump process with $d$ discrete states. The experimenter has control over the protocol parameter $\lambda(t)$ that determines the potential energies of the states, encoded by the vector $\boldU_{\lambda} = (U_1(\lambda), U_2(\lambda), ... , U_d(\lambda)))^T$. Here $\lambda$ is single parameter, but in general it can be multi-dimensional. Although an individual jump process trajectory is stochastic, the time-varying probability distribution over states, represented by the vector $\boldrho(t) = (\rho^1, \rho^2, ... , \rho^d)^T$ with $\sum_i \rho^i = 1$, has deterministic dynamics governed by a master equation 

\begin{equation}
    \dot{\boldrho} = \mathcal{L}_{\lambda} {\boldrho}, \label{eq:master-equation}
\end{equation}
where $\mathcal{L}_\lambda$ is a transition rate matrix for which we impose the following form (similar to \cite{sohl2009minimum})

\begin{equation}
    {[\mathcal{L}_\lambda]^i}_{j} = 
    \begin{cases} 
    c_{ij} e^{\beta (U_j(\lambda) - U_i(\lambda))/2 }
    & {i \neq j} \\
    - \sum_{k \neq j}\ c_{kj}  e^{\beta (U_j(\lambda) - U_k(\lambda))/2 }
    & {i = j} .
  \end{cases} \label{eq:transition-rate-matrix}
\end{equation}
Here $\beta = 1/k_B T$ is the inverse temperature, $k_B$ is the Boltzmann constant, and $c_{ij} = c_{ji}$ is the symmetric non-negative connectivity strength between distinct states $i \neq j$. Transition rate matrices have the property $\sum_i {[{\mathcal{L}_\lambda}]^i}_{j} = 0$, ensuring conservation of total probability. In particular, this matrix ${\mathcal{L}_\lambda}$ satisfies the detailed-balance condition ${[{\mathcal{L}_\lambda}]^i}_{j} \rho_{\mathrm{eq}, \lambda}^j = {[{\mathcal{L}_\lambda}]^j}_{i} \rho_{\mathrm{eq}, \lambda}^i$ for all $i$ and $j$, where $\rho_{\mathrm{eq}, \lambda}^i \propto e^{-\beta U_i(\lambda)}$ is the unique Boltzmann equilibrium distribution for $\boldU_\lambda$.

For time-varying $\lambda(t)$ and $\boldrho(t)$, the ensemble-averaged energy is  $E(t) = \boldU_\lambda^T \boldrho$ and has time derivative 

\begin{equation}
  \dot{E} = \dot{\lambda} \bigg[\frac{d \boldU_\lambda}{d \lambda} \bigg]^T \boldrho + \boldU_\lambda^T \dot{\boldrho} . 
\end{equation}
As is customary in stochastic thermodynamics, the first term in the sum is interpreted as the rate of work applied to the system $\dot{W}$, and the second term the rate of heat in from the heat bath $\dot{Q}$ \cite{stochthermofunctionals}. 

We would like to solve the following optimization problem: if at $t = 0$ we start at the equilibrium distribution $\boldrho_{\mathrm{eq}, i}$ for potential energy $\boldU_{\lambda_i}$, what is the optimal finite-time protocol $\lambda(t)$ that terminates at $\lambda_f$ at final time $t = \tau$, and minimizes the work

\begin{equation}
    W[\lambda(t)] = \int_0^{\tau} \dot{\lambda} \, \bigg\langle \frac{\partial U}{\partial \lambda} \bigg\rangle \dt = \int_0^{\tau}\dot{\lambda} \bigg[\frac{d \boldU_\lambda}{d \lambda} \bigg]^T \boldrho \dt \, ? \label{eq:work}
\end{equation}
We emphasize that this time integral includes any discontinuous jumps of $\lambda$ that may occur at the beginning and end of the protocol, which has been shown to be a common feature for finite-time optimal protocols \cite{schmiedl2007optimal, boundary-layers, blaber2020skewed}. Note that in general, $\boldrho(\tau) \neq \boldrho_{\mathrm{eq}, f}$ the equilibrium distribution corresponding to $\lambda_f$.

The first law of thermodynamics $\Delta E[\lambda(t)] = W[\lambda(t)] + Q[\lambda(t)]$ allows us to write 

\begin{align}
    W[\lambda(t)] &= (\boldU_f^T \boldrho(\tau) - \boldU_{i}^T \boldrho(0)) - \int_0^{\tau} \boldU_\lambda^T \dot{\boldrho} \dt \nonumber \\ 
    &= (\boldU_f - \boldU_{i})^T \boldrho_{\mathrm{eq}, i} + \int_0^{\tau} (\boldU_f -  \boldU_\lambda)^T \mathcal{L}_\lambda \boldrho \dt. \label{eq:work-integral}
\end{align}
Here, $\boldU_{i} = \boldU_{\lambda_i}$, and $\boldU_{f} = \boldU_{\lambda_f}$. In the second line, we use $\boldrho(\tau) = \boldrho(0) + \int_0^{\tau} \dot{\boldrho} \dt$, and invoke \eqref{eq:master-equation}. The first term in the sum is protocol independent, so minimizing $W[\lambda(t)]$ is akin to minimizing the second term

\begin{equation} \label{eq:costfunction}
    J[\lambda(t)] = \int_0^{\tau} (\boldU_f -  \boldU_\lambda)^T \mathcal{L}_\lambda \boldrho \dt, 
\end{equation}
which is now in the form of the fixed-time, free-endpoint Lagrange problem in optimal control theory \cite{liberzon2011calculus}. Compared to a typical Euler-Lagrange calculus of variations problem in classical physics \cite{taylor2005classical, jose2000classical}, here both the initial state $\boldrho(t = 0) = \boldrho_{\mathrm{eq},i}$ and the time interval $[0, \tau]$ are specified, but notably, the final state $\boldrho(t = \tau)$ is unconstrained.

The standard OCT solution derivation begins by expanding the integrand of \eqref{eq:costfunction} with Lagrange multipliers $\boldpi(t) = (\pi_1, \pi_2, ... , \pi_d)^T$

\begin{equation} \label{eq:lagrangian}
    L = (\boldU_f -  \boldU_\lambda)^T \mathcal{L}_\lambda \boldrho + \boldpi^T (\dot{\boldrho} - \mathcal{L}_\lambda \boldrho), 
\end{equation}
so that the desired dynamics \eqref{eq:master-equation} are ensured. A solution $[\boldrho^*(t), \boldpi^*(t), \lambda^*(t)]$ that minimizes $\int_0^{\tau} L \dt$ gives the optimal protocol $\lambda^*(t)$ that minimizes $J[\lambda(t)]$.

A Legendre transform $H = \boldpi^T \dot{\boldrho} - L$ produces the control-theoretic Hamiltonian

\begin{equation}
    H(\boldrho, \boldpi, \lambda) = (\boldpi + \boldU_\lambda - \boldU_f)^T \mathcal{L}_\lambda \boldrho, \label{eq:hamiltonian} 
\end{equation}
where $\boldpi$ may now be interpreted as the conjugate momentum to $\boldrho$. Pontryagin's maximum principle gives necessary conditions for an optimal solution $[\boldrho^*(t), \boldpi^*(t), \lambda^*(t)]$: it must satisfy the canonical equations $\dot \rho^i = \partial{H} / \partial \pi_i$ and $\dot{\pi_i} = - \partial{H} / \partial \rho^i$ for $i = 1, 2, ... , d$, and constraint equation $\partial{H} / \partial \lambda = 0$, with $\partial^2 {H} / \partial \lambda^2 < 0$ along the optimal protocol. Because Eq.~\eqref{eq:hamiltonian} has no explicit time dependence, it remains constant throughout an optimal protocol.  Although this is in a sense analogous to the conserved total energy in a classical system, it does not apparently represent a physical energy of the system \cite{liberzon2011calculus}. 

From Pontryagin's maximum principle, the canonical equations for the Hamiltonian in Eq.~\eqref{eq:hamiltonian} are

\begin{align}
  \dot{\boldrho} &= \mathcal{L}_\lambda \boldrho \label{eq:canonical-discrete-rho} \\ 
  \dot{\boldpi} &= - \mathcal{L}_\lambda^T (\boldpi + \boldU_\lambda - \boldU_f) \label{eq:canonical-discrete-pi},
\end{align}
while the constraint equation coupling the two canonical equations is

\begin{align}
  \bigg( \bigg[\frac{d \boldU_\lambda}{d \lambda} \bigg]^T \mathcal{L}_\lambda + (\boldpi + \boldU_\lambda - \boldU_f)^T \frac{d \mathcal{L}_\lambda}{d \lambda} \bigg) \boldrho \label{eq:constraint-discrete} = 0. 
\end{align}
Because $\boldrho(\tau)$ is unconstrained, the transversality condition fixes the terminal conjugate momentum $\boldpi(\tau) = \boldsymbol 0$ \cite{liberzon2011calculus}\footnote{Alternatively, these equations are obtainable through deriving the Euler-Lagrange equations for the Lagrangian $L(\boldrho, \dot{\boldrho}, \boldpi, \dot{\boldpi}, \lambda, \dot{\lambda}) = (\boldU_f -  \boldU_\lambda)^T \curlyL_\lambda \boldrho + \boldpi^T (\dot{\boldrho} - \curlyL_\lambda \boldrho)$. The transversality condition comes from an extra boundary term when performing the integration by parts to transform $\int_0^{\tau} (\partial L / \partial \dot{\boldrho}) \, \delta \dot{\boldrho} \, dt = \int_0^{\tau} -(\partial L / \partial \dot{\boldrho}) \, \delta \boldrho \, dt + (\partial L / \partial \dot{\boldrho}) \, \delta \boldrho |^{\tau}_{0}$. Because $\boldrho(\tau)$ is unconstrained, the variation at endpoint $\delta \boldrho(\tau)$ may be arbitrary, and thus we have from the boundary term that at optimality, $\partial L / \partial \dot{\boldrho}|_{\tau} = \boldpi(\tau) = \boldsymbol{0}$. Within the calculus of variations, this is known as a natural boundary condition.}.

We have arrived at our first major result in this chapter. For a discrete state Markov jump process satisfying detailed balance, Pontragin's maximum principle allows us to find the work-minimizing optimal protocol $\lambda^*(t)$ by solving the canonical differential Eqs.~\eqref{eq:canonical-discrete-rho} and \eqref{eq:canonical-discrete-pi} coupled by Eq.~\eqref{eq:constraint-discrete}, with the mixed boundary conditions $\boldrho(0) = \boldrho_i$, $\boldpi(\tau) = \boldsymbol{0}$. Notably, no approximations have been used here, and thus the optimal protocols produced within this framework are exact for any time-scale. As will be shown below, efficient algorithms may be written to numerically solve these ordinary differential equations. This will be useful for numerically solving for optimal protocols of a continuous-state stochastic system, as continuous-state Fokker-Planck dynamics may be approximated by a discrete state Markov process with the appropriate master equation \cite{holubec2019physically, zwanzig2001nonequilibrium}. All that remains in our derivation is to take the continuum limit for the corresponding result for a continuous stochastic system undergoing Langevin dynamics. 

\section{Continuous space derivation}

For a continuous-state overdamped system in one dimension, individual trajectories undergo dynamics given by the Langevin equation 

\begin{equation}
    \dot{x} = - \beta D \frac{\partial U}{\partial x} + \eta(t). \label{eq:langevin}
\end{equation}
Here $D$ is the diffusion coefficient, $U(x, \lambda)$ is the $\lambda$-controlled potential, and $\eta(t)$ is Gaussian white noise with statistics $\langle \eta(t)\eta(t') \rangle = 2 D \delta(t - t')$. 

While each individual trajectory is stochastic, the time evolution of the probability density $\rho(x, t)$ of the ensemble is deterministic, given by a Fokker-Planck equation 

\begin{equation}
  \frac{\partial \rho}{\partial t} =  D \bigg[ \frac{\partial^2 \rho }{\partial x^2} + \beta \frac{\partial}{\partial x}\bigg(\rho  \frac{\partial U}{\partial x} \bigg)\bigg] =: \hat{\mathcal{L}}_\lambda \rho , \label{eq:fokker-planck}
\end{equation}
Here, $\hat{\mathcal{L}}_\lambda$ denotes the Fokker-Planck operator, which has a corresponding adjoint operator $\mathcal{L}_\lambda^\dagger$, also known as the backward Kolmogorov operator \cite{risken1996fokker, zwanzig2001nonequilibrium}, that acts on a function $\psi(x, t)$ as

\begin{equation}
  \mathcal{L}^{\dagger}_\lambda \psi := D \bigg[ \frac{\partial^2 \psi }{\partial x^2} - \beta \frac{\partial \psi}{\partial x} \frac{\partial U}{\partial x}\bigg] . \label{eq:adjoint-fokker-planck}
\end{equation}

Again, we want to find a protocol $\lambda(t)$ that minimizes the expected work 

\begin{equation}
    W[\lambda(t)] = \int_0^{\tau} \dot{\lambda} \, \bigg\langle \frac{\partial U}{\partial \lambda} \bigg\rangle \dt , \label{eq:work-continuous}
\end{equation}
beginning at $\lambda(0) = \lambda_i$ and $\rho(x, 0) \propto e^{-\beta U(x, \lambda_i)}$, and ending at $\lambda(\tau) = \lambda_f$ with $\rho(x, \tau)$ unconstrained.

To take the continuum limit of the discrete case, we treat the $d$ states as 1-dimensional lattice sites with spacing $\Delta x$ and reflecting boundaries at $x_\mathrm{b} = \pm (d - 1) \Delta x /2$, and set the connectivity coefficients of Eq.~\eqref{eq:transition-rate-matrix} to $c_{ij} = D (\Delta x)^{-2}$ for all pairs of neighboring sites $\{i,j\}$, s.t. $|i - j| = 1$, and $c_{ij} = 0$ for all else. We define $\rho(x, t) = (\Delta x)^{-1} [\boldrho(t)]^{l(x)}$, $\pi(x, t) = [\boldpi(t)]_{l(x)}$, and $U(x, \lambda) = [\boldU_\lambda]_{l(x)}$, where $l(x) = \lfloor x / \Delta x + d/2 \rfloor$, and take the continuum limit $|x_\mathrm{b}| \rightarrow \infty$ and $\Delta x \rightarrow 0$. Our control-theoretic Hamiltonian then becomes 

\begin{equation}
    H = \int_{-\infty}^{\infty}  (\pi + U - U_f) \, \hat{\mathcal{L}}_\lambda \, \rho \, dx \label{eq:hamiltonian-continuous}
\end{equation}
with $U_f = U(x, \lambda_f)$, while the canonical Eqs.~\eqref{eq:canonical-discrete-rho} and \eqref{eq:canonical-discrete-pi} become

\begin{align}
    \partial_t \rho = \hat{\mathcal{L}}_\lambda \rho  \quad\text{and}\quad  
    \partial_t \pi = -\hat{\mathcal{L}}^{\dagger}_\lambda (\pi + U - U_f) \label{eq:canonical-continuous} .
\end{align}
Finally, under the continuum limit, the constraint Eq.~\eqref{eq:constraint-discrete} becomes

\begin{align}
    \int_{-\infty}^{\infty} \bigg[ \frac{\partial U}{\partial \lambda} \bigg] \bigg( \hat{\mathcal{L}}_\lambda \rho + \beta D  \frac{\partial}{\partial x} \bigg[ \rho \frac{\partial }{\partial x} (\pi + U - U_f) \bigg] \bigg) \, dx = 0, \label{eq:constraint-continuous} 
\end{align}
which may be interpreted as an orthogonality constraint between $\partial U / \partial \lambda$, and a Fokker-Planck operator with modified potential energy $\pi + 2U  - U_f$ acting on $\rho$.

We have now derived an expression that allows us to find the work-minimizing optimal protocol for a continuous-state stochastic system undergoing Langevin dynamics. Just as for the discrete case, solving Eqs.~\eqref{eq:canonical-continuous} and \eqref{eq:constraint-continuous} with initial and terminal conditions $\rho(x, 0) = \rho_{\mathrm{eq}, i}(x)$ and $\pi(x, \tau) = 0$, gives us a principled way to find the optimal protocol $\lambda^*(t)$ that minimizes the work \eqref{eq:work-continuous}. Importantly, these differential equations are much more tractable than the generalized integro-differential equation proposed in \cite{schmiedl2007optimal} for finding the optimal protocol. In particular, these equations are solvable analytically for the control of harmonic potentials, and may be efficiently solved numerically for the control of general anharmonic potentials. 

For the rest of the paper we will consider affine-control potentials of the form 

\begin{equation}
    U(x, \lambda) = U_0(x) + \lambda \, U_1(x) + U_c(\lambda), \label{eq:linear-control-form}
\end{equation}
where $\lambda$ linearly modulates the strength of an auxiliary potential $U_1(x)$ added to the base potential $U_0(x)$, modulo a $\lambda$-dependent constant offset $U_c$. This form is applicable to a wide class of experimental stochastic thermodynamics problems, including molecular pulling experiments \cite{sivak2016thermodynamic, bustamante2021optical, ciliberto2017experiments, ilker2021counterdiabatic, hummer2001free} which can be modeled with potential $U(x, \lambda) = U_\mathrm{sys}(x) + U_\mathrm{ext}(x, \lambda)$ where the external potential of constant stiffness $k$ is $U_\mathrm{ext}(x, \lambda) = k(x - \lambda)^2 / 2$. We see that by expanding the square, this potential is in the form \eqref{eq:linear-control-form} with $U_0(x) = U_\mathrm{sys}(x) + kx^2/2$, $U_1(x)= -kx$, and $U_c(\lambda) = k\lambda^2 / 2$.

By plugging \eqref{eq:linear-control-form} into \eqref{eq:constraint-continuous}, we see that for this class of affine-control potentials the constraint equation is invertible, giving

\begin{equation}
    \lambda[\rho, \pi] = \frac{\lambda_f}{2} + \frac{\int_{-\infty}^{\infty} [{\partial_x}^2 U_1 - \beta (\partial_x U_1) (\partial_x (\pi + U_0))] \, \rho \, dx }{2 \int_{-\infty}^{\infty} (\partial_x U_1)^2 \, \rho \, dx}. \label{eq:constraint-linear}
\end{equation}

Plugging Eqs.~\eqref{eq:linear-control-form} and \eqref{eq:constraint-linear} into \eqref{eq:hamiltonian-continuous} yields $\partial^2 H / \partial \lambda^2 = -2 \int (\partial_x U_1)^2 \rho \, dx < 0$, which demonstrates that the optimal protocol is a minimizing extremum for the work \eqref{eq:costfunction}. A proof for the existence of optimal protocol solutions for Fokker-Planck optimal control is given in \cite{annunziato2013fokker} under loose assumptions. While we currently cannot prove the uniqueness of a solution of Eqs.~\eqref{eq:canonical-continuous} and \eqref{eq:constraint-linear} with our mixed boundary conditions, every solution we have found always outperforms all other protocols we have considered.

We will now illustrate how Eqs.~\eqref{eq:canonical-continuous} and \eqref{eq:constraint-linear} can be used to produce optimal protocols, through particular analytical and numerical examples.

\section{Analytic example}

For the rest of the paper, we set $D = \beta = 1$ for notational simplicity. We start by considering a harmonic potential with $\lambda$ controlling the stiffness of the potential $U(x, \lambda) = \lambda x^2 /2 $, where we identify $U_1 = x^2/2$ and $U_0 = U_\mathrm{c} = 0$. It has been shown \cite{schmiedl2007optimal, aurell2011optimal} that when the probability distribution $\rho$ starts as a Gaussian centered at zero, it remains a Gaussian centered at $0$, with the dynamics of the inverse of the variance $s(t) = \langle x^2 \rangle^{-1}$ given by 

\begin{equation}
    \dot{s} = 2s(\lambda - s), \label{eq:s-dynamics-HO}
\end{equation}
which can be obtained by plugging a zero-mean Gaussian $\rho$ into Eq.~\eqref{eq:canonical-continuous}.

By plugging a truncated polynomial ansantz for the conjugate momentum, $\pi(x, t) = \sum_{k = 0}^n p_k(t) \,x^k / k \, !$ for a finite $n$, into Eq.~\eqref{eq:canonical-continuous} and taking into account our terminal condition $\pi(x, \tau) = 0$, we see that the only surviving terms are the constant and quadratic terms $\pi(x, t) = p_0(t) + p_2(t) x^2/2$, where the coefficients follow dynamics given by

\begin{align}
    \dot{p_0} &=  - (p_2 + \lambda - \lambda_f) \\ 
    \dot{p_2} &=  2 \lambda (p_2 + \lambda - \lambda_f). \label{eq:phi-dynamics-HO}
\end{align}

From our constraint Eq.~\eqref{eq:constraint-linear} we have 

\begin{equation}
    \lambda = \frac{\lambda_f}{2} + \frac{\int_{\infty}^{\infty} (1 - p_2 \,  x^2) \, \rho  \, dx }{ 2 \int_{\infty}^{\infty} x^2 \rho \, dx} = \frac{\lambda_f + s - p_2}{2}. \label{eq:lambda-constraint-HO}
\end{equation}
With this, we eliminate $\lambda(s, p_2)$ from Eqs.~\eqref{eq:s-dynamics-HO} and \eqref{eq:phi-dynamics-HO}, and define $\phi = (s + p_2 - \lambda_f)/2$ to get $\dot{\phi} = -\phi^2$ and $\dot{s} = -2\phi s$. These equations are readily integrable from $t = 0$ to get

\begin{equation}
    \phi(t) = \frac{\phi_{i}}{1 + \phi_{i} t} \ \ \mathrm{and} \ \ s(t) = \frac{\lambda_i}{(1 + \phi_{i} t)^2}, \label{eq:sol-phi-HO}
\end{equation}
where we use $s(0) = \lambda_i$ and define the constant of integration $\phi_{i} = \phi(0)$ yet to be determined. Equating $\phi(\tau) = (s(\tau) + p_2(\tau) - \lambda_f)/2$ allows us to solve

\begin{equation}
    \phi_i = \frac{-(1 + \lambda_f \tau) + \sqrt{1 + 2\lambda_i \tau + \lambda_i \lambda_f \tau^2}}{2\tau + \lambda_f \tau^2}. \label{eq:sol-phii-HO}
\end{equation}
Finally, noting that $\lambda = s - \phi$, we obtain

\begin{equation}
    \lambda(t) = \frac{\lambda_i - \phi_{i}(1 + \phi_{i} t)}{(1 + \phi_{i} t)^2}.  \label{eq:sol-lambda-HO}
\end{equation}

We readily identify Eqs.~\eqref{eq:sol-phii-HO} and \eqref{eq:sol-lambda-HO} as Eqs.~(18) and (19) of \cite{schmiedl2007optimal}. Thus, from our optimal control Eqs.~\eqref{eq:canonical-continuous} and \eqref{eq:constraint-continuous}, we have analytically reproduced the optimal finite-time work-minimizing trajectory for a harmonic trap with variable stiffness. (In SM.I of \cite{zhong2022limited} we provide an analytic derivation of the optimal protocol for the variable trap center case $U(x, \lambda) = (x - \lambda)^2/2$.)

\section{Numerical examples}

\begin{figure*}[t]
    \centering
    \includegraphics[width=0.9\linewidth]{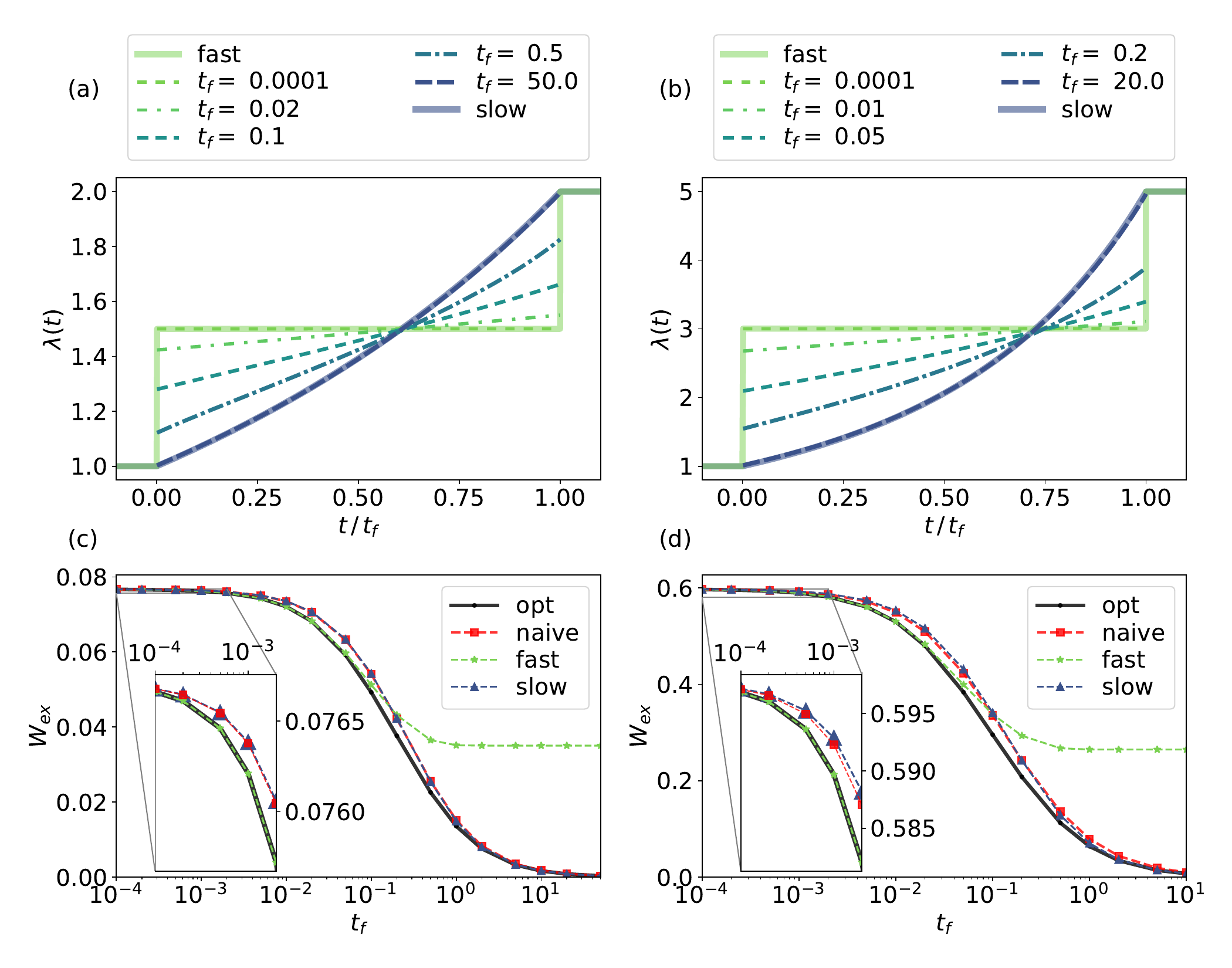}
    \caption[Form and performance for optimal protocols for quartic trap with variable stiffness.]{Form and performance of numerically produced optimal protocols for quartic trap with variable stiffness $U_\lambda(x) = \lambda \, x^4 / 4$ with $\lambda_i = 1, \lambda_f = 2$ on the left column, and $\lambda_i = 1, \lambda_f = 5$. on the right column. (a, b) illustrate optimal protocols for the trap stiffness, across various finite protocol duration values $\tau$. We see that for short times $\tau \ll 1$, the optimal protocol asymptotes to the fast protocol as given in \cite{blaber2021steps}, whereas for long times $\tau \gg 1$, the optimal protocol asymptotes to the slow protocol as given in \cite{sivak2012thermodynamic}. We observe discontinuous jumps at $t = 0$ and $t = \tau$ in our numerically calculated optimal protocols, which is often the case for optimal protocols \cite{schmiedl2007optimal, boundary-layers, blaber2020skewed}. (c) and (d) compare the protocol performance $W_\mathrm{ex}$ among the numerically calculated optimal protocol, the fast protocol, the slow protocol, and the naive protocol. We see that the optimal protocol outperforms all other protocols, with the fast and slow protocols asymptoting in performance to the optimal protocol in their respective small- and large-$\tau$ limits. The form and performance of these optimal protocols are qualitatively similar to those for the harmonic oscillator control \cite{schmiedl2007optimal} (illustrated in Fig. SM.I of \cite{zhong2022limited}).}
    \label{fig:quartic}
\end{figure*}

The harmonic potential problem is exceptional in that we can solve for its optimal protocol analytically. For the vast majority of time-varying potentials, the differential Eqs.~\eqref{eq:canonical-continuous} with constraint \eqref{eq:constraint-linear} do not admit analytic solutions, but can be solved numerically. In this section, we briefly sketch our numerical scheme to solve Eqs.~\eqref{eq:canonical-continuous} and \eqref{eq:constraint-continuous}, and we demonstrate our approach for two classes of quartic potential problems that do not admit analytic solutions: changing the stiffness of a quartic trap, and applying a linear bias to a double-well potential. 

We compare the form and performance of these optimal protocols to three other protocols: naive, fast, and slow. The naive protocol interpolates the starting and ending parameters linearly in time $\lambda(t) = \lambda_i + (t / \tau) (\lambda_f - \lambda_i)$, and generally is not optimal in any regime. The fast protocol, also known as the short-time efficient protocol (STEP) as developed in \cite{blaber2021steps}, is optimal for small-$\tau$ limit, and involves a step to an intermediate value $\lambda^{\mathrm{STEP}}$ for the duration of the protocol. The slow protocol first derived in \cite{sivak2012thermodynamic}, also known as the near-equilibrium protocol, is optimal for large-$\tau$, and is obtained by considering the thermodynamic geometry of protocol parameter space induced by the friction tensor $\xi(\lambda)$, from the linear response of excess work from changes in $\lambda(t)$. With this induced thermodynamic geometry, the slow protocol is a geodesic of $\xi$ given by $\dot{\lambda}(t) \propto \xi(\lambda(t))^{-1/2}$, with $\lambda(0) = \lambda_i$ and $\lambda(\tau) = \lambda_f$. (We provide a more detailed review of the slow and fast protocols, as well as how we produce them for our numerical study in SM.II.A.3 and SM.II.A.4 of \cite{zhong2022limited}.)

Here we briefly describe our discretization and integration scheme. Our lattice-discretization of space and time and approximated Fokker-Planck dynamics largely follow \cite{holubec2019physically}. Just as taking the continuous limit from a discrete-state master equation yields Fokker-Planck dynamics, by discretizing our configuration space onto a lattice, Fokker-Planck dynamics can be approximated by a master equation over lattice states \cite{zwanzig2001nonequilibrium}. Here, we approximate the configuration space by a grid of $d$ points with spacing $\Delta x$ and reflecting boundaries at $x_\mathrm{b} = \pm (d - 1) \Delta x / 2$, akin to the time-dependent Fokker-Planck discretization described in \cite{holubec2019physically}. Our optimal control Eqs.~\eqref{eq:canonical-continuous} and \eqref{eq:constraint-continuous} become the ordinary differential Eqs.~\eqref{eq:canonical-discrete-rho} and \eqref{eq:canonical-discrete-pi}, coupled by \eqref{eq:constraint-discrete}. Time is discretized to $N$ time steps, with either constant or variable timesteps.

Because the transition rate matrix $\mathcal{L}_\lambda$ has non-positive eigenvalues \cite{risken1996fokker, wadia2022solution}, it is numerically unstable to integrate $\boldpi$ forward in time, as any amount of numerical noise becomes exponentially amplified. Rather, we adopt a Forward-Backward sweep method \cite{mcasey2012convergence, lenhart2007optimal}, where approximate solutions for $\boldrho^{(k)}(t)$ and $\boldpi^{(k)}(t)$ are updated iteratively through first obtaining $\boldrho^{(k+1)}$ by solving \eqref{eq:canonical-discrete-rho} and \eqref{eq:constraint-discrete} forwards in time starting with $\boldrho(0) = \boldrho_{i,\mathrm{eq}}$, keeping $\boldpi(t) = \boldpi^{(k)}(t)$  fixed; and then obtaining $\boldpi^{(k+1)}$ by solving \eqref{eq:canonical-discrete-pi} and \eqref{eq:constraint-discrete} backwards in time starting with $\boldpi(\tau) = \boldsymbol{0}$, keeping $\boldrho(t) = \boldrho^{(k+1)}(t)$ fixed. These forward and backward sweeps are iterated until numerical convergence of $\boldrho^*(t), \boldpi^*(t)$, which then is passed to $\eqref{eq:constraint-discrete}$ to obtain the optimal protocol $\lambda^{*}(t)$. (See SM.II of \cite{zhong2022limited} for exact details on our numerical scheme.)

To measure the performance of each protocol $\lambda(t)$, we consider the excess work $W_\mathrm{ex}[\lambda(t)] = W[\lambda(t)] - \Delta F$, where $\Delta F = \log(Z_f) - \log(Z_i)$ is the free energy difference between initial and final equilibrium states, with $Z_\lambda = \int dx \exp (-U_\lambda(x))$ being the partition function. By the Second Law of Thermodynamics, $W_\mathrm{ex} > 0$, and approaches $0$ in the quasistatic $\tau \rightarrow \infty$ limit. (We describe how we numerically compute $W_\mathrm{ex}$ for a given protocol in SM.II.B of \cite{zhong2022limited}.)

Now we present our results for the variable-stiffness quartic trap and linearly biased double-well examples. 

\subsection{Quartic trap with variable stiffness} 

First, we consider the quartic analog of the variable stiffness harmonic oscillator, with the potential given as
\begin{equation}
    U_\lambda(x) = \lambda \frac{x^4}{4}.
\end{equation}
Figs.~\ref{fig:quartic}(a) and \ref{fig:quartic}(b) illustrate the numerically obtained optimal protocols for variable values of protocol time $\tau$, for $\lambda_i = 1, \lambda_f = 2$; and $\lambda_i = 1, \lambda_f = 5$ respectively. We see that the optimal protocols for the variable stiffness quartic trap problem are qualitatively similar to the optimal protocols for the variable stiffness harmonic trap in Section 4 (derived and illustrated in \cite{schmiedl2007optimal}). For both problems, optimal protocols are continuous and monotonic with positive curvature for times $t \in (0, \tau)$, and have discontinuous jumps at $t = 0$ and $t = \tau$. Also plotted are the fast \cite{blaber2021steps} and slow \cite{sivak2012thermodynamic} protocols, which have been derived to be optimal for the small- and large-$\tau$ limits, respectively. We see that the numerically solved optimal protocol asymptotes to these protocols in the respective $\tau$ limits.

Figs.~\ref{fig:quartic}(c) and \ref{fig:quartic}(d) illustrate the excess work $W_\mathrm{ex}$ of various protocols across different time-scales $\tau$. We see that the optimal protocol outperforms all three of the naive, fast, and slow protocols. The performance of the fast protocol converges to the optimal protocol performance for short time-scales $\tau \ll 1$. Likewise, the performance of the slow protocol converges to the optimal protocol performance for long time-scales $\tau \gg 1$. This is expected, and is consistent with how the optimal protocol asymptotes to the fast and slow protocols in the respective time-scales.

\subsection{Linearly biased double-well} 

\begin{figure*}[t]
    \centering
    \includegraphics[width=0.9\linewidth]{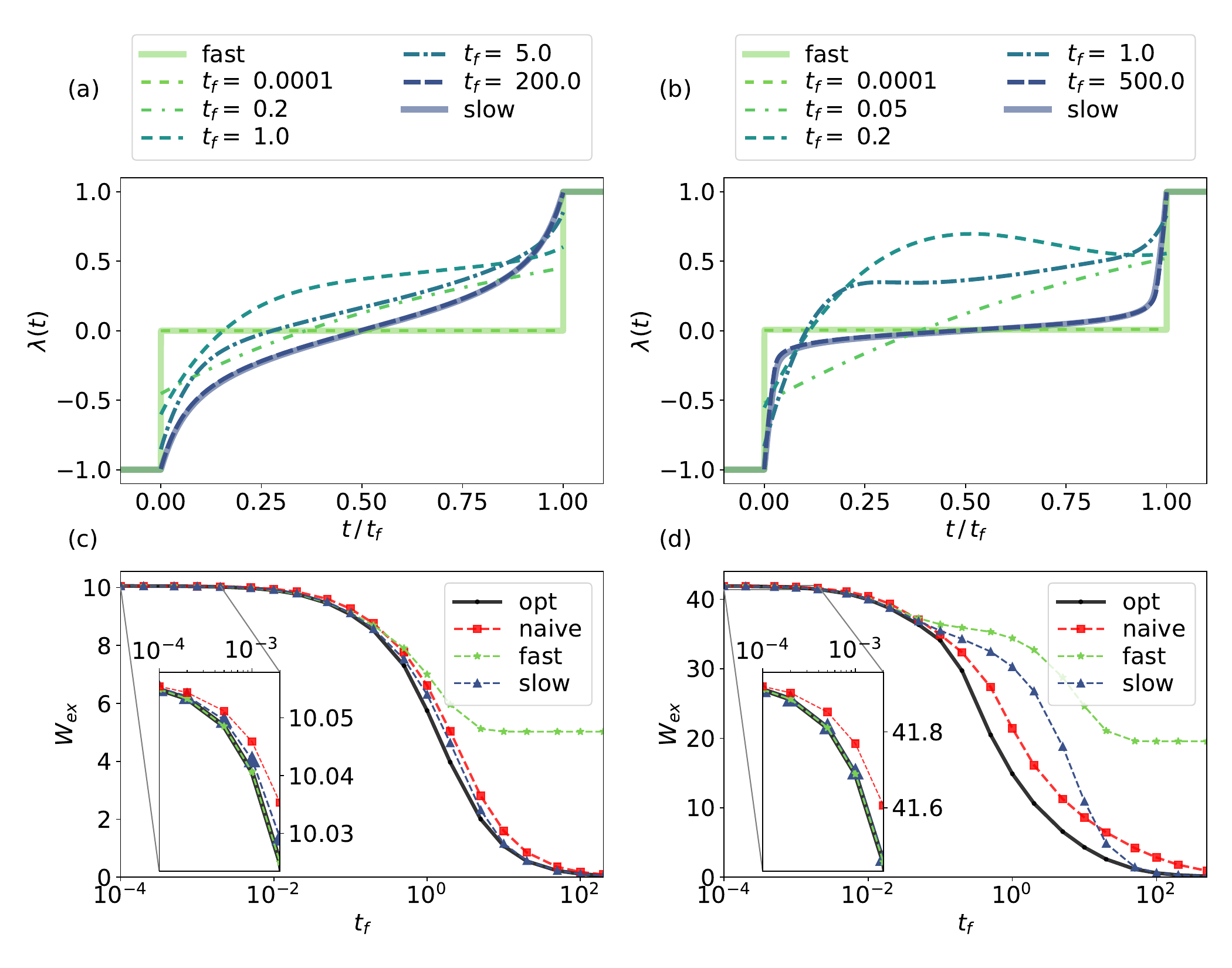}
    \caption[Form and performance for optimal protocols for biased double-well potential, which can be non-monotonic!]{Numerically solved optimal protocols for the linearly biased double-well potential $U_\lambda(x) = E_0 ((x^2 - 1)^2 / 4 - \lambda x)$, $\lambda_i = -1$ and $\lambda_f = 1$; with $E_0 = 4$ in the left column, $E_0 = 16$ in the right. (a, b) illustrate the optimal protocols for the linear bias value, across various finite protocol duration values $\tau$. As with the quartic case in Fig.~\ref{fig:quartic}, here for short times $\tau \ll 1$ and long times $\tau \gg 1$, the optimal protocol asymptotes to the fast and slow protocols respectively. Unlike the slow and fast protocols, for intermediate values of $\tau$ the optimal protocols are not symmetric in $(t, \lambda) \rightarrow (-t, -\lambda)$. For $E_0 = 16$, we observe surprising non-monotonic protocols for $\tau \sim 0.2$. (c, d) depict the protocol performance $W_\mathrm{ex}$ between the numerically calculated optimal protocol and other protocols. Like in the quartic case, we see that the optimal protocol outperforms all other protocols, with the fast and slow protocols asymptoting in performance to the optimal protocol in their respective small- and large-$\tau$ limits. For $E_0 = 16$, the optimal protocol vastly outperforms the other protocols for $\tau \sim 2$.}
    \label{fig:double-well}
\end{figure*}

Here we consider the double-well potential with wells at $x = \pm 1$ with an external linear bias

\begin{equation}
   U_\lambda(x) = E_0 \frac{(x^2 - 1)^2}{4} - \lambda E_0 x .
\end{equation}
Here, $E_0$ sets the energy scale of the ground and external potentials, with a barrier height of $E_0/4$ between the two wells at $\lambda = 0$. This potential is commonly used in the study of bit erasure \cite{proesmans2020optimal, zulkowski2014optimal}, but here we allow only limited control in the form of a linear bias. We note that this problem is qualitatively similar to the \cite{sivakcrooksbarriercrossing}, where a harmonic pulling potential with variable center is applied to a potential with two local minima separated by a barrier. We consider $\lambda_i = -1$ and $\lambda_f = 1$, while varying $E_0$ and $\tau$. Setting the parameter value $\lambda = -1$ biases the potential to the left well, which sufficiently raises the right well above the barrier height and shifts the left well minimum from $x_{\mathrm{well}} = -1$ to $-1.32472$. Setting $\lambda = 1$ gives a symmetric bias to the right well.

Figs.~\ref{fig:double-well}(a) and \ref{fig:double-well}(b) illustrate optimal protocols for $E_0 = 4$ and $E_0 = 16$, which correspond to inter-well barrier heights of $1 \, k_\mathrm{B} T$ and $4 \, k_\mathrm{B} T$ respectively. Just as before, the optimal protocol asymptotes to the fast and slow protocols in the small- and large-$\tau$ limits. We note here that the optimal protocols obtained for various values of $E_0$ and $\tau$ have intriguing properties. First of all, both the fast and slow protocols are symmetric under inversion $(\lambda(t), t) \rightarrow (-\lambda(t), \tau - t)$, which arises from the symmetry $U_\lambda(x) = U_{-\lambda}(-x)$ with $\lambda_f = -\lambda_i$, and the construction of these protocols. We see though that the optimal protocol obtained by solving \eqref{eq:canonical-continuous} and \eqref{eq:constraint-continuous} do not follow this this symmetry for intermediate values of timescale $\tau$. This discovery of barrier crossing optimal protocols breaking symmetry was first made in \cite{automaticdifferentiation}. At first this symmetry-breaking may seem counter-intuitive, but this can be understood by noting that $\lambda_i$ and $\lambda_f$ play completely different roles in our optimal control problem: $\lambda_i$ specifies the initial condition $\rho(x, 0)$, while $\lambda_f$ specifies $U_f(x)$ in the cost function. 

\begin{figure*}[t]
    \centering
    \includegraphics[width=0.9\linewidth]{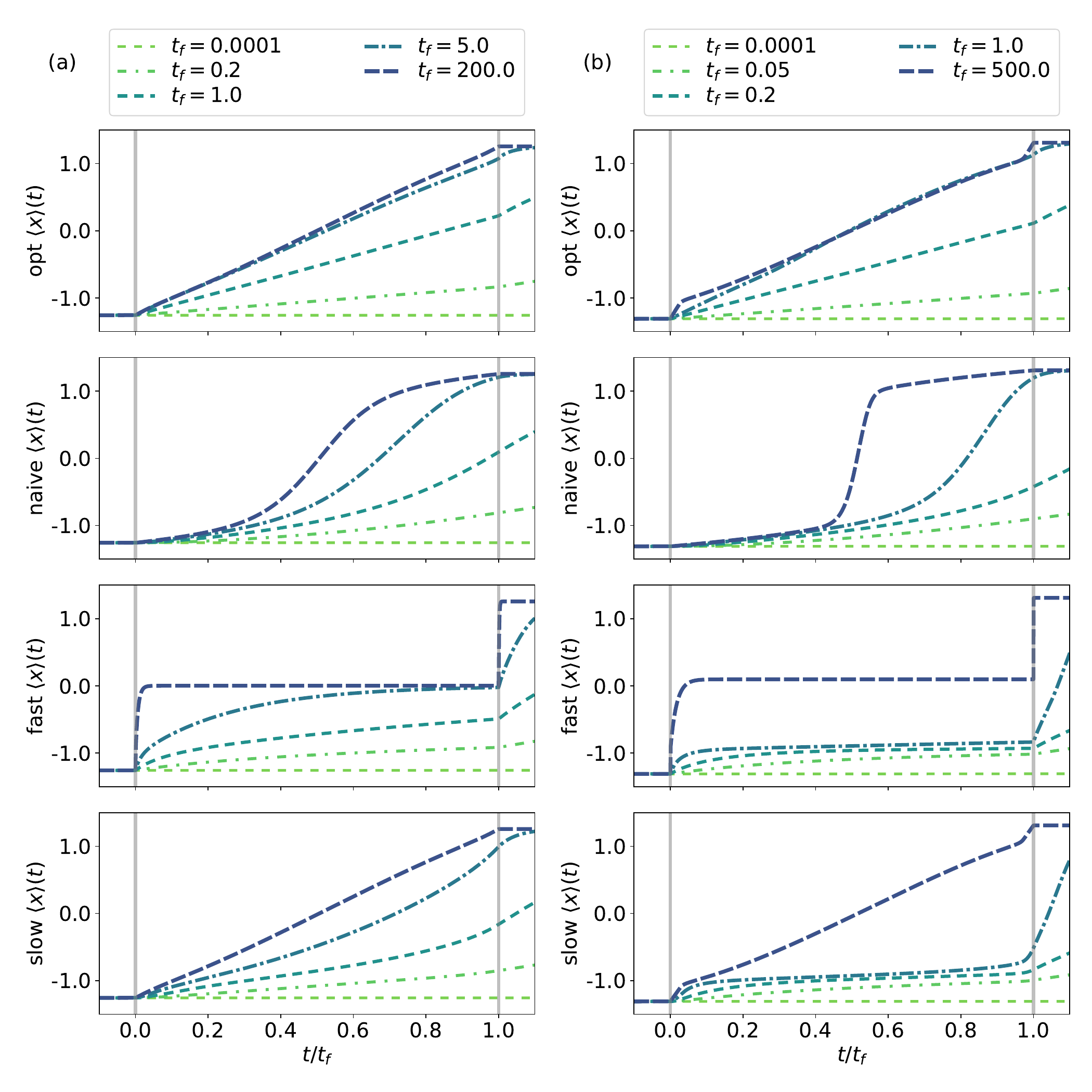}
    \caption[The evolution of the mean position $\langle x \rangle(t)$ for the linearly biased double-well under different protocols and protocol durations.]{The evolution of the mean position $\langle x \rangle(t)$ for the linearly biased double-well problem $U_\lambda(x) = E_0 ((x^2 - 1)^2 / 4 - \lambda x)$, across various protocol duration values $\tau$. Here, $\lambda_i = -1$ and $\lambda_f = 1$, with (a) $E_0 = 4$, and (b) $E_0 = 16$. The first row depicts the optimal protocol, the second the naive protocol, the third the fast protocol, and the fourth the slow protocol. For the optimal protocol, $\langle x \rangle(t)$ increases monotonically with near-constant velocity, which we argue is a generic property of limited-control optimal controls. In comparison, the naive, fast, and slow protocols evolve the mean $\langle x \rangle(t)$ with much more variable velocity. The deviation from constant-velocity roughly corresponds to larger $W_\mathrm{ex}$ values, as depicted in Figs.~\ref{fig:double-well}(c) and \ref{fig:double-well}(d).}
    \label{fig:mean-evolution}
\end{figure*}

Furthermore, not only do we find non-symmetric protocols, we discover that for $E_0 = 16$, the optimal protocol $\lambda(t)$ is non monotonic at certain intermediate timescales $\tau \sim 0.2$. This result is surprising, given that the underlying stochastic system \eqref{eq:langevin} is overdamped --- it
has no momentum degrees of freedom that could incentivize overshoots. To our knowledge, no optimal or approximately-optimal protocols for a single parameter $\lambda$ have been reported to exhibit this sort of non-monotonic behavior. In this regime, the optimal protocol cannot be interpreted as a geodesic for an underlying thermodynamic metric, as the latter can only produce monotonic protocols.  

To explain this overshoot, we consider the mean position of the probability density under the optimal protocol $\langle x \rangle = \int \rho(x, t) \, x \, dx$ as a function time $t$. This is shown in Figs.~\ref{fig:mean-evolution}(a) and \ref{fig:mean-evolution}(b), where we see $\langle x \rangle$ increases at a nearly constant rate under the optimal protocol. This may be interpreted as the limited-control optimal protocol allowing barrier-crossing to occur at an approximately constant velocity. On the other hand, when full control over the potential is allowed, the distribution mean $\langle x \rangle$ always maintains a constant speed under the full-control optimal protocol (see SM.III of \cite{zhong2022limited} for a derivation drawn from optimal transport theory). This suggests that insofar as a limited-control optimal protocol should approximate the full-control optimal protocol, it drives the mean of the probability distribution to travel with near-constant velocity, even if requiring an overshoot as is the case for the $E_0 \sim 16, \tau \sim 0.2$ regime.

Figs.~\ref{fig:double-well}(c) and \ref{fig:double-well}(d) illustrate the performance of these protocols. Just as we found for the harmonic potential, the OCT protocol outperforms all three other considered protocols, with performance of fast and slow protocols approaching the optimal protocol performance in their respective $\tau$ limits. We see that for barrier height $E_0 = 16$, the optimal protocol vastly outperforms all other protocols at intermediate $\tau$ values. For instance, at $\tau = 2$ the optimal protocol gives $W_\mathrm{ex} = 10.61$, which is significantly smaller than the naive protocol $W_\mathrm{ex} = 16.12$ and slow protocol $W_\mathrm{ex} = 26.77$ values. This shows the existence of truly far from equilibrium regimes, for which protocols derived assuming either fast or near-equilibrium approximations deviate significantly from the true, fully non-equilibrium optimal protocol, in both form and performance.

\section{Discussion}

It is typically the case in experimental and engineering contexts that only a finite set of degrees of freedom of a system is controllable. We have shown that the problem of finding work-minimizing optimal protocols is naturally framable as an optimal control theory (OCT) problem. Using tools and techniques from OCT, we have devised a method to derive optimal protocols in the case where there is only limited control of the form of the system's potential. Our framework allows us to reproduce known analytic results for the control of a harmonic oscillator, as well as to efficiently calculate optimal protocols numerically for a large class of limited-control potentials.

Previous work on dissipation-minimizing optimal protocols revealed thermodynamic geometry on protocol parameter space through the friction tensor \cite{sivak2012thermodynamic, wadia2022solution}, and on probability density space through the $L^2$-Wasserstein metric \cite{watanabe2022finite, dechant2019thermodynamic, nakazato2021geometrical, chen2019stochastic}. We have found that the protocol optimization problem has a deep Hamiltonian structure, typical of OCT problems \cite{liberzon2011calculus}. It is interesting to ponder what insights may be gleaned from the study of optimal protocols for non-equilibrium processes when both Riemmanian and symplectic structures are considered together.

It is straightforward to generalize our results configuration and parameter spaces that are multi-dimensional, which suggests a number of natural extensions. First, by allowing time-varying control of temperature $\beta^{-1} = k_\mathrm{B} T$ and asserting time-periodicity for the protocol, we can construct optimal finite-time heat engines arbitrarily far from equilibrium, building off of \cite{frim2022optimal, watanabe2022finite, ye2022optimal}. Cyclical protocols may also be considered for when the state space and/or configuration space are non-Euclidean manifolds \cite{frim2021engineered}; e.g., for the external control of rotory motor proteins like $F_{o}F_1$ \cite{lucero2019optimal}. Recent surprising results demonstrate that if detailed-balance breaking transition rates were allowed in the control of Markov jump processes, finite-time transitions between different probability distributions may be conducted with arbitrarily small entropy production \cite{remlein2021optimality, dechant2022minimum, yoshimura2022information, van2022thermodynamic}; our framework is easily adaptable to these kinds systems through the replacement of every instance of potential energy difference with a forcing matrix that need not be symmetric $[U_i(\lambda) - U_j(\lambda)] \rightarrow F_{ij}(\lambda)$ in Eq.~\eqref{eq:transition-rate-matrix}, and it would be interesting to observe whether calculated work-minimizing protocols would contain similar features. Finally, it would be intriguing to extend our framework to the study of underdamped systems where both position and velocity degrees of freedom $(x, v)$ make up the configuration space \cite{gomez2008optimal, muratore2014extremals}, as because the kinetic term of the underlying Klein-Kramers equation cannot be controlled, control is intrinsically limited to just the spatial degrees of freedom.

When the configuration space has many degrees of freedom, the curse of dimensionality kicks in, where the memory required to store the probability distribution is exponential in the number of dimensions of the configuration space \cite{kappen2005path}. In this case, it may be more computationally tractable to sample individual stochastic trajectories to compute the friction tensor \cite{sivak2012thermodynamic, rotskoff2015optimal} or gradients of the protocol \cite{automaticdifferentiation} in order to calculate optimal protocols. It will be of interest to study the effectiveness of configuration space dimensionality reduction techniques (e.g., density functional theory \cite{te2020classical}, Zwanzig-Mori projection operators \cite{zwanzig2001nonequilibrium}) to make the calculation of optimal protocols through our framework computationally tractable for high dimensional configuration spaces. 

We have shown that optimal control theory is a natural and powerful framework for the design and study of  thermodynamically optimal protocols. In the spirit of \cite{roach2018application}, it is our hope that through considering the optimal control of non-equilibrium probability densities considered here and elsewhere \cite{annunziato2013fokker, palmer2011hamiltonian, bakshi2020open}, we may better understand how it is that biological systems, which operate far from equilibrium, function efficiently across vastly different length- and time-scales.


\newcommand{\boldeta}{\eta} 
\newcommand{\boldlambda}{{\boldsymbol{\lambda}}}
\renewcommand{\boldx}{{\boldsymbol{x}}}
\newcommand{\trajx}{X}
\newcommand{\boldc}{c}
\renewcommand{\dt}{{\,\mathrm{d}t}}
\renewcommand{\ds}{{\,\mathrm{d}s}}
\renewcommand{\dx}{{\,\mathrm{d}x}}
\renewcommand{\dee}{{\mathrm{d}}}
\newcommand{\gammadot}{{\dot{\gamma}}}

\newcommand{\trajp}{P}
\renewcommand{\boldeta}{\eta} 
\renewcommand{\boldlambda}{{\boldsymbol{\lambda}}}
\renewcommand{\boldx}{{\boldsymbol{x}}}
\renewcommand{\trajx}{X}
\renewcommand{\boldc}{c}
\renewcommand{\tee}{{\mathrm{t}}}
\renewcommand{\tf}{{\mathrm{\tau}}}
\renewcommand{\A}{{\mathrm{A}}}
\renewcommand{\B}{{\mathrm{B}}}
\renewcommand{\C}{{\mathrm{C}}}
\renewcommand{\D}{{\mathrm{D}}}
\renewcommand{\U}{{\mathrm{U}}}
\renewcommand{\W}{{\mathrm{W}}}
\renewcommand{\boldU}{{\boldsymbol{U}}}
\renewcommand{\boldpi}{{\boldsymbol{\pi}}}
\renewcommand{\boldpsi}{{\boldsymbol{\psi}}}
\renewcommand{\curlyL}{{\mathcal{L}}}
\renewcommand{\curlyW}{{\mathcal{W}}}
\renewcommand{\bolddx}{{\Delta \boldsymbol{x}}}
\renewcommand{\dX}{{\mathrm{d}X}}
\renewcommand{\dP}{{\mathrm{d}P}}
\renewcommand{\dXtilde}{{\mathrm{d}\tilde{X}}}
\renewcommand{\dtau}{{\,\mathrm{d}\tau}}
\renewcommand{\dt}{{\,\mathrm{d}t}}

\chapter{Time-Asymmetric Fluctuation Theorem and Efficient Free Energy Estimation}

The free-energy difference $\Delta F$ between two high-dimensional systems is notoriously difficult to compute, but very important for many applications such as drug discovery. We demonstrate that an unconventional definition of work introduced by Vaikuntanathan and Jarzynski (2008) satisfies a microscopic fluctuation theorem that relates path ensembles that are driven by protocols unequal under time-reversal. It has been shown before that counterdiabatic protocols---those having additional forcing that enforces the system to remain in instantaneous equilibrium, also known as escorted dynamics or engineered swift equilibration---yield zero-variance work measurements for this definition. We show that this time-asymmetric microscopic fluctuation theorem can be exploited for efficient free energy estimation by developing a simple (i.e., neural-network free) and efficient adaptive time-asymmetric protocol optimization algorithm that yields $\Delta F$ estimates that are orders of magnitude lower in mean squared error than the generic linear interpolation protocol with which it is initialized.

\subsection{Note}

This chapter is based off of: Adrianne Zhong*, Ben Kuznets-Speck*, and Michael R DeWeese. ``Time-asymmetric fluctuation theorem and efficient free-energy estimation''. In: \emph{Physical Review E} 110.3 (2024), p. 034121. (*denoting equal contribution). 


In parlance of the first chapter, we can consider the augmented Langevin equation
\begin{equation}
    \dot{X}_t = - \mu \boldnabla  U_{\gamma(t)}  - \mu \boldnabla V_t + \sqrt{2 \mu \kBT} \, \xi_t \nonumber 
\end{equation}
which has the Fokker-Planck equation
\begin{equation}
    \partial_t \rho_t = \mu \boldnabla \cdot \{ \rho_t \boldnabla (U_{\gamma(t)} + \kBT \ln \rho_t + V_t) \}. . \nonumber
\end{equation}
If $\gamma(t)$ and $V_t$ were carefully chosen to as to satisfy a particular PDE
\begin{equation}
    \partial_t \ln \rhoeq_{\gamma(t)} = \mu  \nabla^2 V_t + \mu \boldnabla \ln \rhoeq_{\gamma(t)} \cdot \boldnabla V_t \nonumber
\end{equation}
(equivalently, the continuity equation $\partial_t \rhoeq_{\gamma(t)} = \mu \boldnabla \cdot (\rhoeq_{\gamma(t)} \nabla V_t)$), then the solution of the Fokker-Planck equation is $\rho_t = \rhoeq_{\gamma(t)}$. 

This potential $V_t(x)$ is called a \emph{counterdiabatic potential} in that it provides a force that steers the thermodynamic state along the trajectory of equilibrium thermodynamic states $\rho_t = \rhoeq_{\gamma(t)}$, which in this chapter turns out to be useful for computing free energy differences between different potentials, which is of interest of physicists, chemists, and pharmaceutical scientists alike! 

While calculating optimal protocols through optimal control theory on the Fokker-Planck equation (as presented in the previous chapter) runs into the curse of dimensionality for high-dimensional systems, in this chapter we present an adaptive protocol learning algorithm based on reinforcement learning, that does not suffer from this curse. In short, after various stochastic Langevin trajectories are simulated for a given protocol, the protocol is updated so that the trajectories that required less work are likelier to occur. In this way a controlled Langevin equation can be viewed as a stochastic policy. It works very well!

\section{Chapter introduction}

Free energy differences $\Delta F = F_B - F_A$ between pairs of potential energy functions $U_A(\boldx)$ and $U_B(\boldx)$ are sought after by physicists, chemists, and pharmaceutical scientists alike \cite{cournia2017relative, tawa1998calculation, kelly2006aqueous, chipot2007free, shirts2010free, cournia2021free}. Here, $\boldx \in \mathbb{R}^d$ is the configuration space coordinate, and the free energy for each potential is defined as $F_{A,B} = -\beta^{-1} \ln \int e^{-\beta U_{A,B}(\boldx)}\mathrm{d}\boldx$, where $\beta = (k_\mathrm{B} T)^{-1}$ is inverse temperature. For high dimensional systems, $\Delta F$ can only be calculated numerically through sampling methods, which can be computationally costly and slow to converge \cite{chipot2007free}. Here we present an adaptive method that greatly reduces the variance of $\Delta F$ estimates, based on a new fluctuation theorem we derive.

One class of estimators takes work measurements as input from protocols $U(\boldx, t)$ that ``switch'' $U(\boldx, 0) = U_A(\boldx) \rightarrow U(\boldx, t_f) = U_B(\boldx)$ in finite time $t_f$. Because the work, traditionally defined for a trajectory $\trajx(t)|_{t\in[0,t_f]}$\footnote{In this paper, to avoid confusion we use lower case $x$ to denote configurations, and upper case $X(t)$ to denote (stochastic) trajectories.} 
as  
\begin{equation}
W_\mathrm{trad}[\trajx(t)] = \int_0^{t_f} \frac{\partial U}{\partial t} (\trajx(t), t) \, \mathrm{d}t \label{eq:work-trad}
\end{equation}
satisfies the Jarzynski equality
\begin{equation}
    \langle e^{-\beta W} \rangle = e^{-\beta \Delta F}, \label{eq:jarzynski-equality}
\end{equation}
the Jarzynski estimator $\widehat{\Delta F}_\mathrm{Jar} = -\beta^{-1} \ln \, \big( n_s^{-1} \sum_{i=1}^{n_s} e^{-\beta W^i_\mathrm{trad}}\big)$ may be applied to work measurements $\{W^i_\mathrm{trad}, | i =1, .., n_s\}$. Unfortunately this estimator can be slow to converge, because the average is often dominated by rare events.

Estimators that use bi-directional work measurements (i.e., those that also consider $U_B \rightarrow U_A$ switching processes) generally have lower variance than uni-directional work estimators \cite{shirts2005comparison}. In particular, Shirts et al. in \cite{shirts2003equilibrium} showed that if forward  $\{W^i_F \, | \, i =1, .., n_s\}$ and reverse work measurements $\{W^i_R \, | \, i =1, .., n_s\}$, assumed here to be equal in number for simplicity, are collected from forward and reverse protocols satisfying Crooks Fluctuation Theorem
\begin{equation}
    \mathcal{P}_F(+W) = \mathcal{P}_R(-W) \, e^{\beta (W - \Delta F)}, \label{eq:fluctuation-theorem}
\end{equation}
then the Bennett acceptance ratio estimator $\widehat{\Delta F}_\mathrm{BAR}$ \cite{bennett1976efficient}, defined implicitly as the $\Delta F$ satisfying
\begin{equation}
  \sum_{i = 1}^{n_s} \frac{1}{1 + e^{-\beta(W^i_F - \Delta F)}} -   \sum_{j = 1}^{n_s}  \frac{1}{1 + e^{-\beta(W^j_R + \Delta F)}} = 0, \label{eq:BAR-estimator}
\end{equation}
is the lowest-variance asymptotically-unbiased estimator. Bi-directional measurements of $W_\mathrm{trad}$ for a pair of time-reversal-symmetric forward and reverse protocols satisfy Eq.~\eqref{eq:fluctuation-theorem} \cite{crooks1998nonequilibrium, crooks1999entropy}, but measurements can also be collected from \textit{mixtures} of different measurement-protocol pairs
\begin{equation}
    \mathcal{P}_F(\cdot) = \sum_i \alpha_i \mathcal{P}^{i}_F(\cdot) \quad\mathrm{and}\quad \mathcal{P}_R(\cdot) = \sum_i \alpha_i \mathcal{P}^{i}_R(\cdot) \label{eq:mixtures}
\end{equation}
with $\sum_i \alpha_i = 1$, as long as each $(\mathcal{P}^{i}_F, \mathcal{P}^{i}_R)$ pair satisfies Eq.~\eqref{eq:fluctuation-theorem}.

In this paper, we consider the non-standard definition of work introduced in \cite{vaikuntanathan2008escorted}, for which remarkably, there exists finite-time protocols that yield zero-variance work measurements.  Importantly, these zero-variance protocols include a separate counterdiabatic forcing term that effectuates a faster-than-quasistatic time evolution. We explicitly show that this non-standard definition of work satisfies the fluctuation theorem Eq.~\eqref{eq:fluctuation-theorem} for measurements that are produced from separate forward and reverse protocols that are \textit{unequal} under time-reversal. We demonstrate that the time-asymmetric fluctuation theorem for this unconventional work may be exploited for efficient free energy estimation by proposing an algorithm that iteratively improves time-asymmetric protocols and uses measurements collected across \textit{all} iterations. On three examples of increasing complexity, we show that $10^3$ measurements made under our adaptive protocol algorithm give $\Delta F$ estimates that are a factor of $\sim 10^2 - 10^4$ lower in mean squared error than the same number of { }measurements made with the generic time-symmetric linear interpolation protocol with which it was initialized. 

The first version of this paper was posted on ArXiv in April 2023 \cite{zhong2023time-preprint}. Near-simultaneously, the preprint \cite{vargas2023transport} was posted on ArXiv, in which the authors independently derived the same theoretical results as we found for overdamped dynamics, and demonstrated through impressive numerical results the utility of the time-asymmetric fluctuation theorem.

\section{Time-asymmetric work}

For our setting we consider a time-varying potential energy $U_0(\boldx, t)$ for $t \in [0, t_f]$, that begins at $U_0(\boldx, 0) = U_A(\boldx)$ and ends at $U_0(\boldx, t_f) = U_B(\boldx)$. To this we add an additional potential $U_1(\boldx,t)$ that satisfies $U_1(\boldx, 0) = U_1(\boldx, t_f) = 0$. In the overdamped limit, a trajectory $\trajx(t)$ evolves according to the Langevin equation
\begin{equation} \label{eq:forward-langevin}
\dot \trajx(t) = -\nabla ( U_0 + U_1 ) + \sqrt{2\beta^{-1} } \,  \boldeta(t)\quad\mathrm{with}\quad \trajx(0) \sim \rho_A(\cdot).
\end{equation}
Here, $\rho_A(\boldx) = e^{-\beta [U_A(\boldx) - F_A]}$ is the equilibrium distribution for $U_A(\boldx)$, and $\boldeta(t)$ is an instantiation of standard $d$-dimensional Gaussian white noise specified by $\langle \eta_i(t) \rangle = 0$ and $\langle \eta_i(t)\eta_j(t') \rangle = \delta_{ij} \delta(t - t')$.\footnote{Typically there is a factor of the friction coefficient $\gamma$ multiplying the left side of \eqref{eq:forward-langevin}. For notational simplicity we have set it to one, which given that it is equal for all dimensions may always be done through a time-rescaling.} (We consider underdamped dynamics in  (Chapter Appendix \thechapter.C.)

In \cite{vaikuntanathan2008escorted} the authors introduced an unconventional work definition, which in our setting is the trajectory functional
\begin{equation} \label{eq:time-asymmetric-work} 
  W[\trajx(t)] = \int_{0}^{t_f} \frac{\partial U_0}{\partial t} - \nabla U_0 \cdot \nabla U_1 + \beta^{-1} \nabla^2 U_1 \, \mathrm{d}t 
\end{equation}
($\nabla^2$ is the scalar Laplace operator), and demonstrated that, remarkably, $W[\trajx(t)] = \Delta F$ for \textit{every} trajectory $\trajx(t)$, if $U_1(\boldx, t)$ gives the counterdiabatic force for $U_0(\boldx, t)$, meaning
\begin{equation}
  \frac{\partial \rho_0}{\partial t} = \nabla \cdot (\rho_0 \nabla U_1) \quad\mathrm{for}\quad \rho_0(\boldx, t) := e^{-\beta [U_0(\boldx, t) - F_0(t)]}. \label{eq:sufficient-condition}
\end{equation}
Here $\rho_0(\boldx, t)$ is the instantaneous equilibrium distribution corresponding to $U_0(\boldx, t)$, with time-dependent free energy $F_0(t) = -\beta^{-1} \ln \int e^{-\beta U_0(\boldx, t)} \mathrm{d}\boldx$ satisfying $F_0(0) = F_A$ and $F_0(t_f) = F_B$. Counterdiabatic driving has been studied before in various contexts \cite{del2013shortcuts, guery2019shortcuts, ilker2022shortcuts, iram2021controlling, frim2021engineered}. Under these conditions, the time-dependent probability distribution for Eq.~\eqref{eq:forward-langevin} is always in instantaneous equilibrium with $U_0(\boldx, t)$. 

Indeed, expanding Eq.~\eqref{eq:sufficient-condition} yields
\begin{equation}
  \frac{\partial U_0}{\partial t} - \nabla U_0 \cdot \nabla U_1 + \beta^{-1} \nabla^2 U_1 = \frac{\mathrm{d}F_0}{\mathrm{d}t}, \label{eq:integrand-reduction}
\end{equation}
which, when plugged into Eq.~\eqref{eq:time-asymmetric-work}, shows that the time-asymmetric work $W[\trajx(t)] = \int_0^{t_f} \dot{F_0}(t) \, \mathrm{d}t = F_0(t_f) - F_0(0) = \Delta F$ for \textit{every} trajectory $\trajx(t)$. With optimally chosen $U_0(\boldx, t)$ and $U_1(\boldx, t)$, the free energy difference may be obtained from simulating a single finite-time trajectory. Unfortunately, Eq.~\eqref{eq:integrand-reduction} is typically infeasible to solve for multidimensional systems, and to formulate the PDE, $\dot{F_0}(t)$, and therefore $\Delta F$, must already be known.

\section{Time-asymmetric microscopic fluctuation theorem}

In the late 1990s, Crooks \cite{crooks1998nonequilibrium, crooks1999entropy} discovered that the microscopic fluctuation theorem
\begin{equation}
  W[\trajx(t)] = \Delta F + \beta^{-1} \ln \frac{\mathcal{P}[\trajx(t)]}{\tilde{\mathcal{P}}[\tilde{\trajx}(t)]} \label{eq:work-MFT-definition}
\end{equation}
is satisfied by the traditional work $W = W_\mathrm{trad}$. Here $\mathcal{P}[\trajx(t)]$ is the probability of observing a trajectory $\trajx(t)$, and $\tilde{\mathcal{P}}[\tilde{\trajx}(t)]$ is the probability of observing its time-reversed trajectory $\tilde{\trajx}(t) = \trajx(t_f - t)$ in a ``reverse'' path ensemble driven by the protocol $\tilde{U}(\boldx, t) = U(\boldx, t_f - t)$. In this section, we derive the microscopic fluctuation theorem satisfied by the unconventional work definition Eq.~\eqref{eq:time-asymmetric-work}.

In our overdamped setting, the probability of realizing a trajectory $\trajx(t)$ from the dynamics Eq.~\eqref{eq:forward-langevin} may be formally expressed, up to a normalization factor, as 
\begin{equation}
\mathcal{P}[\trajx(t)] = \rho_A(\trajx(0)) e^{-\beta S[\trajx(t)]}, \label{eq:forward-path-probability}
\end{equation}
where
\begin{equation} 
    S[\trajx(t)] = (\mathrm{I}) \int_0^{t_f} \frac{|\dot{\trajx} + \nabla(U_0 + U_1)|^2}{4} \mathrm{d}t, \label{eq:forward-path-action}
\end{equation}
is the Onsager-Machlup action functional (see Chapter Appendix~\ref{appendix:OM-action} for a quick review, also \cite{adib2008stochastic}). We use  ($\mathrm{I}$) to indicate that the integral is taken in an Itô sense (also reviewed in Chapter Appendix~\ref{appendix:stochastic-integrals}). After Eqs.~\eqref{eq:time-asymmetric-work} and \eqref{eq:forward-path-probability} are plugged into Eq.~\eqref{eq:work-MFT-definition}, straightforward manipulations under the rules of stochastic calculus (see Chapter Appendix~\ref{appendix:MFT-derivation}) yield

\begin{align}
    \tilde{\mathcal{P}}[\tilde{\trajx}(t)] &= e^{-\beta \{ U_A(\trajx(0)) - F_A + S[\trajx(t)] + W[\trajx(t)] - \Delta F\} } \nonumber \\  
    &= \rho_B(\tilde{\trajx}(0)) e^{-\beta \tilde{S}[\tilde{\trajx}(t)]  } , \label{eq:MFT}
\end{align}
where $\rho_B(\boldx) = e^{-\beta[U_B(\boldx) - F_B]}$ is the equilibrium distribution for $U_B(\boldx)$, and
\begin{equation}
\tilde{S}[\tilde{\trajx}(t)] =  (\mathrm{I}) \int_0^{t_f} \frac{|\dot{\tilde{\trajx}} + \nabla (\tilde{U}_0 - \tilde{U}_1)|^2}{4} \mathrm{d}t\label{eq:reverse-path-action} 
\end{equation}
has the form of a path action, with $\tilde{U}_{0,1}(\boldx , t) = U_{0,1}(\boldx , t_f - t)$ denoting the time-reversed potential energies. 
Eq.~\eqref{eq:MFT} gives the probability of observing the path $\tilde{\trajx}(t)$ under the Langevin equation
\begin{equation}
\dot{\tilde{\trajx}}(t) = -\nabla (\tilde{U}_0 - \tilde{U}_1) + \sqrt{2 \beta^{-1}} \boldeta(t) \quad\mathrm{with}\quad \tilde{\trajx}(0) \sim \rho_B(\cdot), \label{eq:reverse-langevin}
\end{equation}
which differs from Eq.~\eqref{eq:forward-langevin} by a minus sign on the $U_1$ term. In other words, the reverse path ensemble that satisfies Eq.~\eqref{eq:MFT} for the time-asymmetric work Eq.~\eqref{eq:time-asymmetric-work} is one that is driven by a protocol $\tilde{U}_0 - \tilde{U}_1$ that is \textit{different} from the time-reversal of the forward protocol $U_0 + U_1$. One can also verify that its associated definition of work 
\begin{equation}
  \tilde{W}[\tilde{\trajx}(t)] = \int_{0}^{t_f} \frac{\partial \tilde{U}_0}{\partial t} - \nabla \tilde{U}_0 \cdot \nabla (-\tilde{U}_1) + \beta^{-1} \nabla^2 (-\tilde{U}_1) \, \mathrm{d}t, \label{eq:reverse-work} 
\end{equation}
satisfies $\tilde{W}[\tilde{\trajx}(t)] = -W[\trajx(t)]$, so the same optimal $U_0(\boldx, t)$ and $U_1(\boldx, t)$ satisfying Eq.~\eqref{eq:sufficient-condition} also give $\tilde{W}[\tilde{\trajx}(t)] = -\Delta F$ for every trajectory. 

Through standard methods (see Chapter Appendix~\ref{appendix:MFT-FT}), the fluctuation theorem Eq.~\eqref{eq:fluctuation-theorem} follows directly from the microscopic fluctuation theorem Eq.~\eqref{eq:work-MFT-definition}. Thus, the time-asymmetric work (Eq.~\eqref{eq:time-asymmetric-work}) holds a deeper significance than how it may first appear -- it relates the forward and reverse path ensembles given by Eqs.~\eqref{eq:forward-langevin} and \eqref{eq:reverse-langevin} that are driven by time-asymmetric protocols. 

Though much more involved, the time-asymmetric fluctuation theorem may also be derived for underdamped dynamics through similar techniques. We include our derivation in {Chapter Appendix \thechapter.C}.

Significantly, the time-asymmetric fluctuation theorem may be exploited for efficient free energy estimation. In particular, by considering optimizing two \textit{different} protocols---one for the forward dynamics and the other for the reverse dynamics---the variance of $\Delta F$ estimates may be lowered by orders of magnitude. We now propose our algorithm demonstrating this.

\section{Algorithm}

In this section we present an on-the-fly adaptive importance-sampling protocol optimization algorithm, inspired by \cite{jie2010connection}, that uses the previously collected bi-directional samples (i.e., from both forward and reverse protocols) to iteratively discover lower-variance time-asymmetric protocols. Exploiting the mathematical structure of the Onsager-Machlup action, our algorithm requires minimal computational overhead, solely the inclusion of easily-computable auxiliary variables in each trajectory's time-evolution.

Concretely, we consider the objective function 
\begin{align}
    J = J_F + J_R = \langle W \rangle_F + \langle \tilde{W} \rangle_R. \label{eq:objective-function} 
\end{align}
Jensen's inequality applied to Eq.~\eqref{eq:jarzynski-equality} implies $\langle W \rangle_F \geq \Delta F$ and $\langle \tilde{W} \rangle_R \geq -\Delta F$, with equality only for zero-variance optimal protocols. 

Our simulations are performed using the Euler-Mayurama method to discretize Eqs.~\eqref{eq:forward-langevin} and \eqref{eq:reverse-langevin}. Instead of directly discretizing Eq.~\eqref{eq:time-asymmetric-work}, we measure for every trajectory the expression derived from Eq.~\eqref{eq:MFT}
\begin{align}
    W[\trajx(t)] &= U_B(\trajx(t_f)) - U_A(\trajx(0)) + \beta^{-1} \ln \frac{\mathcal{P}[\trajx(t) | \trajx(0)]}{\tilde{\mathcal{P}}[\tilde{\trajx}(t) | \tilde{\trajx}(0)]} 
\end{align}
    with the correct discrete-path probabilities, so as to preserve the fluctuation theorem \eqref{eq:work-MFT-definition}. In our setting this may be written as
\begin{equation}
    W[\trajx(t)] = \{ U_B(\trajx(t_f)) + \tilde{S}[\tilde{\trajx}(t)] \} - \{ U_A(\trajx(0)) + S[\trajx(t)] \}.   \label{eq:shadow-work}
\end{equation}

From now on we will use Einstein notation, where repeated upper and lower Greek indices signify summation. Let $\{U_\mu(\boldx, t) \, | \, \mu = 1, ..., M \}$ denote a set of time-dependent basis functions. We used a linear parameterization of the forward and reverse protocols $U_F = U_0 + U_1$ and $U_R = U_0 - U_1$
\begin{equation}
U_{F,R}(\boldx, t) = \begin{cases} U_A(\boldx) &\quad\mathrm{for}\quad t = 0 \\ 
\theta^\mu_{F,R} U_\mu(\boldx, t) &\quad\mathrm{for}\quad t \in (0, t_f) \\ 
U_B(\boldx) &\quad\mathrm{for}\quad t = t_f \\ 
\end{cases} \label{eq:linear-basis}
\end{equation}
 with parameters $\theta = (\theta_F, \theta_R) \in \mathbb{R}^{2 M}$, which explicitly shows that the forward and reverse protocols are not constrained to be equal under time-reversal. Having unequal $\theta_F \neq \theta_R$ yields $U_1 \neq 0$, which gives a counterdiabatic driving that is otherwise absent when forward and reverse protocols are constrained to be equal. By minimizing the work under this parameterization we find the parameters that yield an approximation to the true  counterdiabatic force (i.e., $U_1$ that satisfies the PDE Eq.~\eqref{eq:integrand-reduction}). Under this parameterization, the Onsager-Machlup path action Eq.~\eqref{eq:forward-path-action} and the time-asymmetric work Eq.~\eqref{eq:shadow-work} are quadratic in $\theta$
\begin{align}
  S[\trajx(t); \theta] &= \theta^\mu_F \theta^\nu_F \mathsf{a}_{\mu \nu} + \theta^\mu_F \mathsf{b}_{\mu} + \theta\mbox{-}\mathrm{independent \ terms}, \label{eq:quadratic-action}  \\
  W[\trajx(t); \theta] &= - (\theta^\mu_F \theta^\nu_F \mathsf{a}_{\mu \nu} + \theta^\mu_F \mathsf{b}_{\mu} + \mathsf{c}) +  (\theta^\mu_R \theta^\nu_R \mathsf{\tilde{a}}_{\mu \nu} + \theta^\mu_R \mathsf{\tilde{b}}_{\mu} + \mathsf{\tilde{c}}) , \label{eq:quadratic-work}
\end{align}
where
%
\begin{equation}
    \begin{split}
    \mathsf{a}_{\mu \nu}[\trajx(t)]  &= (\mathrm{I}) \int_0^{t_f} \frac{\nabla U_\mu \cdot \nabla U_\nu}{4} \, \mathrm{d}t, \\
    \mathsf{b}_{\mu}[\trajx(t)]  &= (\mathrm{I}) \int_0^{t_f} \frac{\dot{\trajx} \cdot \nabla U_\mu}{2} \, \mathrm{d}t, \\
      \mathsf{c}[\trajx(t)] &= U_A(\trajx(0)),
\end{split}
    \quad\quad\quad
  \begin{split}
  \mathsf{\tilde{a}}_{\mu \nu}[\trajx(t)] &= (\mathrm{BI}) \int_0^{t_f} \frac{\nabla {U}_\mu \cdot \nabla {U}_\nu}{4} \, \mathrm{d}t,\\
  \mathsf{\tilde{b}}_{\mu}[\trajx(t)]  &=  - (\mathrm{BI}) \int_0^{t_f} \frac{\dot{\trajx} \cdot \nabla {U}_\mu}{2} \, \mathrm{d}t, \\ 
      \mathrm{and} \quad\quad \mathsf{\tilde{c}}[\trajx(t)] &= U_B(\trajx(t_f)) \label{eq:abc}
  \end{split}
\end{equation}
%
are $\theta$-\textit{independent} functionals of the time-discretized trajectory $\trajx(t)$.\footnote{The astute reader might point out that under the rules of stochastic calculus, how they are written out $\mathsf{a} = \mathsf{\tilde{a}}$ for all $\boldx(t)$. However, they are different under a finite time-discretization, and thus must both be kept track.} Here, (BI) refers to a Backwards Itô integral, needed to write terms of the reverse ensemble $\tilde{S}[\tilde{\trajx}(t)]$ as a functional of $\trajx(t)$. (Eqs.~\eqref{eq:quadratic-action}--\eqref{eq:abc} apply for the reverse path ensemble $\tilde{S}[\tilde{\trajx}(t)], \tilde{W}[\tilde{\trajx}(t)]$, through the transformation $t \rightarrow t_f - t$, $\{F, R\} \rightarrow \{R, F \}$.) These variables $\mathsf{a}, \mathsf{\tilde{a}} \in \mathbb{R}^{M \times M}, \mathsf{b}, \mathsf{\tilde{b}} \in \mathbb{R}^M$, and $\mathsf{c}, \mathsf{\tilde{c}} \in \mathbb{R}$ are akin to the eligibility trace variables (sometimes called ``Malliavin weights'') used in reinforcement learning policy-gradient algorithms \cite{williams1992simple, peters2008reinforcement, warren2012malliavin, das2019variational, das2022direct}, which are dynamically evolved with each trajectory $\trajx(t)$.

In the following two paragraphs we consider only the forward ensemble for simplicity. When a trajectory $\trajx^i(t)$ is sampled, we calculate not only its work $W^i = W[X^i(t)]$ but also its functional values $(\mathsf{a}^{i}, \mathsf{b}^{i}, \mathsf{c}^{i}, \mathsf{\tilde{a}}^{i}, \mathsf{\tilde{b}}^{i}, \mathsf{\tilde{c}}^{i}$) (Eq.~\eqref{eq:abc}), which are saved alongside the protocol parameters $\theta^{i} = (\theta_F^{i}, \theta_R^{i})$ employed to sample the trajectory. Given $n_s$ sampled trajectories, the collected data $\{(W^i, \mathsf{a}^{i}, \mathsf{b}^{i}, \mathsf{c}^{i}, \mathsf{\tilde{a}}^{i}, \mathsf{\tilde{b}}^{i}, \mathsf{\tilde{c}}^{i}, \theta^i)  \}_{i = 1}^{n_s}$ may be used to construct a $\theta$-dependent importance-sampling estimator for $\langle W \rangle_F$
\begin{equation}
  \hat{J}_F(\theta) = \frac{\sum_{i=1}^{n_s} \mathsf{r}^i_F (\theta) \, \mathsf{w}^i_F(\theta) }{ \sum_{i=1}^{n_s} \mathsf{r}^i_F(\theta) }, \label{eq:importance-sampling-W-estimator}
\end{equation}
where the sum is over collected forward samples $i$, $\mathsf{r}^i_F(\theta)$ is the likelihood ratio (i.e., the Radon–Nikodym derivative)
\begin{equation}
    \mathsf{r}^i_F(\theta) = \frac{\mathcal{P}[\trajx^i(t) \  \mathrm{from} \ \theta]}{\mathcal{P}[\trajx^i(t) \  \mathrm{from} \  \theta^{i}]} = e^{-\beta (S[ \trajx^i(t); \theta] - S[ \trajx^i(t); \theta^i] )}.
\end{equation}
Here, $S[X^i(t); \theta]$ and $\mathsf{w}^i(\theta) = W[\trajx^i(t); \theta]$, defined in Eqs.~\eqref{eq:quadratic-action} and~\eqref{eq:quadratic-work}, are the path action and time-asymmetric work for the trajectory $\trajx^i(t)$ as if it were sampled under $\theta$ instead of the protocol $\theta^{i}$ it was actually sampled under. \textit{Protocols may now be optimized by using Eq.~\eqref{eq:importance-sampling-W-estimator} as a surrogate objective function, which yields a newly optimized protocol $\theta^*$ that can be used to iteratively sample even lower-variance trajectories.}

Of course, the quality of the importance-sampling estimate Eq.~\eqref{eq:importance-sampling-W-estimator} degrades the further away the input $\theta$ is from the set of $\theta^{i}$ under which samples were collected. One common heuristic of this degradation is the effective sample size \cite{mcbook}
\begin{equation}
    n_F^{\mathrm{eff}}(\theta) = \frac{ \big(\sum_{i=1}^{n_s} \mathsf{r}^i_F(\theta)\big)^2 }{ \sum_{i=1}^{n_s} \mathsf{r}^i_F(\theta)^2 }, 
\end{equation}
ranging from $1$ (uneven $\mathsf{r}^i_F$ values, high degradation) to $n_s$ (equal $\mathsf{r}^i_F$ values, low degradation).

We now state our algorithm (pseudocode given in the  {Alg.~1}): at each iteration, an equal number of independent forward and reverse trajectories are simulated through Eqs.~\eqref{eq:forward-langevin} and \eqref{eq:reverse-langevin} using the $U_F, U_R$ specified by current parameters $\theta_\mathrm{curr}$, with the time-asymmetric work $W$ and auxiliary variables ($\mathsf{a}, \mathsf{b}, \mathsf{c}, \mathsf{\tilde{a}}, \mathsf{\tilde{b}}, \mathsf{\tilde{c}}$) of each trajectory dynamically calculated; then the protocol is updated through solving the nonlinear constrained optimization problem
\begin{equation}
    \theta_\mathrm{next} = \mathrm{argmin}_\theta \big\{ \hat{J}(\theta) \, | \, \{n_F^{\mathrm{eff}}(\theta),n_R^{\mathrm{eff}}(\theta)\} \geq f n_s \big\}, \label{eq:optimization}
\end{equation}
for which there are efficient numerical solvers (e.g. SLSQP \cite{kraft1988software} pre-implemented in SciPy \cite{2020SciPy-NMeth}). Here $\hat{J}(\theta) = \hat{J}_F(\theta) + \hat{J}_R(\theta)$, $n_F^{\mathrm{eff}}(\theta)$ and $n_R^{\mathrm{eff}}(\theta)$ are constructed with the $n_s$ forward and $n_s$ reverse samples collected over all iterations, and $f \in [0, 1)$ is a hyperparameter specifying the constraint strength: the fraction of total samples we are constraining $n^\mathrm{eff}_{F,R}$ to be greater or equal than. Finally, $\widehat{\Delta F}_\mathrm{BAR}$ is calculated with the bi-directional work measurements collected across all iterations using Eq.~\eqref{eq:BAR-estimator}, which is permitted by the satisfaction of Eq.~\eqref{eq:mixtures}.

\section{Numerical examples}

In this section we report the performance of our algorithm for three test model systems.

We chose our basis set in order to represent protocols of the form
\begin{equation}
    U(\boldx, t) = \lambda_{A}(t) \, U_A(\boldx) + \lambda_{B}(t) \, U_B(\boldx) + \lambda_C(t) \, U_C(\boldx) \label{eq:protocol-form}
 \end{equation}
where $U_C(\boldx)$ is an additional quasi-counterdiabatic potential 
\begin{equation}
    U_C(\boldx) = -\boldc \cdot \boldx \quad\mathrm{with}\quad \boldc = \frac{\langle \boldx \rangle_B - \langle \boldx \rangle_A}{t_f} 
\end{equation}
that provides a spatially-uniform forcing proportional to the difference in equilibrium mean positions $\langle \boldx \rangle_{A,B} = \int \boldx \, \rho_{A,B}(\boldx) \, \mathrm{d}\boldx$,\footnote{The first two terms are a generalization of the protocol form $U_\lambda(\boldx) = (1 - \lambda) \, U_A(\boldx) + \lambda \, U_B(\boldx)$ commonly considered in stochastic thermodynamics. Typically, linear combinations of just $U_A(\boldx)$ and $U_B(\boldx)$ are not effective in transporting $\rho_A(\boldx) \rightarrow \rho_B(\boldx)$, motivating our inclusion of the linear potential $U_C(\boldx)$. We found that empirically, by including this easy-to-compute $U_C(x)$, we achieved much higher performance than using only $U_A(x)$ and $U_B(x)$.} that is needed to, e.g., spatially translate an equilibrium distribution (see Appendix A in \cite{li2017shortcuts}). The basis set is given by
\begin{equation}
    \bigg\{ U_{\ell}(\boldx) \, p_m\bigg(\frac{2t}{t_f} - 1 \bigg)  \, \bigg| \, \ell \in \{A, B, C\}, \, 0 \leq m \leq m_\mathrm{max} \bigg\},  
\end{equation}
where $p_m(\cdot)$ denotes the $m$-th Legendre polynomial.

For all numerical examples, $m_\mathrm{max} = 4$ was used; we found that using a larger $m_\mathrm{max}$ did not substantially improve performance, while drastically increasing computational runtime (see Discussion section below, where we discuss how computational runtime scales with number of parameters). The algorithm was initialized with 120 bi-directional samples drawn from a generic naive linear interpolation protocol $\lambda_A(t) = 1 - t/t_f, \lambda_B(t) = t/t_f, \lambda_C(t) = 0$ for both $U_F$ and $U_R$. At each iteration Eq.~\eqref{eq:optimization} was solved for $n_\mathrm{mb} = 20$ independently subsampled minibatches of size $n_s^\mathrm{mb} = 80$ with $f = 0.3$; the protocol was then updated to the minibatch-averaged $\theta_\mathrm{next}$; finally, 20 additional bi-directional samples were drawn with the new protocol. In total, 44 iterations were performed, giving 1000 bi-directional samples.

\begin{figure}[h]
	\centering
	\includegraphics[width=.6\textwidth]{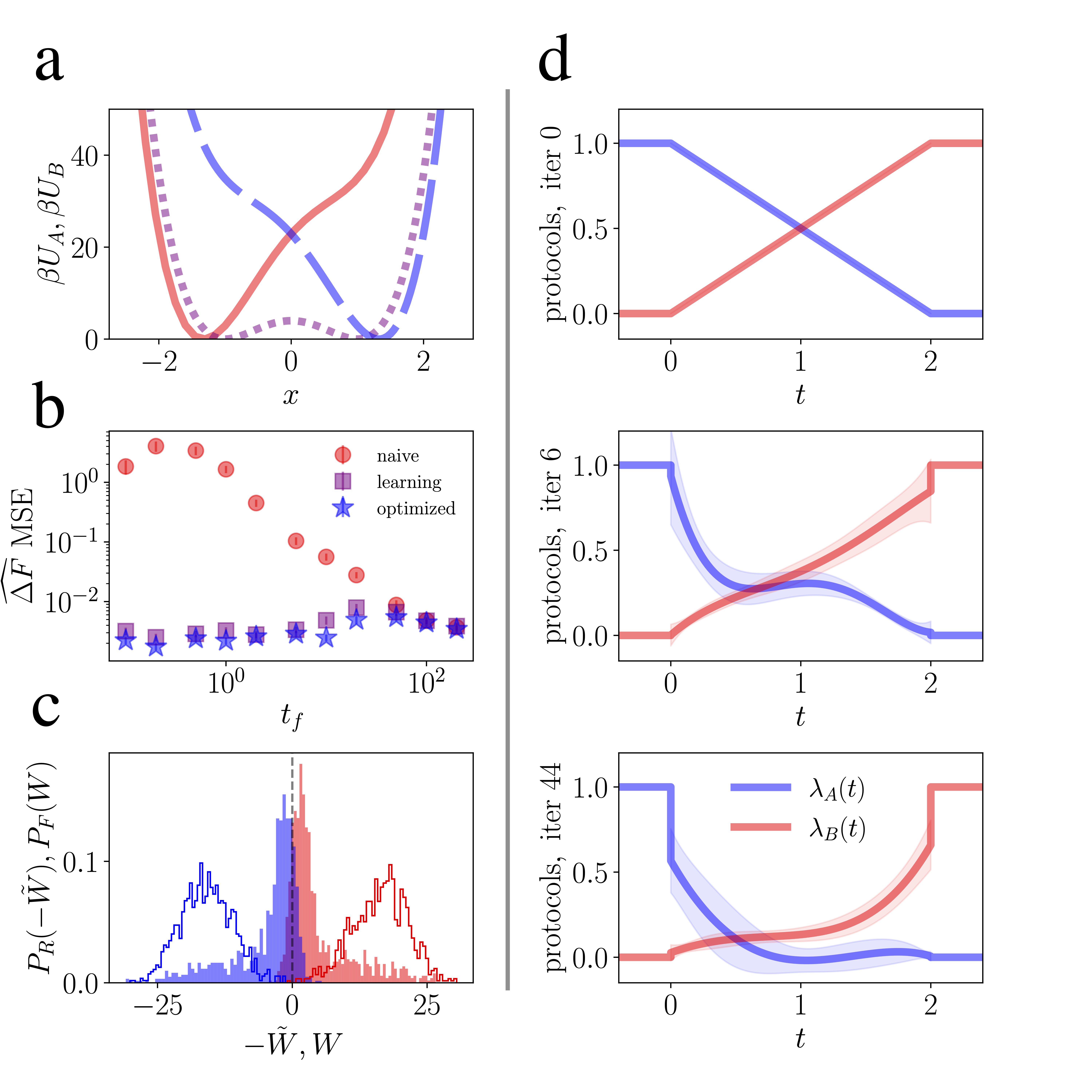}
	\caption[Optimal time-asymmetric protocols for the biased double well, their optimization, and their performance in $\Delta F$ estimation.]{ (a) The potentials $U_A(\boldx)$ ( {solid} red) and $U_B(\boldx)$ ( {dashed} blue) are obtained by linearly biasing a double well ( {dotted} purple). (b) $\widehat{\Delta F}_\mathrm{BAR}$ mean squared error from 1000 bi-directional measurements drawn solely from the naive protocol (red  {circles}), cumulatively from protocols that are adaptively optimized (``learning'') with our algorithm (purple  {squares}), and solely from the last-iteration (``optimized'') protocols (blue  {stars}) for various protocol times $t_f$. (c) Single-trial $\mathcal{P}_F(W)$ (red,  {right of the dashed vertical line}) and $\mathcal{P}_R(-\tilde{W})$ (blue,  {left of the dashed vertical line}) work distributions for 1000 measurements from the naive protocol, (unfilled) and adaptively optimized protocols (filled) for $t_f = 2$, the analytic ground truth $\Delta F = 0$ shown as a grey dashed line. Measurements made with protocol optimization have significantly more overlap, leading to lower estimator error. (d) Forward protocols $\lambda_A(t)$ ({blue, going from} $\lambda_A(0) = 1$  {to} $\lambda_A(t_f) = 0$) and $\lambda_B(t)$ ( {red, going from} $\lambda_B(0) = 0$  {to} $\lambda_B(t_f) = 1$) at various iterations of protocol optimization for $t_f = 2$. Shaded region represents variability across $100$ independent trials. In the optimized last-iteration protocol, $\lambda_A(t) + \lambda_B(t)$ (giving the energy scale) is greatly reduced at intermediate times, while $\lambda_B(t) - \lambda_A(t)$ (giving the linear bias) is time-asymmetrically shifted. The reverse protocols (not shown here) are similar,  {See Fig.~}\ref{fig:SI-DW}.}
	\label{dw_mw_fig}
\end{figure}

\subsection{Linearly-Biased double well}

The first system we consider is a one-dimensional quartic double-well with a linear bias (Fig.~\ref{dw_mw_fig}(a)). The potentials are $U_A(x) = E_0[ (x^2 - 1)^2/4 - x]$, $U_B(x) = E_0[ (x^2 - 1)^2/4 + x]$ (cf. \cite{zhong2022limited} for optimal protocols minimizing $\langle W_\mathrm{trad}\rangle_F$). We set $U_C(x) = 0$ because $U_B(x) - U_A(x)$ is already linear in $x$. We use $\beta = 1, E_0 = 16$, and a timestep $\Delta t = 1 \times 10^{-3} \tau$ where $\tau = 1$ is the natural timescale (here the length scale, inverse temperature, and friction coefficient are all unity $\ell = \beta = \gamma = 1$).

Fig.~\ref{dw_mw_fig}(b) displays the $\widehat{\Delta F}_\mathrm{BAR}$ estimator mean squared error for 1000 bi-directional work measurements collected solely from the naive protocol (red), the 1000 measurements collected cumulatively over on-the-fly protocol optimization (purple), and 1000 measurements collected solely from the last-iteration (blue). Each dot represents the empirical average over 100 independent trials. Note that the mean squared error is up to 1600 times lower under protocol optimization compared to under the naive protocol (obtained at $t_f = 0.2$). For $t_f \gtrsim 10$ the algorithm does not converge within the 1000 measurements  {(see Fig.~}\ref{fig:SI-DW} {)}, leading to less improvement. Fig.~\ref{dw_mw_fig}(c) shows that bi-directional work measurements collected under the protocol optimization algorithm have significantly more overlap than measurements collected from the naive protocol, leading to reduced estimator error. Fig.~\ref{dw_mw_fig}(d) gives snapshots on how the optimal protocol is adaptively learned. Note that the optimized time-asymmetric protocols feature discontinuous jumps occurring at $t = 0$ and $t= t_f$, which has been observed to be ubiquitous for time-symmetric protocols ~\cite{schmiedl2007optimal,then2008computing, bonancca2018minimal, naze2022optimal, blaber2021steps, zhong2022limited, whitelam2023train, rolandi2023optimal, engel2023optimal, esposito2010finite} (also see \cite{zhong2024beyond} for a recently proposed explanation for this phenomenon).

\begin{figure}[h]
	\centering
	\includegraphics[width=.7\textwidth]{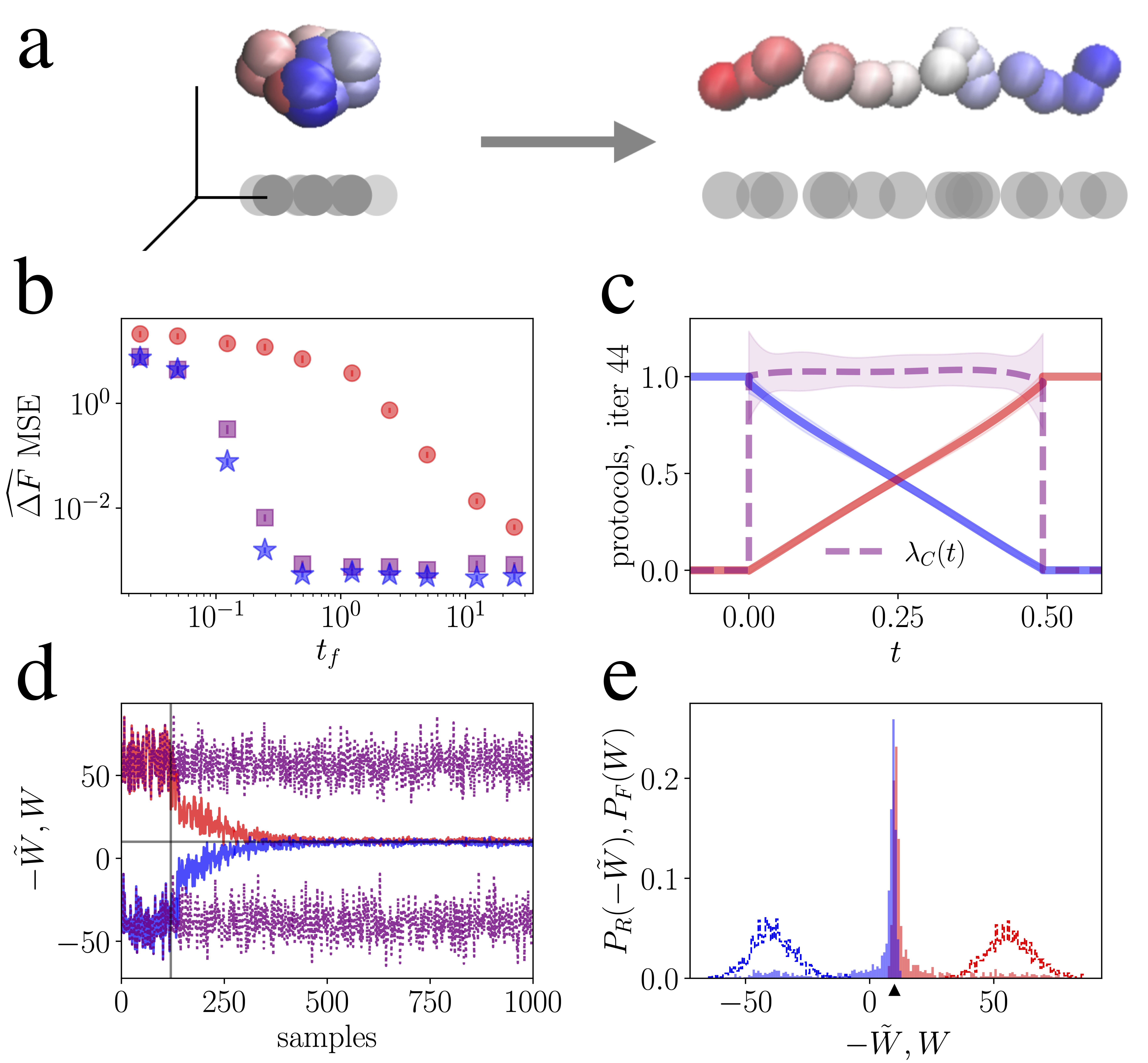}
	\caption[Optimal time-asymmetric protocols for the Rouse polymer model.]{(a) A Rouse polymer is stretched from a collapsed state to an extended one. (b) $\widehat{\Delta F}_\mathrm{BAR}$ mean squared error verses protocol time, points colored as in Fig.~\ref{dw_mw_fig}. (c) For moderate protocol times ($t_f = 0.5 \, \tau_\mathrm{R}$ displayed here) the optimized protocol $(\lambda_A(t), \lambda_B(t), \lambda_C(t))$ learned in 44 iterations is the counterdiabatic protocol Eq.~\eqref{eq:rouse-cd}.  {Here, }$\lambda_C(t)$ { is drawn with the dashed purple lines.} (d) Single-trial bi-directional work samples from the naive protocol ($W$ and $-\tilde{W}$  {dotted} purple) and adaptively-optimized protocols ($W$ red  {starting above zero}, $-\tilde{W}$ blue  {starting below zero}) for $t_f = 0.5 \, \tau_\mathrm{R}$. Vertical line demarcates start of protocol optimization. The analytic ground truth value $\Delta F$ is shown as a horizontal line. (e) Work distributions corresponding to the samples in (d). The ground truth is indicated by the triangular arrow. Cumulative measurements made under protocol optimization (filled) have dramatically greater overlap than measurements made under the naive protocols (unfilled), leading to lower estimator error.}
	\label{Rouse_mw_fig}
\end{figure}

\subsection{Rouse polymer}

Next we consider a $(N+1)$-bead Rouse polymer 
(Fig.~\ref{Rouse_mw_fig}(a)) with stiffness $k$ and intrinsic energy given by $U_\mathrm{Rouse}(x_0, x_1, ..., x_{N}) = \sum_{n=0}^{N-1} (k/2) (x_{n+1} - x_n)^2$ from harmonic bonds between adjacent beads \cite{doi1988theory}.\footnote{Because of the harmonic nature of the interbead potential, the dynamics of the Rouse polymer separates in its 3 spatial dimensions; without loss of generality, we can consider just a single spatial dimension.} We estimate $\Delta F$ between a collapsed state (fixing $x_0 = x_N = 0$) and an extended state (fixing $x_0 = 0$, $x_N = \lambda_f$), so our configuration space is $\boldx \in \mathbb{R}^{N - 1}$ with potential energies $U_A(x_1, ..., x_{N-1}) = U_\mathrm{Rouse}(0, x_1, ..., x_{N-1}, 0)$ and $U_B(x_1, ..., x_{N-1}) = U_\mathrm{Rouse}(0, x_1, ..., x_{N-1}, \lambda_f)$. Equilibrium averages $\langle x_n \rangle_A = 0, \langle x_n \rangle_B = n \lambda_f  / N$ give $U_C(\boldx) = - (\lambda_f / N t_f) \sum^N_{n=1}  n x_n$, and the analytic ground truth free energy  {(i.e., the true free energy difference value against which we can compare algorithm estimates)} is given by $\Delta F = F_B - F_A = k \lambda_f^2 / (2N)$. It may be verified that for this problem the time-varying potential energies
\begin{gather}
  U_0(\boldx, t) = \bigg( 1 - \frac{t}{t_f} \bigg) U_A(\boldx) + \bigg(\frac{t}{t_f} \bigg)U_B(\boldx) \nonumber \\ U_1(\boldx, t) = U_C(\boldx) \label{eq:rouse-cd}
\end{gather}
solve Eq.~\eqref{eq:integrand-reduction} and are thus counterdiabatic. 

We use $\beta  = k = 1$, $N = 20$, and timestep $\Delta t = 2.5 \, \times 10^{-5} \tau_\mathrm{R}$ where $\tau_\mathrm{R} = \beta N^2/\pi^2$ is the Rouse relaxation time \cite{doi1988theory}. Initial conditions for $\rho_A(\boldx)$ and $\rho_B(\boldx)$ were drawn from a normal-modes basis as described in {Chapter Appendix \thechapter.D}. Fig.~\ref{Rouse_mw_fig}(b) shows an improvement of up to $8300$ (for $t_f = 0.5 \, \tau_\mathrm{R}$) in $\widehat{\Delta F}_\mathrm{BAR}$ mean squared error between naive and optimized protocols. The counterdiabatic solution Eq.~\eqref{eq:rouse-cd} corresponds to $\lambda_A(t) = (1 - t / t_f), \lambda_B(t) = t / t_f$, and $\lambda_C(t) = 1$, which what the algorithm learns for $t_f = \, \tau_\mathrm{R}$ as depicted in Fig.~\ref{Rouse_mw_fig}(c). (This was generally the case for $t_f \geq 0.5 \, \tau_\mathrm{R}$. For $t_f < 0.5 \, \tau_\mathrm{R}$ the algorithm learns a sub-optimal local solution that still provides some improvement,  {see Fig.~}\ref{fig:SI-Rouse}.) Fig.~\ref{Rouse_mw_fig}(d) shows single-trial traces of bi-directional work measurements for the naive protocol (purple) and adaptively-optimized protocols (red for $W$, blue for $-\tilde{W}$), for $t_f = 0.5 \tau_\mathrm{R}$. The protocol converges in $\sim 20$ iterations (requiring $\sim 500$ measurements), and then consistently gives work measurements closely centered at the ground truth free energy (gray horizontal line). Histograms of these traces (filled) are shown in Fig.~\ref{Rouse_mw_fig}(e), exhibiting a remarkable increase in the overlap compared with their naive counterparts (unfilled).

\subsection{Worm-like chain with attractive linker}

We now consider a $(N+1)$-bead worm-like chain model (WLC) in 2 dimensions with an added Lennard-Jones interaction between the first and last beads (similar to the 3rd example of \cite{kuznets2023inferring}). Fixing $(x_0, y_0) = (0, 0)$, the configuration space is $\vec{\phi} \in \mathbb{R}^N$, where $\phi_n$ is the angle of the $n$th bond with respect to the $x$-axis, with $(x_n(\vec{\phi}), y_n(\vec{\phi})) = (\sum_{m=1}^n \cos \phi_m, \sum_{m=1}^n \sin \phi_m)$. The angular potential $U_\phi = k \sum^{N-1}_{n=1} [1 - \cos(\phi_{n+1}-\phi_{n})]$ penalizes the bending of adjacent bonds, and $U_\mathrm{LJ} = 4 \epsilon_\mathrm{LJ} [(\sigma_\mathrm{LJ}/ r_N )^{12}-(\sigma_\mathrm{LJ} / r_N )^6]$ specifies the interaction between first and last beads, where $r_N = \sqrt{ x_N^2 + y_N^2}$ is the end-to-end distance. We take $k = 6, \beta = 1, \epsilon_\mathrm{LJ} = 8, \sigma_\mathrm{LJ} = 4$, and $N = 15$. 

\begin{figure}[h]
	\centering
	\includegraphics[width=\textwidth]{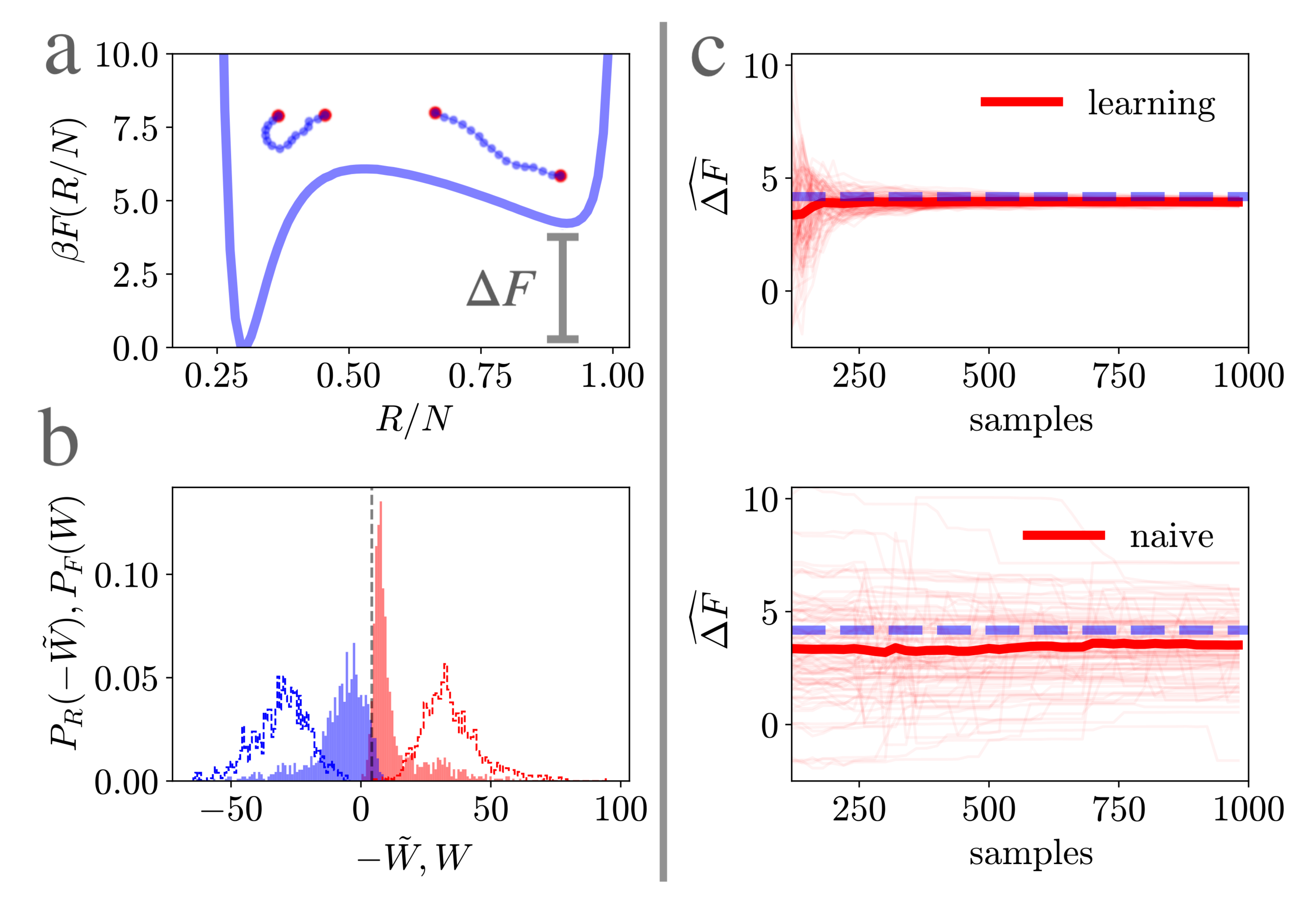}
	\caption[Optimal time-asymmetric protocols for the worm-like chain with attractive linker]{ Worm-like chain with an attractive linker. (a) Numerically obtained ground truth free energy surface relative to its value at $R_A = 2^{1/6} \sigma_\mathrm{LJ}$; the left well corresponds to the ends of the chain bound to one another and the right well corresponds to a nearly straight configuration, with a free energy difference $\Delta F \approx 4.18$. (b) Work distributions before (unfilled) and during (filled) optimization for $t_f = 0.71$, the ground truth shown as a grey dashed line.  {(Same coloring as Fig. 1.)} (c) The $\widehat{\Delta F}_\mathrm{BAR}$ estimator updated over the $1000$ samples converges to the ground truth value (dashed blue line) much more quickly under protocol optimization than under the naive protocol. At $1000$ samples the protocol optimization free energy estimate was $\widehat{\Delta F} = 3.94 \pm 0.11$, while for the naive protocol was $\widehat{\Delta F} = 3.52 \pm 1.48$ (c.f. ground truth value of $\Delta F = 4.18$). It took $200$ total samples for the mean squared error to drop below $1.00 \, (k_\mathrm{B} T)$ under protocol optimization.}
	\label{WLC_mw_fig}
\end{figure}

\begin{figure}
\centering
\includegraphics[width=\textwidth]{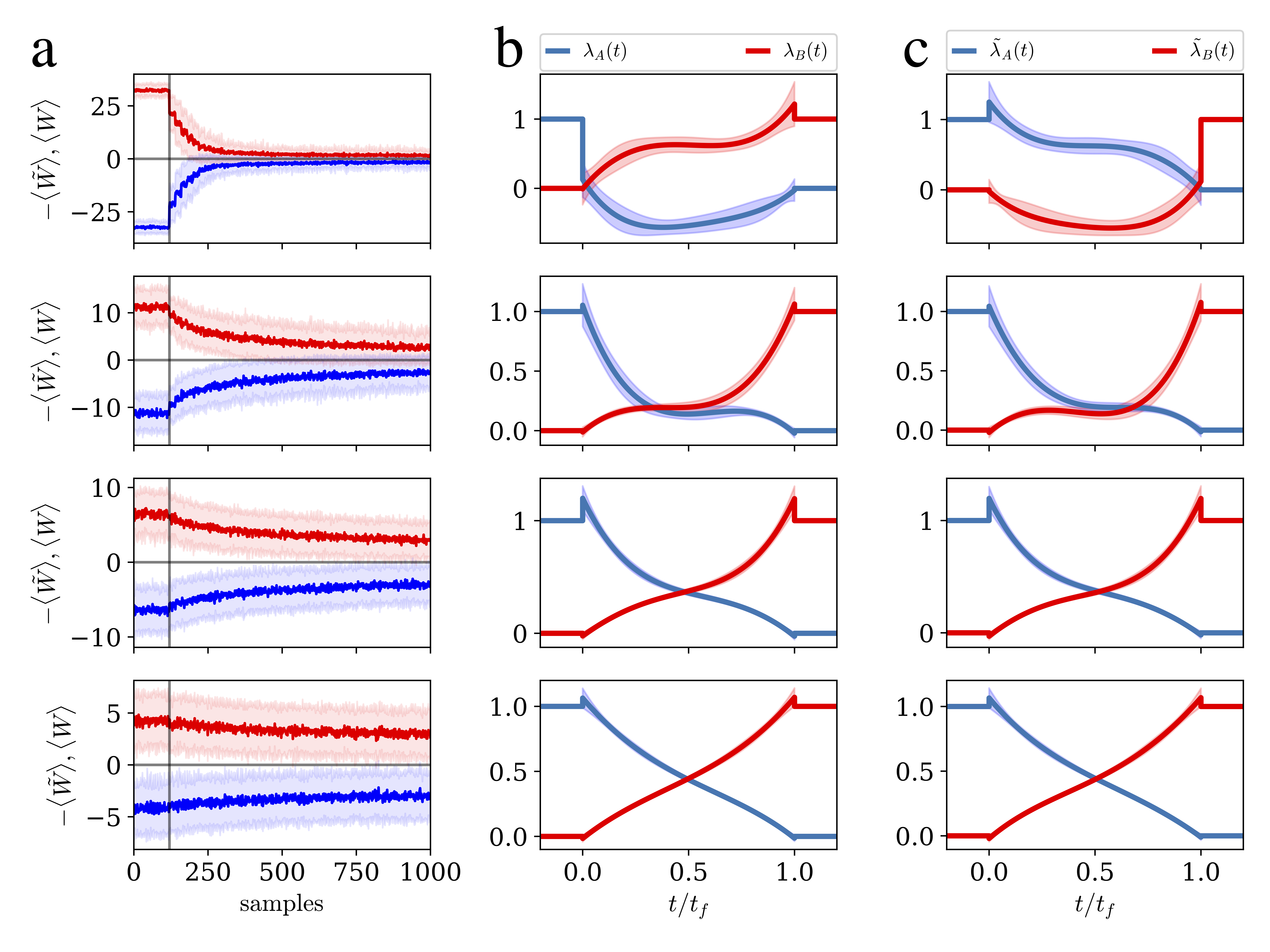}
    \caption[Convergence of the protocol optimization for the linearly-biased double well.]{(a) For the linearly-biased double well, ensemble-averaged work across 100 trials, as a function of sample number.  {Same coloring as Fig. 2(d).} Parameter optimization begins at 120 samples, and happens every 20 samples. The rows correspond to protocol times $t_f = 0.2, 5.0, 20.0,$ and $50.0$ respectively. Convergence is slower for larger $t_f$; for $t_f = 50.0$ the protocol may not have converged within 1000 samples. (b) The forward protocols $U_F(\cdot, t) = \lambda_A(t) U_A(\cdot) + \lambda_B(t) U_B(\cdot)$ after 1000 samples.  {Same coloring as Fig. 1(d).} Rows correspond to the same $t_f$. (c) The reverse protocols $U_R(\cdot, t) = \tilde{\lambda}_A(t) U_A(\cdot) + \tilde{\lambda}_B(t) U_B(\cdot)$ after 1000 samples. The reverse protocols appear to satisfy $\tilde{\lambda}_A(t) = \lambda_B(t_f - t)$ and $\tilde{\lambda}_B(t) = \lambda_A(t)$, which is due to the symmetry of $U_A(\cdot)$ and $U_B(\cdot)$ in the problem.  }
    \label{fig:SI-DW}
\end{figure}

\begin{figure}
\centering
\includegraphics[width=\textwidth]{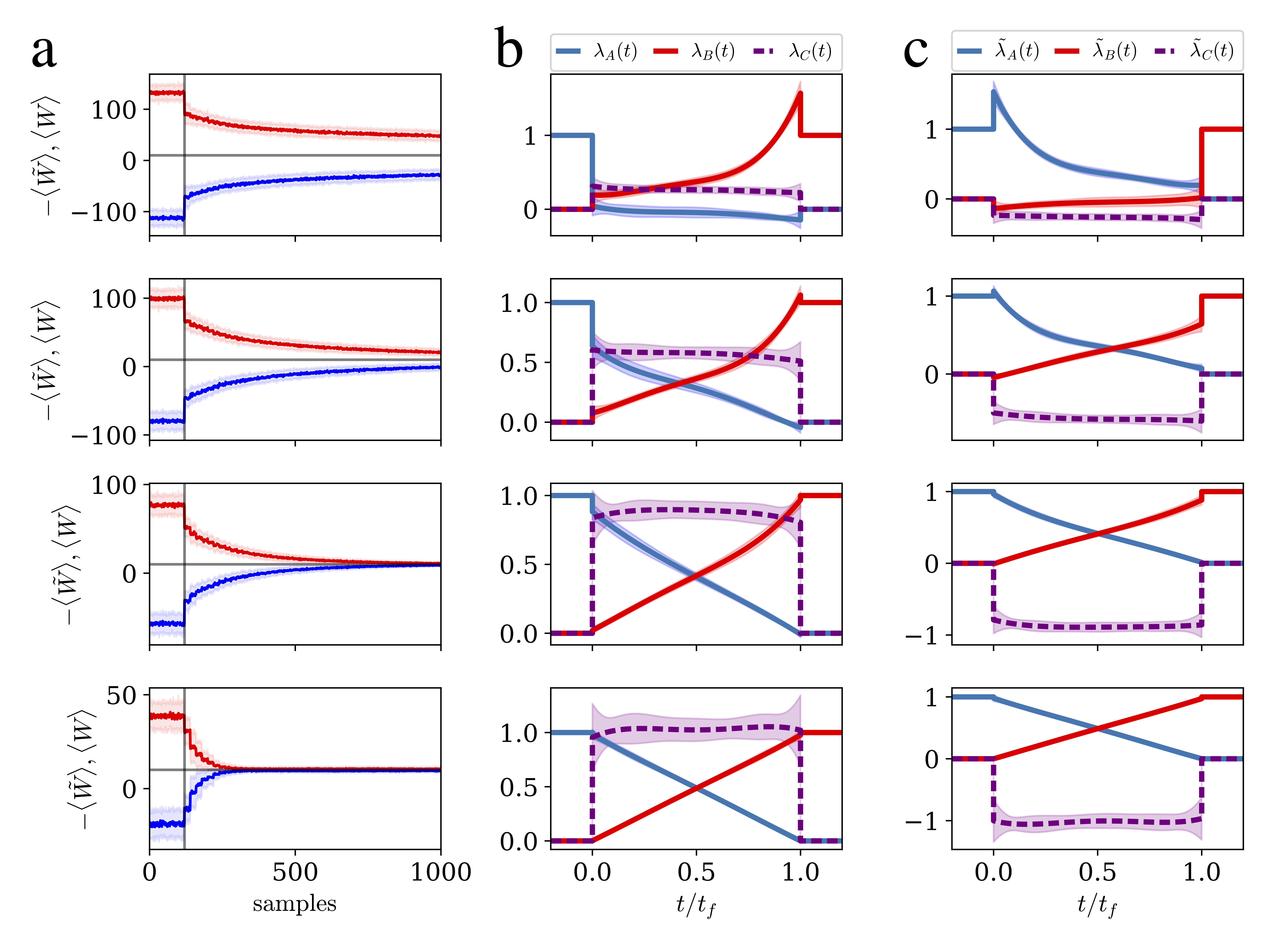}
    \caption[Convergence of the protocol optimization for the Rouse polymer.]{(a) For the Rouse polymer, ensemble-averaged work across 100 trials, as a function of sample number. Parameter optimization begins at 120 samples, and happens every 20 samples.  {Same coloring as Fig. 2(d).} The rows correspond to protocol times $t_f = 0.05 \, \tau_\mathrm{R} , 0.12 \, \tau_\mathrm{R} , 0.25 \, \tau_\mathrm{R},$ and $ 1.23 \, \tau_\mathrm{R}$ respectively. Unlike for the double well, convergence here is slower for smaller $t_f$. (b) The forward protocols $U_F(\cdot, t) = \lambda_A(t) U_A(\cdot) + \lambda_B(t) U_B(\cdot) + \lambda_C(t) U_C(\cdot) $ after 1000 samples.   {Same coloring as Fig. 2(c).} Rows correspond to the same $t_f$. For $t_f = 0.05 \, \tau_\mathrm{R}$ and $t_f = 0.12 \, \tau_\mathrm{R}$, the protocol has not yet converged to the counterdiabatic solution $\lambda_A(t) = (1 - t / t_f), \lambda_B(t) = t / t_f, \lambda_C = 1$. (c) The reverse protocols $U_R(\cdot, t) = \tilde{\lambda}_A(t) U_A(\cdot) + \tilde{\lambda}_B(t) U_B(\cdot) + \tilde{\lambda}_C(t) U_C(\cdot)$ after 1000 samples. The reverse protocols appear to satisfy $\tilde{\lambda}_A(t) = \lambda_B(t_f - t)$, $\tilde{\lambda}_B(t) = \lambda_A(t)$, and $\tilde{\lambda}_C(t) = -\lambda_C(t)$, which is due to the symmetry of $U_A(\cdot), U_B(\cdot)$, and $U_C(\cdot)$ in the problem. }
    \label{fig:SI-Rouse}
\end{figure}

Fig.~\ref{WLC_mw_fig}(a) displays the conditioned free energy $F(R/N) := - \beta^{-1} \ln \rho_\mathrm{eq}(r_N = R)$, where $\rho_\mathrm{eq}$ is the equilibrium probability of observing the end-to-end distance under $U = U_\phi + U_\mathrm{LJ}$ (constructed from $10^7$ equilibrium samples of $U_\phi$, obtained with the Metropolis-adjusted Langevin Algorithm \cite{roberts1996exponential}, that were reweighted by $U_\mathrm{LJ}$).\footnote{From the change of coordinates $R = \sqrt{x_N^2 + y_N^2}, \vartheta = \tan^{-1}(y_N / x_N)$ and the radial symmetry of $U_\phi$ and $U_\mathrm{LJ}$, one has $\rho_\mathrm{eq}(R) \propto (2\pi R)^{-1} \rho_\mathrm{eq}(x_N = R, y_N = 0)$.} $F(R/N)$ exhibits a deep well for $R_A \approx 2^{1/6} \sigma_\mathrm{LJ}$ (trapped/bent state) and a shallow well at large $R_B \approx 0.9 \, N$ (free/relaxed state), separated by a barrier; their difference in value $\Delta F \approx 4.18$ may be calculated by estimating the $\Delta F$ between $U_A(\vec{\phi}) = U_\phi + U_\mathrm{LJ} + (k_\mathrm{ext}/2)(r_N-\lambda_i)^2$ and $U_B(\vec{\phi}) = U_\phi + U_\mathrm{LJ} + (k_\mathrm{ext}/2)(r_N - \lambda_f)^2$ for $\lambda_i = 2^{1/6}\sigma_\mathrm{LJ}, \lambda_f = 0.9 \, N$, and $k_\mathrm{ext} \gg 1$. 

We calculate the $\Delta F$ between $U_A$ and $U_B$ for $k_\mathrm{ext} = 200$. We use timestep $\Delta t = 1.41 \times 10^{-4} \, \tau_\mathrm{LJ}$ where $\tau_\mathrm{LJ} = \sqrt{\epsilon_\mathrm{LJ}/\sigma^2}$ is the Lennard-Jones timescale. We use $U_C(\vec{\phi}) = -\sum_n c_n r_n$, radially pulling on each individual bead, constructed with $c_n = (\langle r_n \rangle_B - \langle r_n \rangle_A)/t_f$ from equilibrium samples of $\rho_A$ and $\rho_B$. Fig.~\ref{WLC_mw_fig}(b) displays single-trial work histograms for $t_f = 0.71 \, \tau_\mathrm{LJ}$, showing that work measurements made under our protocol optimization algorithm are much closer to the ground truth $\Delta F \approx 4.18$ (numerically obtained via the Metropolis-adjusted Langevin Algorithm, see previous paragraph), as opposed to using the naive protocol. Fig.~\ref{WLC_mw_fig}(c) shows the updating $\widehat{\Delta F}_\mathrm{BAR}$ estimator over 100 independent trials converges substantially faster to the ground truth. With 1000 bi-directional samples under the naive protocol the mean squared error is $1.62 \ (k_\mathrm{B} T)$; under protocol optimization, only 160 samples (i.e., just after two iterations of protocol optimization) are required to have a smaller mean squared error. Over various $t_f$, the mean squared error is up to 120 times lower under protocol optimization compared to under the naive protocol,  {see Fig.~}\ref{fig:SI-WLC-performance}.

\section{Discussion}

In this chapter, we derived the time-asymmetric microscopic fluctuation theorem for the unconventional work introduced by \cite{vaikuntanathan2008escorted}. We then demonstrated its practical utility for free energy estimation by presenting an adaptive time-asymmetric protocol optimization algorithm, whose effectiveness we illustrated with three toy models of varying complexity. Time-asymmetric protocols have been considered before \cite{li2017shortcuts, li2019stochastic, blaber2020skewed}, but to our knowledge we are the first to use $\widehat{\Delta F}_\mathrm{BAR}$ on bi-directional work measurements from adaptive time-asymmetric protocols. A clear next step is to test our algorithm on more physically realistic systems. This work should be straightforward to implement with JAX-MD \cite{schoenholz2020jax}. In principle our algorithm should work with underdamped dynamics \cite{li2019stochastic}, and it should also be possible to adaptively optimize the protocol time $t_f$ and sampling ratio $n_F/n_R$. Another future direction to pursue is to differentially weight early versus later samples in the estimator to account for differences in the variance of work measurements, as the algorithm more closely approximates a counterdiabatic protocol. 

The fast convergence in our method comes from exploiting of the quadratic structure of the Onsager-Machlup path action to construct $\hat{J}(\theta)$, which allows all samples to be used in each optimization step. Typically the most computationally expensive step in a molecular dynamics simulation is calculating potential energy gradients $\nabla U$ to evolve $\trajx(t)$, which does not need to be repeated to evolve $(\mathsf{a}_{\mu \nu}, \mathsf{b}_\mu, ...)$. A valid critique of our algorithm is that the number of auxiliary variables included with each trajectory scales quadratically with the number of basis functions, becoming prohibitively large when considering, for example, a separate control force on each particle of a many particle system. However, we have shown that a small number of basis functions to represent Eq.~\eqref{eq:protocol-form} already produces a substantial improvement in efficiency for our three examples. That said, it is straightforward to add additional basis functions (cf. Eq. (2) of \cite{naden2014linear}), which may be useful for more complex and realistic systems. It would be interesting to apply recent methods \cite{singh2023variational} to learn the optimal set of additional basis functions, that apply force along specific coordinates: bonds, angles, native contacts and other collective variables to further improve performance for larger scale systems. 

As mentioned at the end of the Introduction, our results were independently derived in~\cite{vargas2023transport} within a machine learning context. It is noteworthy that stochastic thermodynamics has shown to be a useful theoretical framework not only for non-equilibrium statistical physics, but also for machine learning in flow-based diffusion models \cite{sohl2015deep, song2020score, doucet2022score, albergo2023stochastic}. In particular, we recognize significant ties between our work, and that of ``Stochastic Normalizing Flows'' \cite{wu2020stochastic}, wherein authors also consider constructing counterdiabatic protocols under the name ``deterministic invertible functions''. It can be shown that counterdiabatic protocols are perfect stochastic normalizing flows, and they report (after sufficient neural-network training) excellent numerical results for sampling and free energy estimation. The primary difference is that in their work they fix $U_0 = (1 - t/t_f) U_A + (t/t_f) U_B$ and use a neural-network ansatz, whereas here we use an adaptive importance sampling algorithm with a linear spatio-temporal basis ansatz for both $U_0$ and $U_1$. Likewise, we note \cite{bernton2019schr} explores time-asymmetric Markovian processes for sampling, building off of entropy-regularized optimal transport wherein solutions of the continuous-time Schrödinger bridge problem involve asymmetrically-controlled diffusion processes \cite{chen2021stochastic} (see also \cite{de2021diffusion, berner2022optimal}). This is intriguing, as solving for optimal time-\textit{symmetric} protocols has been shown to be equivalent to solving the continuous-time formulation \cite{benamou2000computational} of standard optimal transport \cite{aurell2011optimal, chen2019stochastic, nakazato2021geometrical, zhong2022limited, chennakesavalu2023unified}. In light of all this, we suspect deep theoretical connections between stochastic thermodynamics and machine learning may be further uncovered through the time-asymmetric fluctuation theorem. 

Documented code for this project may be found at \url{https://github.com/adrizhong/dF-protocol-optimization}.

The authors would like to thank Nilesh Tripuraneni, Hunter Akins, Chris Jarzynski, Gavin Crooks, Steve Strong, and the participants of the Les Houches ``Optimal Transport Theory and Applications to Physics'' and Flatiron ``Measure Transport, Diffusion Processes and Sampling'' workshops for useful discussions; Jorge L. Rosa-Raíces for helpful comments on an earlier manuscript version; and Evie Pai for lending personal computing resources. This research used the Savio computational cluster resource provided by the Berkeley Research Computing program at the University of California, Berkeley (supported by the University of California Berkeley Chancellor, Vice Chancellor for Research, and Chief Information Officer). AZ is supported by the Department of Defense (DoD) through the National Defense Science \& Engineering Graduate (NDSEG) Fellowship Program. BKS is supported by the Kavli Energy Nanoscience institute through the Philomathia Foundation Fellowship. MRD thanks Steve Strong and Fenrir LLC for supporting this project. This work was supported in part by the U.S. Army Research Laboratory and the U.S. Army Research Office under contract W911NF-20-1-0151.


\begin{subappendices} 
\section{Microscopic fluctuation theorem} \label{appendix:MFT}

\subsection{The Onsager-Machlup Action} \label{appendix:OM-action}

For overdamped Langevin dynamics for $\trajx(t) \in \mathbb{R}^d$, 
\begin{equation}
    \dot{\trajx}(t) = -\nabla U(\trajx(t), t) + \sqrt{2\beta^{-1} } \, \boldeta(t) \quad\mathrm{with}\quad \trajx(0) \sim \rho(\cdot) ,\label{eq:SI-langevin}
\end{equation}
where $\boldeta(t)$ is an instantiation of standard Gaussian white noise with statistics $\langle \eta_i(t) \rangle = 0$ and $\langle \eta_i(t) \eta_j(t') \rangle = \delta_{ij} \delta(t - t')$, and $\rho(\cdot)$ is its initial distribution, the formal expression for the probability of a path's realization is (up to a multiplicative factor) 
\begin{equation}
    \mathcal{P}[\trajx(t)] = \rho(\trajx(0)) e^{-\beta S[\trajx(t)]}. \label{eq:SI-path-probability}
\end{equation}
Here $S[\trajx(t)]$ is the Onsager-Machlup Path Action functional 
\begin{equation}
  S[\trajx(t)] = (\mathrm{I}) \int_0^{t_f} \frac{|\dot{\trajx}(t) + \nabla U(\trajx(t), t) |^2 }{4} \, \mathrm{d} t
\end{equation}
which comes from the path discretization into $N$ timesteps with timestep $\Delta t = t_f / N$: $\trajx(t) \rightarrow [\trajx_0, \trajx_1,  ... \, , \trajx_N]$, with $\trajx_n \approx \trajx(t_n)$, $N = t_f / \Delta t$, and $t_n = n \Delta t$, generated from Euler-Maruyama dynamics

\begin{gather} 
\trajx_0 \sim \rho(\cdot) \\ 
\trajx_{n+1} = \trajx_n - \nabla U(\trajx_n, t_n) \, \Delta t + \sqrt{2 \beta^{-1}} \, \Delta B_n,
\end{gather} 
where $\Delta B_n \sim \mathcal{N}(0, \Delta t \, I_{d})$ is a $d$-dimensional Gaussian random variable (i.e., Brownian increment) \cite{adib2008stochastic}.

The probability of the realization of a particular path is then

  \begin{align}
    \mathcal{P}(\trajx_0, \trajx_1, ... \, , \trajx_N ) &= \mathcal{P}(\trajx_0) \mathcal{P}(\trajx_1 | \trajx_0) \mathcal{P}(\trajx_2 | \trajx_1) ... \mathcal{P}(\trajx_N | \trajx_{N-1}) \nonumber  \\ 
    &= \rho(\trajx_0) \prod_{n = 0}^{N-1} (4 \pi \beta^{-1} \Delta t)^{-d/2} \exp \bigg( -\frac{|\trajx_{n+1} - \trajx_n + \nabla U(\trajx_n, t_n) \Delta t|^2}{4 \beta^{-1} \Delta t} \bigg) \nonumber \\  
    &\propto \rho(\trajx_0) \exp \bigg( - \beta \sum_{n=0}^{N - 1} \frac{ | ( \frac{\trajx_{n+1} - \trajx_n}{\Delta t}) + \nabla U(\trajx_n, t_n) |^2 }{4}\Delta t \bigg),
  \end{align}
where the normalization factor $ (4\pi \beta^{-1} \Delta t)^{-Nd/2}$ is hidden in the last line. 

Taking $N \rightarrow \infty$ with $\Delta t = t_f / N \rightarrow 0$, the sum within the exponential becomes

\begin{align}
    &\sum_{n=0}^{N - 1} \frac{ | ( \frac{\trajx_{n+1} - \trajx_n}{\Delta t}) + \nabla U(\trajx_n, t_n) |^2 }{4}\Delta t \nonumber \\ 
    &\quad\longrightarrow\quad (\mathrm{I}) \int_0^{t_f} \frac{|\dot{\trajx}(t) + \nabla U(\trajx(t), t)|^2}{4} \mathrm{d} t = S[\trajx(t)],
\end{align}
which yields the formal expression Eq.~\eqref{eq:SI-path-probability}. 

\subsection{Stochastic Integrals and Itô's formula} \label{appendix:stochastic-integrals}

Here we briefly review the rules of stochastic calculus. For a stochastic path (i.e., a ``Brownian motion'') $\trajx(t)|_{t \in [0, t_f]}$ from Eq.~\eqref{eq:SI-langevin} and some vector-valued function $b(\boldx,t)$, the three following choices for the time-discretization of the integral $\int_0^{t_f} b(\trajx(t), t) \cdot \dot{\trajx}(t) \, \mathrm{d} t$ :

\begin{gather*}
  \sum_{n=0}^{N - 1} b(\trajx_n, t_n) \cdot \Delta \trajx_n, \\
  \sum_{n=0}^{N - 1} b(\trajx_{n + \frac{1}{2}}, t_{n + \frac{1}{2}}) \cdot \Delta \trajx_n, 
\end{gather*}
and 

\begin{align*}
  \sum_{n=0}^{N - 1} b(\trajx_{n+1}, t_{n+1}) \cdot \Delta \trajx_n 
\end{align*}
(here $\Delta \trajx_n = (\trajx_{n+1} - \trajx_n)$, $\trajx_{n + \frac{1}{2}} = (\trajx_n + \trajx_{n+1}) / 2$, and $t_{n + \frac{1}{2}} = (t_n + t_{n+1}) / 2$) do \textit{not} necessarily converge to the same value under the $N \rightarrow \infty$, $\Delta t = t_f / N \rightarrow 0$ limit. This is in contrast to the case where $\trajx(t)|_{t \in [0, t_f]}$ is continuously differentiable, e.g., the solution of a well-behaved \textit{deterministic} differential equation, in which case the above three time-discretizations do converge to the same integral value under the limit \cite{pugh2002real}. 

Therefore, for trajectories $\trajx(t)|_{t \in [0, t_f]}$ obtained through the stochastic differential equation Eq.~\eqref{eq:SI-langevin}, we must define each of these as distinct integrals

\begin{gather*}
   (\mathrm{I}) \int_0^{t_f} b(\trajx(t), t) \cdot \dot{\trajx} \, \mathrm{d} t := \lim_{N \rightarrow \infty} \sum_{n=0}^{N - 1} b(\trajx_n, t_n) \cdot \Delta \trajx_n,  \\
  (\mathrm{S}) \int_0^{t_f} b(\trajx(t), t) \cdot \dot{\trajx} \, \mathrm{d} t := \lim_{N \rightarrow \infty} \sum_{n=0}^{N - 1} b(\trajx_{n + \frac{1}{2}}, t_{n + \frac{1}{2}}) \cdot \Delta \trajx_n, 
\end{gather*}
and 

\begin{gather*}
  (\mathrm{BI}) \int_0^{t_f} b(\trajx(t), t) \cdot \dot{\trajx} \, \mathrm{d} t := \lim_{N \rightarrow \infty} \sum_{n=0}^{N - 1} b(\trajx_{n+1}, t_{n+1}) \cdot \Delta \trajx_n ,
\end{gather*}
which are the Itô, Stratonovich, and Backwards Itô integrals, respectively. They are related to one another by Itô's lemma \cite{oksendal2013stochastic}

\begin{align}
  &(\mathrm{I}) \int_0^{t_f} b(\trajx(t), t) \cdot \dot{\trajx} + \beta \, \nabla \cdot b(\trajx(t), t) \, \mathrm{d} t \nonumber \\
  &\quad= (\mathrm{S}) \int_0^{t_f} b(\trajx(t), t) \cdot \dot{\trajx} \, \mathrm{d} t \nonumber \\
  &\quad\quad= (\mathrm{BI}) \int_0^{t_f} b(\trajx(t), t) \cdot \dot{\trajx} - \beta \, \nabla \cdot b(\trajx(t), t) \, \mathrm{d} t. \label{eq:SI-itos-formula}
\end{align}
The Stratonovich integration convention (i.e., with the time-symmetric midpoint-rule discretization) is particularly convenient because ordinary calculus rules (e.g., the chain rule, product rule, etc.) apply. 

Note that the three separate time-discretizations of integrals of the form $\int_0^{t_f} f(\trajx(t), t) \, \mathrm{d} t$ :

\begin{gather*}
  \sum_{n=0}^{N - 1} f(\trajx_n, t_n) \,\Delta t , \\ 
  \sum_{n=0}^{N - 1} f(\trajx_{n + \frac{1}{2}}, t_{n + \frac{1}{2}}) \, \Delta t ,
\end{gather*}
and 
\begin{align*}
  \quad \sum_{n=0}^{N - 1} f(\trajx_{n+1}, t_{n+1}) \, \Delta t
\end{align*}
\textit{do} converge to the same value under the $N \rightarrow \infty$ with $\Delta t = t_f / N \rightarrow 0$ limit, thus $(\mathrm{I}) \int_0^{t_f} f(\trajx(t), t) \, \mathrm{d} t = (\mathrm{S}) \int_0^{t_f} f(\trajx(t), t) \, \mathrm{d} t = (\mathrm{BI}) \int_0^{t_f} f(\trajx(t), t) \, \mathrm{d} t$.

\subsection{Microscopic fluctuation theorem derivation} \label{appendix:MFT-derivation}

Here, we use the stochastic calculus reviewed above to derive Eq.~\eqref{eq:work-MFT-definition}, i.e., the equivalence of its first and second lines. We start by manipulating the expression within the exponent

  \begin{gather}
U_A(\trajx(0)) + S[\trajx(t)] + W[\trajx(t)] \nonumber \\ = U_B(\trajx(t_f)) + (\mathrm{I}) \int_0^{t_f} \bigg\{ \frac{|\dot{\trajx} + \nabla(U_0 + U_1)|^2}{4} -  (\dot{\trajx} + \nabla U_1) \cdot \nabla U_0 + \beta^{-1} \nabla^2 (U_1 - U_0) \bigg\} \, \mathrm{d} t  \nonumber \\ 
= U_B(\trajx(t_f)) + \frac{1}{4} (\mathrm{I}) \int_0^{t_f} |\dot{\trajx}|^2 \, \mathrm{d} t + \frac{1}{4} (\mathrm{I}) \int_0^{t_f} |\nabla (U_1 - U_0 ) |^2 \, \mathrm{d} t + \nonumber \\ \frac{1}{2} (\mathrm{I}) \int_0^{t_f} \dot{\trajx} \cdot \nabla (U_1 - U_0 ) + 2 \beta \nabla^2 (U_1 - U_0) \, \mathrm{d} t \nonumber  \\
= U_B(\trajx(t_f)) + \frac{1}{4} (\mathrm{BI}) \int_0^{t_f} |\dot{\trajx}|^2 \, \mathrm{d} t + \frac{1}{4} (\mathrm{BI}) \int_0^{t_f} |\nabla (U_1 - U_0 ) |^2 \, \mathrm{d} t + \nonumber \\ \frac{1}{2} (\mathrm{BI}) \int_0^{t_f} \dot{\trajx} \cdot \nabla (U_1 - U_0 )  \, \mathrm{d} t  \nonumber  \\
= U_B(\trajx(t_f)) + (\mathrm{BI}) \int_0^{t_f} \frac{|\dot{\trajx} + \nabla (U_1 - U_0) |^2}{4} \, \mathrm{d} t \nonumber  \\
= U_B(\tilde{\trajx}(0)) + (\mathrm{I}) \int_0^{t_f} \frac{|-\dot{\tilde{\trajx}} + \nabla (\tilde{U}_1 - \tilde{U}_0) |^2}{4} \, \mathrm{d} t \nonumber  \\
= U_B(\tilde{\trajx}(0)) + \tilde{S}[\tilde{\trajx}(t)], \quad\mathrm{where}\quad \tilde{S}[\tilde{\trajx}(t)] = (\mathrm{I}) \int_0^{t_f} \frac{|\dot{\tilde{\trajx}} + \nabla (\tilde{U}_0 - \tilde{U}_1) |^2}{4} \, \mathrm{d} t. 
\end{gather}
The first equality comes from using Ito's lemma $U_0(X(t_f), t_f) - U_0(X(0), 0) = (\mathrm{I}) \int_0^{t_f} \partial_t U_0 + \dot{X} \cdot \nabla U_0 \, + \beta^{-1} \nabla^2 U_0 \, \mathrm{d} t$. The second equality follows from standard algebraic manipulation. The third equality comes from converting the Forward Itô integrals $(\mathrm{I})$ to Backward Itô integrals $(\mathrm{B I})$ using Itô's formula Eq.~\eqref{eq:SI-itos-formula}. The fourth equality results from standard algebraic manipulation. The fifth equality comes from the time-reversal transformation $t \rightarrow t_f - t$, with the Backward Itô integral becoming a forward Itô integral under time-reversal.

Finally, we plug in the above to the first line of Eq.~\eqref{eq:MFT} to obtain

\begin{align}
    \tilde{\mathcal{P}}[\tilde{\trajx}(t)] &= e^{-\beta \{ U_A(\trajx(0)) - F_A + S[\trajx(t)] + W[\trajx(t)] - \Delta F\} } \nonumber \\ 
    &= e^{-\beta\{ U_B (\tilde{\trajx}(0)) - F_B + \tilde{S}[\tilde{\trajx}(t)] \} } \nonumber \\ 
    &= \rho_B(\tilde{\trajx}(0)) e^{-\beta \tilde{S}[\tilde{\trajx}(t)]  } ,
\end{align}
thus completing our derivation.

\section{Deriving the fluctuation theorem from the microscopic fluctuation theorem} \label{appendix:MFT-FT}

In this section we derive the Crooks Fluctuation Theorem
\begin{equation}
    \frac{\mathcal{P}_F(+W)}{\mathcal{P}_R(-W)} = e^{+\beta(W - \Delta F)} 
\end{equation}
from the microscopic fluctuation theorem
\begin{equation}
    \mathcal{P}[\trajx(t)] e^{-\beta W[\trajx(t)]} = \tilde{\mathcal{P}}[\tilde{\trajx}(t)] e^{-\beta \Delta F}.\label{eq:SI-MFT-2}
\end{equation}

We begin by recalling that $\mathcal{P}_F(\cdot)$, giving the probability of observing a particular work value in the forward ensemble, is defined as 

\begin{align}
    \mathcal{P}_F(w) &= \langle \delta(W - w) \rangle_F \nonumber \\
    &= \int D \trajx(t) \mathcal{P}[\trajx(t)] \delta(W[\trajx(t)] - w),
\end{align}
where we write $w$ to distinguish the argument of $\mathcal{P}_F(\cdot)$ from the path-functional work $W = W[\trajx(t)]$. Here $D \trajx(t)$ denotes an integral over all paths $\trajx(t)|_{t\in[0, t_f]}$, $\mathcal{P}[\trajx(t)]$ is the probability of its realization, and $\delta(\cdot)$ is the Dirac-delta function. Plugging in Eq.~\eqref{eq:SI-MFT-2} to the above expression, we get 

\begin{align}
    \mathcal{P}_F(w) &= \int D \trajx(t) \tilde{\mathcal{P}}[\tilde{\trajx}(t)] e^{+\beta\{W[\trajx(t)] - \Delta F \} }\delta(W[\trajx(t)] - w) \nonumber \\ 
    &=  e^{+\beta (w - \Delta F) }  \int D \trajx(t) \tilde{\mathcal{P}}[\tilde{\trajx}(t)] \delta(W[\trajx(t)] - w)  \nonumber \\ 
    &= e^{+\beta (w - \Delta F) }  \int D\tilde{\trajx}(t) \tilde{\mathcal{P}}[\tilde{\trajx}(t)] \delta(-\tilde{W}[\tilde{\trajx}(t)] - w)  \nonumber \\ 
    &= e^{+\beta(w - \Delta F)} \mathcal{P}_R(-w),
\end{align}
where in the second line we pull out the exponential using the Dirac delta function, in the third line we consider the coordinate change $\trajx(t) \rightarrow \tilde{\trajx}(t)$ (using also $\tilde{W}[\tilde{\trajx}(t)] = -W[\trajx(t)]$, see Eq.~\eqref{eq:reverse-work} and the text that follows it in the main text), and in the fourth line we have recognized that the path integral expression is equivalent to the probability of observing the work value $-w$ in the reverse path ensemble.

%
\begin{algorithm}[H]
\small 
\caption{Time-Asymmetric Protocol Optimization via Adaptive Importance Sampling}\label{alg:algorithm}
\begin{algorithmic}[1]
  \State \textbf{inputs} $\beta$, $U_{A, B}(\boldx)$; stepsize $\mathrm{d}t$, no. timesteps $N$, basis fns $\{U_\mu(\boldx, n)|_{n = 1, .., N}\}$, initial guess $\theta_\mathrm{init}$
  \State \textbf{parameters} Samples per iteration $N_\mathrm{s}$, minibatches per iteration $N_\mathrm{mb}$, minibatch size $n^\mathrm{mb}_s$, constraint strength $f$
  \State \textbf{given} Methods $\Call{DrawSampleA}$, $\Call{DrawSampleB}$ returning eq. samples from $\rho_A$, $\rho_B$
  \State \textbf{output} Iteratively updated $\widehat{\Delta F}$ estimate
  \State 
  \Function{RunTrajF}{parameters $\theta = (\theta_F, \theta_R)$} \hfill \Comment{Euler-Maruyama method}
    \State Obtain $\boldx_0 \gets \Call{DrawSampleA}$
    \State Initialize $\boldx, \mathsf{a}_{\mu\nu}, \mathsf{b}_{\mu}, \mathsf{\tilde{a}}_{\mu\nu}, \mathsf{\tilde{b}}_{\mu} \gets \boldx_0, 0, 0, 0, 0$
    \For {$n = 1, ..., N$}
      \State Evaluate $\nabla \mathsf{U}_\mu \gets \nabla U_\mu(\boldx, n) $ for each $\mu$
      \State Calculate $\mathrm{d}\boldx \gets -\theta^\mu_F \nabla \mathsf{U}_\mu \mathrm{d}t + \sqrt{2 \beta^{-1}} \, \mathrm{d} \mathsf{B}$, where $\mathrm{d} \mathsf{B} \sim \mathcal{N}(0, \mathrm{d}t \times I_{d})$ 
      \State Evaluate $\nabla \mathsf{\tilde{U}}_\mu \gets \nabla U_\mu(\boldx + \mathrm{d}\boldx, n) $ for each $\mu$
      \State Evolve $\boldx \gets \boldx + \mathrm{d} \boldx$
      \State Evolve $\mathsf{a}_{\mu\nu}, \mathsf{b}_{\mu} \gets \mathsf{a}_{\mu\nu} +  \nabla \mathsf{U}_\mu \cdot \nabla \mathsf{U}_\nu \, \mathrm{d} t / 4,  \mathsf{b}_{\mu} + \nabla \mathsf{U}_\mu \cdot \mathrm{d} \boldx / 2$
      \State Evolve $\mathsf{\tilde{a}}_{\mu\nu}, \mathsf{\tilde{b}}_{\mu} \gets 
      \mathsf{\tilde{a}}_{\mu\nu} +  \nabla \mathsf{\tilde{U}}_\mu \cdot \nabla \mathsf{\tilde{U}}_\nu \, \mathrm{d} t / 4,   \mathsf{\tilde{b}}_{\mu} - \nabla \mathsf{\tilde{U}}_\mu \cdot \mathrm{d} \boldx / 2$ \hfill \Comment{$\mathrm{d}\tilde{\boldx} = - \mathrm{d} \boldx$}
    \EndFor
    \State Evaluate $\mathsf{c}, \mathsf{\tilde{c}} \gets U_A(\boldx_0), U_B(\boldx)$
    \State Calculate $W \gets - (\theta^\mu_F \theta^\nu_F \mathsf{a}_{\mu\nu} + \theta^\mu_F \mathsf{b}_{\mu} +  \mathsf{c} ) + (\theta^\mu_R \theta^\nu_R \mathsf{\tilde{a}}_{\mu\nu} + \theta^\mu_R \mathsf{\tilde{b}}_{\mu}  +  \mathsf{\tilde{c}})$
    \State \Return $W, \mathsf{a}_{\mu \nu}, \mathsf{b}_\mu, \mathsf{c}, \mathsf{\tilde{a}}_{\mu \nu}, \mathsf{\tilde{b}}_{\mu}, \mathsf{\tilde{c}}, \theta$
  \EndFunction
  \State 
  \Function{RunTrajR}{parameters $\theta = (\theta_F, \theta_R)$} 
    \State Obtain $\tilde{\boldx}_0 \gets \Call{DrawSampleB}$
    \State Initialize $\tilde{\boldx}, \mathsf{\tilde{a}}_{\mu\nu}, \mathsf{\tilde{b}}_{\mu}, \mathsf{a}_{\mu\nu}, \mathsf{b}_{\mu} \gets \tilde{\boldx}_0, 0, 0, 0, 0$
    \For {$n = 1, ..., N$}
      \State Evaluate $\nabla \mathsf{\tilde{U}}_\mu \gets \nabla U_\mu(\tilde{\boldx}, N + 1 - n) $ for each $\mu$ \hfill \Comment{$\nabla \tilde{U}_\mu(\cdot, n) = \nabla U_\mu(\cdot, N + 1 - n)$} 
      \State Calculate $\mathrm{d}\tilde{\boldx} \gets -\theta^\mu_R \nabla \mathsf{\tilde{U}}_\mu \mathrm{d}t + \sqrt{2 \beta^{-1} } \, \mathrm{d} \mathsf{B}$, where $\mathrm{d} \mathsf{B} \sim \mathcal{N}(0, \mathrm{d}t \times I_{d})$ 
      \State Evaluate $\nabla \mathsf{{U}}_\mu \gets \nabla U_\mu(\tilde{\boldx} + \mathrm{d}\tilde{\boldx}, N + 1 - n)$ for each $\mu$
      \State Evolve $\tilde{\boldx} \gets \tilde{\boldx} + \mathrm{d} \tilde{\boldx}$
      \State Evolve $\mathsf{\tilde{a}}_{\mu\nu}, \mathsf{\tilde{b}}_{\mu} \gets \mathsf{\tilde{a}}_{\mu\nu} +  \nabla \mathsf{\tilde{U}}_\mu \cdot \nabla \mathsf{\tilde{U}}_\nu \, \mathrm{d} t / 4,   \mathsf{\tilde{b}}_{\mu} + \nabla \mathsf{\tilde{U}}_\mu \cdot \mathrm{d} \tilde{\boldx} / 2$
      \State Evolve $\mathsf{a}_{\mu\nu}, \mathsf{b}_{\mu} \gets \mathsf{a}_{\mu\nu} +  \nabla \mathsf{U}_\mu \cdot \nabla \mathsf{U}_\nu \, \mathrm{d} t / 4,  \mathsf{b}_{\mu} - \nabla \mathsf{U}_\mu \cdot \mathrm{d} \tilde{\boldx} / 2$
    \EndFor
    \State Evaluate $\mathsf{\tilde{c}}, \mathsf{c} \gets U_B(\tilde{\boldx}_0), U_A(\tilde{\boldx})$
    \State Calculate ${\tilde{W}} \gets - (\theta^\mu_R \theta^\nu_R \mathsf{\tilde{a}}_{\mu\nu} + \theta^\mu_R \mathsf{\tilde{b}}_{\mu}  +  \mathsf{\tilde{c}}) + (\theta^\mu_F \theta^\nu_F \mathsf{a}_{\mu\nu} + \theta^\mu_F \mathsf{b}_{\mu} +  \mathsf{c} ) $
    \State \Return  ${\tilde{W}}, \mathsf{\tilde{a}}_{\mu \nu}, \mathsf{\tilde{b}}_\mu, \mathsf{\tilde{c}}, \mathsf{a}_{\mu \nu}, \mathsf{b}_\mu, \mathsf{c}, \theta$
  \EndFunction
  \State
  \algstore{myalg}
  \end{algorithmic}
\end{algorithm}

\begin{algorithm}[H]\small
\begin{algorithmic}[1]
\algrestore{myalg}

\newpage
  \Function{UpdateTheta}{forward samples $\mathcal{S}_F$, reverse samples $\mathcal{S}_R$} 
    \State Initialize $\mathcal{S}_{\theta,\mathrm{mb}} \gets \{ \}$
    \RepeatN{$N_\mathrm{mb}$} \hfill \Comment{Use larger $N_\mathrm{mb}$ for larger $|\mathcal{S}_F|$} 
      \State Randomly select $\mathcal{S}^{\mathrm{mb}}_F \subset \mathcal{S}_F$ of size $n^\mathrm{mb}_s$ without replacement
      \State Randomly select $\mathcal{S}^{\mathrm{mb}}_R \subset \mathcal{S}_R$ of size $n^\mathrm{mb}_s$ without replacement
      \State $\theta^* \gets \mathrm{argmin}_\theta \{ \hat{J}_F(\theta; \mathcal{S}^{\mathrm{mb}}_F ) +  \hat{J}_R(\theta; \mathcal{S}^{\mathrm{mb}}_R) \, | \, n_F^\mathrm{eff}(\theta; \mathcal{S}^{\mathrm{mb}}_F ) \geq f n^\mathrm{mb}_s, n_R^\mathrm{eff}(\theta; \mathcal{S}^{\mathrm{mb}}_R ) \geq f n^\mathrm{mb}_s \}$ 
      \State $\mathcal{S}_{\theta,\mathrm{mb}}  $.insert($\theta^*$)
    \End
    \State \Return mean($\mathcal{S}_{\theta,\mathrm{mb}} $)
  \EndFunction
  \State
  \Procedure{Main}\,
    \State Initialize parameters $\theta \gets \theta_\mathrm{init}$ and sample arrays $\mathcal{S}_F, \mathcal{S}_R \gets \{ \} , \{ \} $
    \Repeat
      \RepeatN{$N_\mathrm{s}$} \hfill \Comment{Use $N_\mathrm{mb} + N_\mathrm{s}$ on first iteration}
        \State $\mathcal{S}_F$.insert(\Call{RunTrajF}{$\theta$})
        \State $\mathcal{S}_R$.insert(\Call{RunTrajR}{$\theta$})
      \End
      \State Update estimate $\widehat{\Delta F} \gets \widehat{\Delta F}_\mathrm{BAR}(\mathcal{S}_F, \mathcal{S}_R)$ 
      \State $\theta \gets$ \Call{UpdateTheta}{$\mathcal{S}_F, \mathcal{S}_R$} 
    \Until out of computer time
  \EndProcedure
  \end{algorithmic}
\end{algorithm}

\begin{figure}
\centering
\includegraphics[width=\textwidth]{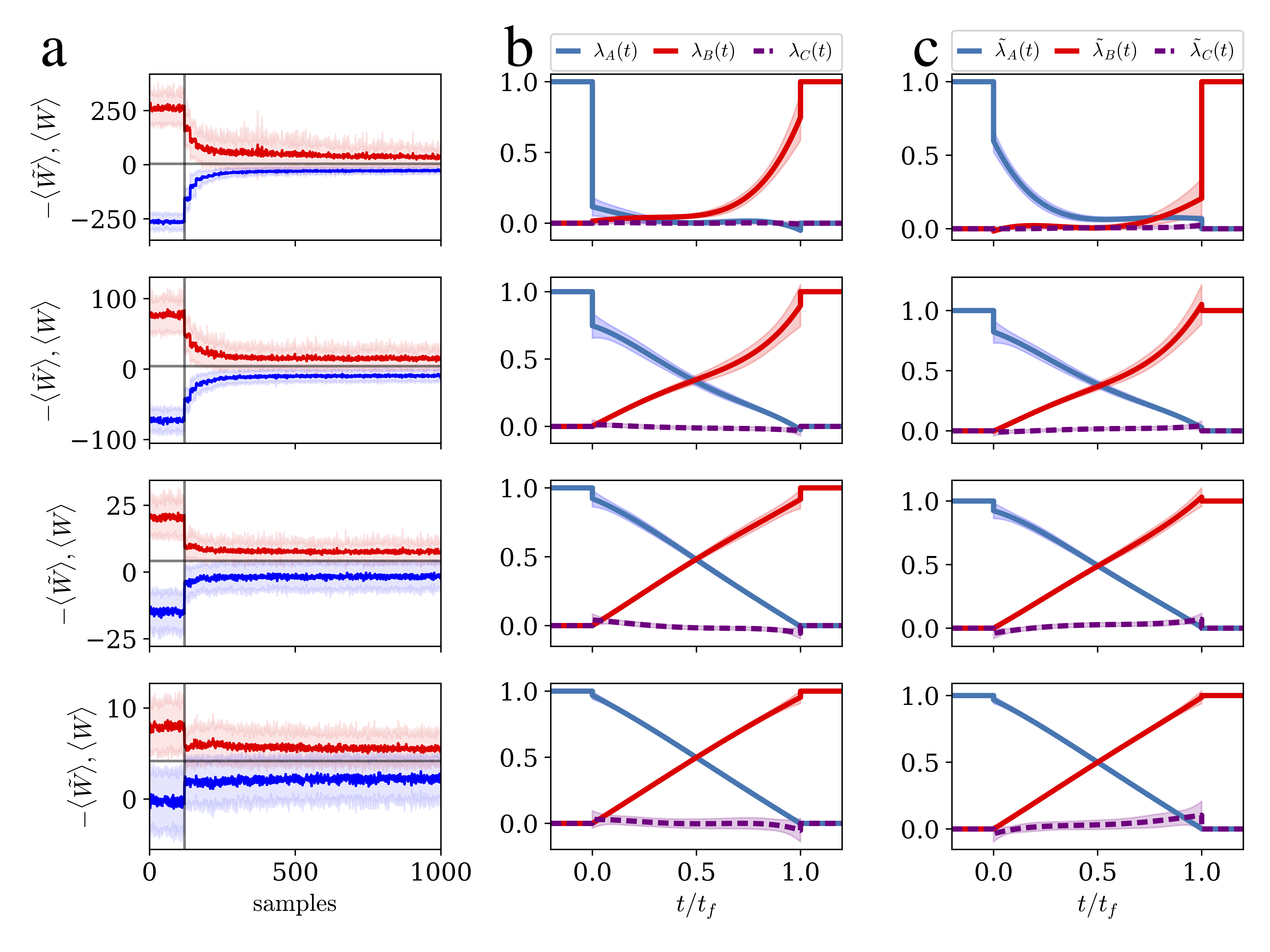}
    \caption[Convergence of the protocol optimization for the worm-like chain with attractive linker.]{(a) For the worm-like chain with attractive linker, ensemble-averaged work across 100 trials, as a function of sample number. Parameter optimization begins at 120 samples, and happens every 20 samples. The rows correspond to protocol times $t_f = 0.07 \, \tau_\mathrm{LJ} , 0.28 \, \tau_\mathrm{LJ} , 1.41 \, \tau_\mathrm{LJ},$ and $7.07 \, \tau_\mathrm{LJ}$ respectively. It appears convergence is reached rapidly, within the 1000 samples for all cases.  {Same coloring as Fig. 2(d).} (b) The forward protocols $U_F(\cdot, t) = \lambda_A(t) U_A(\cdot) + \lambda_B(t) U_B(\cdot) + \lambda_C(t) U_C(\cdot) $ after 1000 samples.  {Same coloring as Fig. 2(c).} Rows correspond to the same $t_f$. For small $t_f$, the protocol has reduced magnitude. This corresponds to lowering the potential, or raising the temperature (i.e., smaller $\beta U(\cdot, t)$). (c) The reverse protocols $U_R(\cdot, t) = \tilde{\lambda}_A(t) U_A(\cdot) + \tilde{\lambda}_B(t) U_B(\cdot) + \tilde{\lambda}_C(t) U_C(\cdot)$ after 1000 samples. Due to the intrinsic asymmetry of the problem between pulled and collapsed states, the resulting reverse protocols do not obey the symmetries observed in the double-well and Rouse polymer problems.}
    \label{fig:SI-WLC}
\end{figure}
%

\begin{figure}
\centering
\includegraphics[width=0.75\textwidth]{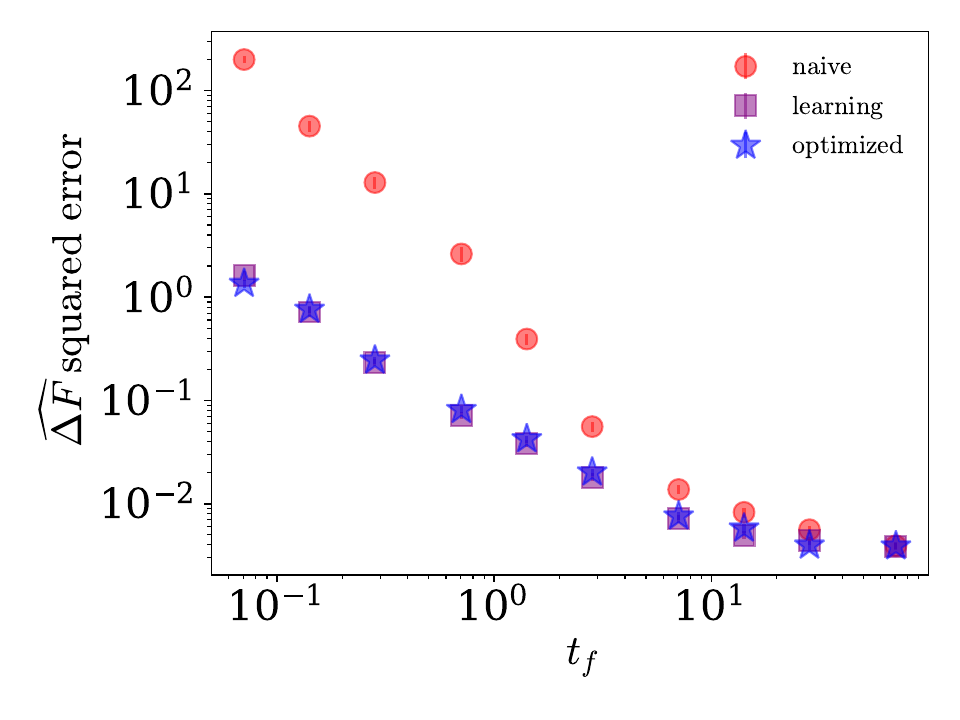}
    \caption[Performance plot for optimal protocols for the worm-like chain with attractive linker.]{Performance plot for the worm-like chain with attractive linker,  {same coloring as Fig. 1(b)}. That the mean squared error for protocol learning is near equal to the optimized protocol implies convergence occurs quickly within protocol optimization, cf Fig. \ref{fig:SI-WLC}(a). At $t_f = 0.07\, \tau_\mathrm{LJ}$, the MSE is $123.3$ times lower under protocol optimization than under the naive protocol.  }
    \label{fig:SI-WLC-performance}
\end{figure}

\section{Time-asymmetric fluctuation theorem for underdamped dynamics}

In this section, we generalize the time-asymmetric fluctuation theorem to underdamped dynamics. To begin, we review three types of dynamics: overdamped, underdamped, and deterministic. 

At inverse temperature $\beta$, overdamped Langevin dynamics for a stochastic trajectory $X(t) \in \mathbb{R}^d$ are given by the overdamped Langevin equation
\begin{equation}
  \dot{\trajx} = - \mu \nabla_x U_0(X(t),t) + \sqrt{2 \mu \beta^{-1} } \eta(t) 
\end{equation}
where $\mu$ is the mobility, and $\eta(t)$ is an instantiation of standard $d$-dimensional Gaussian white noise, i.e. with statistics $\langle \eta_i(t) \rangle = 0, \langle \eta_i(t) \eta_j(t') \rangle = \delta_{ij} \delta(t - t')$. Note that for generality the mobility is \textit{not} set to one, in contrast to the main text.

On the other hand, underdamped dynamics for position and momentum variables $X(t) \in \mathbb{R}^d, P(t) \in \mathbb{R}^d$ are given by the underdamped Langevin equation

\begin{gather}
  \dot{\trajx} = \frac{P(t)}{m} \nonumber \\ 
  \dot{\trajp} = -\nabla_x U_0(X(t), t) - \gamma P(t) + \sqrt{2 \gamma \beta^{-1} } \zeta(t), 
\end{gather}
where $m$ is the mass, $\gamma$ is the friction coefficient, and $\zeta(t)$ is also an instantiation of standard $d$-dimensional Gaussian white noise with $\langle \zeta_i(t) \rangle = 0, \langle \zeta_i (t) \zeta_j(t') \rangle = \delta_{ij} \delta(t - t')$ \cite{sivak2013using}. Importantly, this underdamped dynamics has the Hamiltonian
\begin{equation}
    H(x, p, t) = \frac{|p|^2}{2m} + U_0(x, t).
\end{equation}
Note that taking the limit $\gamma \rightarrow 0$ reproduces standard Hamiltonian mechanics. 

Finally, we explicitly write out the deterministic dynamics under a flow field $b_1(x, p, t) \in \mathbb{R}^{2d}$ that applies to both position and momentum variables

\begin{align}
  \dot{\trajx} &= b_1^x(X(t), P(t), t) \nonumber \\ 
  \dot{\trajp} &= b_1^p(X(t), P(t), t). 
\end{align}
This vector flow field $b_1$ is a generalization of the time-asymmetric (gradient) force provided by $-\nabla U_1$ in the main text.

\subsection{Generalized Langevin dynamics}

For our derivation, we define a hybridized Langevin equation that combines all of the above three dynamics

\begin{gather}
  \dot{\trajx} = b_1^x(X(t), P(t), t) +  \frac{P(t)}{m} + \big\{ -\mu \nabla_x U_0(X(t),t) + \sqrt{2 \mu \beta^{-1} } \eta(t) \big\} \nonumber \\ 
  \dot{\trajp} = b_1^p(X(t), P(t), t)  -\nabla_x U_0(X(t), t) +  \big\{ - \gamma P(t) + \sqrt{2 m \gamma \beta^{-1} } \zeta(t) \big\} \nonumber \\ \nonumber \\
  \mathrm{with} \quad\quad X(0), P(0) \sim \rho_A, \label{eq:SI-gen-langevin}
\end{gather}
where $\rho_A$ is the equilibrium distribution corresponding to the Hamiltonian $H_A(x, p) = H(x, p, 0)$ at time $t= 0$, to be specified below. By considering $m$, $\mu$, and $\gamma$ as independent parameters, overdamped dynamics are reproduced by taking the limit $\gamma \rightarrow 0, m \rightarrow \infty$ with the assumption $b_1^x(x, p, t)$ has no $p$-dependence, while underdamped dynamics are reproduced under the limit $\mu \rightarrow 0$, with further taking $\gamma \rightarrow 0$ yielding deterministic Hamiltonian dynamics. 

\subsection{Time-asymmetric work and path action}

In this general setting, we consider protocols that ``switch'' the Hamiltonian between 
\begin{align}
    &H(x, p, 0) = H_A(x, p) = \frac{|p|^2}{2m} + U_A(x) \nonumber \\
    &\rightarrow H(x, p, t_f) = H_B(x, p) =\frac{|p|^2}{2m} + U_B(x),  \nonumber 
  \end{align} 
by switching the potential energy $U(x, 0) = U_A(x) \rightarrow U(x, t_f) = U_B(x)$. The equilibrium distributions for $H_A$ and $H_B$ are given by  $\rho_{A, B}(x, p) = e^{-\beta [H_{A, B}(x, p) - F_{A, B}]}$, with $F_{A, B} = -\beta^{-1} \ln \int e^{-\beta H_{A, B}} \, \mathrm{d}x \, \mathrm{d}p$ denoting the equilibrium free energy. For ease of notation, from here on out we will denote a phase-space trajectory with variable $Z(t) := (X(t), P(t))$ (not to be confused with the partition function $\exp(-\beta F)$ often denoted by the same variable). 

The unconventional work for a stochastic trajectory $Z(t) |_{t\in [0, t_f]}$ is given by 
\begin{equation}
    W[Z(t)] = \int_0^{t_f} \frac{\partial H}{\partial t} + b_1 \cdot \nabla H - \beta^{-1} \nabla \cdot b_1  \, \dt , \label{eq:SI-unconventional-work}
\end{equation}
(we use notation $b_1 \cdot \nabla H = b_1^x \cdot \nabla_x H + b_1^p \cdot \nabla_p H$ and $\nabla \cdot b_1 = \nabla_x \cdot b_1^x + \nabla_p \cdot b_1^p$) \cite{vaikuntanathan2008escorted}, while the probability of observing the trajectory under Eq.~\eqref{eq:SI-gen-langevin} is 
\begin{equation}
    \mathcal{P} [Z(t)] = \rho_A(Z(0)) \, e^{- \beta S[Z(t)]} 
\end{equation}
with the path action given by the Itô integral
\begin{align*}
      &S[Z(t)] = (\mathrm{I)} \int_0^{t_f} \frac{|\dot{\trajx} - b_1^x - \frac{P}{m} + \mu \nabla_x U_0|^2}{4 \mu} + \frac{|\dot{\trajp} - b_1^p + \nabla_x U_0 + \gamma P |^2}{4 m \gamma } \, \dt. 
\end{align*}
This may be seen by considering the probability of obtaining the particular realization of the noise terms $\eta(t)$ and $\zeta(t)$ that produce the trajectory $Z(t) = (X(t), P(t))$.

\subsection{Derivation of microscopic fluctuation theorem} 

As with the detailed derivation for the overdamped case in Chapter Appendix~\ref{appendix:MFT-derivation}, we manipulate the sum 
\begin{gather}
    H_A(Z(0)) + S[Z(t)] + W[Z(t)] \nonumber \\
   = H_B(Z(t_f)) + (\mathrm{I)} \int_0^{t_f} \bigg\{ \frac{|\dot{\trajx} - b_1^x - \frac{P}{m} + \mu \nabla_x U_0|^2}{4 \mu} + \frac{|\dot{\trajp } - b_1^p + \nabla_x U_0 + \gamma P |^2}{4 m \gamma } - \nonumber \\ 
    \quad\quad\quad(\dot{Z} - b_1) \cdot \nabla H  \ -  \beta^{-1} (\nabla \cdot b_1 + \mu \nabla^2_x H + m \gamma \nabla^2_p H )  \bigg\} \, \dt  \nonumber
    \\ 
   = H_B(Z(t_f)) + (\mathrm{I}) \int_0^{t_f} \frac{|\dot{X}|^2}{4 \mu} + \frac{|\dot{P}|^2}{4 m \gamma} \, \dt +  \nonumber \\
   (\mathrm{I}) \int_0^{t_f} \frac{|-b_1^x - \frac{P}{m} + \mu \nabla_x U_0|^2 }{4 \mu} + \frac{|-b_1^p + \nabla_x U_0 + \gamma P|^2 }{4 m \gamma} \, \dt \ + \nonumber \\ 
   \frac{1}{2} (\mathrm{I}) \int_0^{t_f} \dot{X} \cdot \bigg(  \frac{ -b_1^x - \frac{P}{m} + \mu \nabla_x U_0}{\mu}\bigg) + \dot{P} \cdot \bigg( \frac{-b_1^p + \nabla_x U_0 + \gamma P}{m \gamma} \bigg) \, \dt \ + \nonumber \\ 
   (\mathrm{I}) \int_0^{t_f}  ( b_1^x - \dot{X}) \cdot \nabla_x U_0 + (b_1^p - \dot{P}) \cdot\frac{P}{m}  - \beta^{-1} ( \nabla_x \cdot (b_1^x +  \mu \nabla_x U_0) + \nabla_p \cdot (b_1^p + m \gamma \frac{P}{m} ) )  \,  \dt. 
  \end{gather}
The first equality comes from plugging in the work and path action definitions and applying the total time derivative with Itô's lemma $H_B(Z(t_f)) - H_A(Z(0)) = (\mathrm{I}) \int_0^{t_f}  \partial_t H + Z \cdot \nabla H + \beta^{-1} (\mu \nabla_x^2 H + m \gamma \nabla_p^2 H) \, \dt$ (here the divergence terms come from the Gaussian white noise on $X(t)$ and $P(t)$, see Chapter Appendix \thechapter.A), while the second equality is obtained by expanding out each of the squared terms.

Continuing with our derivation, we insert $\nabla_x \cdot p = \nabla_p \cdot \nabla_x U_0= 0$ into the integral, as well as apply the following two algebraic manipulations

\begin{gather} 
     (\mathrm{I}) \int_0^{t_f} \frac{|-b_1^x - \frac{P}{m} + \mu \nabla_x U_0|^2 }{4 \mu} + \frac{|-b_1^p + \nabla_x U_0 + \gamma P|^2 }{4 m \gamma} \, \dt +  \nonumber \\ (\mathrm{I}) \int_0^{t_f}  b_1^x \cdot \nabla_x U_0 + b_1^p \cdot \frac{P}{m} \,  \dt \nonumber \\
    = (\mathrm{I}) \int_0^{t_f} \bigg\{ \frac{|-( b_1^x + \frac{P}{m}) + \mu \nabla_x U_0|^2 }{4 \mu} +  (b_1^x + \frac{P}{m}) \cdot \nabla_x U_0 - \nonumber \\ (\nabla_x U_0  - b_1^p) \cdot \frac{P}{m} + \frac{|(-b_1^p + \nabla_x U_0) + \gamma P|^2 }{4 m \gamma} \bigg\} \, \dt  \nonumber \\ 
    = (\mathrm{I}) \int_0^{t_f} \frac{|b_1^x + \frac{P}{m} + \mu \nabla_x U_0|^2 }{4 \mu} + \frac{|-b_1^p + \nabla_x U_0 - \gamma P|^2 }{4 m \gamma} \, \dt, 
\end{gather}
and 
\begin{gather}
    \frac{1}{2} (\mathrm{I}) \int_0^{t_f} \dot{X} \cdot \bigg(  \frac{ -b_1^x - \frac{P}{m} + \mu \nabla_x U_0}{\mu}\bigg) + \dot{P} \cdot \bigg( \frac{-b_1^p + \nabla_x U_0 + \gamma P}{m \gamma} \bigg) \, \dt - \nonumber \\ (\mathrm{I}) \int_0^{t_f} \dot{X} \cdot \nabla_x U_0 + \dot{P} \cdot \frac{P}{m} \, \dt \nonumber \\ 
    = \frac{1}{2} (\mathrm{I}) \int_0^{t_f} (- \dot{X}) \cdot \bigg(  \frac{ b_1^x + \frac{P}{m} + \mu \nabla_x U_0}{\mu}\bigg) + \dot{P} \cdot \bigg( \frac{-b_1^p + \nabla_x U_0 - \gamma P}{m \gamma} \bigg) \, \dt, 
\end{gather}
ultimately yielding

\begin{gather}
  H_A(Z(0)) + S[Z(t)] + W[Z(t)] \nonumber 
  \\
    = H_B(Z(t_f)) + (\mathrm{I}) \int_0^{t_f} \frac{|\dot{X}|^2}{4 \mu} + \frac{|\dot{P}|^2}{4 m \gamma} \, \dt + \nonumber \\ 
    (\mathrm{I}) \int_0^{t_f} \frac{|b_1^x + \frac{P}{m} + \mu \nabla_x U_0|^2 }{4 \mu} + \frac{|-b_1^p + \nabla_x U_0 - \gamma P|^2 }{4 m \gamma} \, \dt \ + \nonumber \\ 
    \frac{1}{2} (\mathrm{I}) \int_0^{t_f} \bigg\{ (- \dot{X}) \cdot \bigg(  \frac{ b_1^x + \frac{P}{m} + \mu \nabla_x U_0}{\mu}\bigg) + \dot{P} \cdot \bigg( \frac{-b_1^p + \nabla_x U_0 - \gamma P}{m \gamma} \bigg) \ - \nonumber \\
   2 \beta^{-1}  \bigg[  \mu \nabla_x \cdot \bigg( \frac{b_1^x + \mu \nabla_x U_0 + \frac{P}{m} }{\mu} \bigg) - m \gamma \nabla_p \cdot \bigg(\frac{ -b_1^p - \gamma P + \nabla_x U_0 }{m \gamma} \bigg) \bigg] \bigg\}  \,  \dt \nonumber 
   \\ 
   = H_B(Z(t_f)) + (\mathrm{BI}) \int_0^{t_f} \frac{|\dot{X}|^2}{4 \mu} + \frac{|\dot{P}|^2}{4 m \gamma} \, \dt + \nonumber \\ 
    (\mathrm{BI}) \int_0^{t_f} \frac{|b_1^x + \frac{P}{m} + \mu \nabla_x U_0|^2 }{4 \mu} + \frac{|-b_1^p + \nabla_x U_0 - \gamma P|^2 }{4 m \gamma} \, \dt \ + \nonumber \\ 
    \frac{1}{2} (\mathrm{BI}) \int_0^{t_f} (-\dot{X}) \cdot \bigg(  \frac{ b_1^x + \frac{P}{m} + \mu \nabla_x U_0}{\mu}\bigg) + \dot{P} \cdot \bigg( \frac{-b_1^p + \nabla_x U_0 - \gamma P}{m \gamma} \bigg) \, \dt  \nonumber 
    \\
    = H_B(Z(t_f)) + (\mathrm{BI}) \int_0^{t_f} \frac{|-\dot{\trajx} + b_1^x + \frac{P}{m} + \mu \nabla_x U_0|^2}{4 \mu} + \frac{|\dot{\trajp } - b_1^p + \nabla_x U_0 - \gamma P |^2}{4 m \gamma } \, \dt
   \nonumber  \\ 
    = H_B(\tilde{Z}(0)) + (\mathrm{I}) \int_0^{t_f} \frac{|\dot{\tilde{\trajx}} + \tilde{b}_1^x - \tilde{P} / m + \mu \nabla_x \tilde{U}_0|^2}{4 \mu} + \frac{|\dot{\tilde{\trajp}} + \tilde{b}_1^p + \nabla_x \tilde{U}_0 + \gamma \tilde{P} |^2}{4 m \gamma } \, \dt  \nonumber \\ 
    =: H_B(\tilde{Z}(0)) + \tilde{S}[\tilde{Z}(t)]. 
  \end{gather}
Here, in the second equality we use Itô's lemma to transform from forward to backwards Itô integrals; in the third equality we express the terms back into the squared expression; and in the fourth equality we perform the time-reversal change of variables $\tilde{X}(t) = X(t_f - t), \tilde{P}(t) = -P(t_f - t)$ with  $\tilde{b}^x_1(x, p, t) = b^x_1(x, -p, t_f - t), \tilde{U}_0(x, t) = U_0(x, t_f - t)$, and $\tilde{b}^p_1(x, p, t) = - b^p_1(x, -p, t_f - t)$, which transforms the backwards Itô integral back into a forward Itô integral. 

Here, $\tilde{S}[\tilde{Z}(t)]$ is the path action for the generalized Langevin dynamics 

    \begin{gather}
  \dot{\tilde{\trajx}} = -\tilde{b}_1^x(\tilde{X}(t), \tilde{P}(t), t) +  \frac{\tilde{P}(t)}{m} + \big\{ -\mu \nabla_x \tilde{U}_0(\tilde{X}(t),t) + \sqrt{2 \mu \beta^{-1} } \eta(t) \big\} \nonumber \\ 
  \dot{\tilde{\trajp}} = -\tilde{b}_1^p(\tilde{X}(t), \tilde{P}(t), t)   -\nabla_x \tilde{U}_0(\tilde{X}(t), t) - \gamma \tilde P(t) + \sqrt{2 m \gamma \beta^{-1} } \zeta(t), 
\end{gather}
which from inspection differs from Eq.~\eqref{eq:SI-gen-langevin} by having minus signs in front of the $b_1^x$ and $b_1^p$ terms. After specifying the initial conditions $\tilde X(0),  \tilde P(0) \sim \rho_B$, we have a path ensemble for which the probability of observing a particular trajectory $\tilde{Z}(t)|_{t=0}^{t_f}$ satisfies 
\begin{align}
    \tilde{\mathcal{P}}[\tilde{Z}(t)] &= \rho_B(\tilde{Z}(0)) e^{-\beta \tilde{S}[\tilde{Z}(t)]} \nonumber \\
    &= e^{-\beta\{ H_B(\tilde{Z}(0)) - F_B + \tilde{S}[\tilde{Z}(t)] \} } \nonumber \\
    &= e^{-\beta\{ H_A(Z(0)) + S[Z(t)] + W[Z(t)] - F_B \} } \nonumber \\
    &= \rho_A(Z(0)) e^{-\beta \{ S[Z(t)] + F_A - F_B + W[Z(t)] \} } \nonumber \\ 
    &= P[Z(t)] e^{ -\beta\{ W[Z(t)] - \Delta F \} }, 
\end{align}
namely the time-asymmetric microscopic fluctuation theorem for the generalized dynamics Eq.~\eqref{eq:SI-gen-langevin}.
\\
\\
\section{Initial samples for Rouse polymer from normal-modes decomposition} \label{appendix:normal-modes}

For the Rouse polymer, random samples may be drawn by exploiting a normal-modes decomposition. We can write 

    \begin{equation}
      U_{A,B}(x_1, x_2, ..., x_{N-1}) = \bar{U} \bigg(x_1 - \frac{\lambda_{i,f}}{N} , x_2 - \frac{2 \lambda_{i,f}}{N}, ..., x_{N-1} - \frac{(N - 1)\lambda_{i,f}}{N} \bigg) + \frac{k \lambda_{i,f}^2}{2N},
    \end{equation}
where $\bar{U}(\vec{y}) = \vec{y}^T K \vec{y} / 2$ with
\begin{equation}
K = 
 \begin{pmatrix} 
    2k & -k & & & &  \\
    -k & 2k &  -k & & \huge{0} &\\
       & -k & & \ddots & & \\
       &    &  \ddots & & -k &  \\ 
       &  \huge{0}  &   & -k & 2k & -k \\
       & &  & & -k & 2k
    \end{pmatrix}.
\end{equation}

Then, we can do an eigenmode decomposition of $K$, writing

$$
\bar{U}(\vec{y}) = \hat{U}(\vec{z}) = \sum_{n=1}^{N-1} \frac{\kappa_n z_n^2}{2},
$$
where 

\begin{align}
\kappa_n &= 2\bigg[ 1 - \cos\bigg( \frac{\pi n }{N }\bigg) \bigg], \\
z_n &= \sqrt{\frac{2}{N}} \, \sum_{m=1}^{N-1} \sin\bigg( \frac{2 \pi n m}{N} \bigg) y_m.
\end{align}

Finally, for an individual initial condition, we draw the normal random variable $z_n \sim \mathcal{N}(\mu = 0, \sigma^2 = (\beta \kappa_n)^{-1})$ for each $n$, as $\hat{\rho}(\vec{z}) \propto \prod \exp (-\beta \kappa_n z_n^2 / 2)$; then we convert from $\vec{z}$ to $\vec{y}$ coordinates via 
\begin{equation}
    y_n = \sqrt{\frac{2}{N}} \, \sum_{m=1}^{N-1}  \sin\bigg( \frac{2 \pi n m}{N} \bigg) z_m,
\end{equation}
before finally adding $x_n = y_n + n \lambda_{i,f} /N$ to get our initial condition. 

Incidentally, the $k \lambda_{i,f}^2 / 2N$ in the expression comparing $U_{A,B}(\vec{x})$ to $U(\vec{y})$ is, up to an additive constant, the free energy $F_{A,B}$.

\end{subappendices} 

\chapter{Equivalence between Thermodynamic Geometry and Optimal Transport}

A fundamental result of thermodynamic geometry is that the optimal, minimal-work protocol that drives a nonequilibrium system between two thermodynamic states in the slow-driving limit is given by a geodesic of the friction tensor, a Riemannian metric defined on control space. For overdamped dynamics in arbitrary dimensions, we demonstrate that thermodynamic geometry is equivalent to $L^2$ optimal transport geometry defined on the space of equilibrium distributions corresponding to the control parameters. We show that obtaining optimal protocols past the slow-driving or linear response regime is computationally tractable as the sum of a friction tensor geodesic and a counterdiabatic term related to the Fisher information metric. These geodesic-counterdiabatic optimal protocols are exact for parameteric harmonic potentials, reproduce the surprising non-monotonic behavior recently discovered in linearly-biased double well optimal protocols, and explain the ubiquitous discontinuous jumps observed at the beginning and end times. 

\subsection{Note} 

This chapter is based off of: Adrianne Zhong, and Michael R. DeWeese. ``Beyond linear response: Equivalence between thermodynamic geometry and optimal transport.'' \emph{Physical Review Letters} 133, no. 5 (2024): 057102.

This is the pinnacle chapter of the thesis! It feels quite full circle for me, as I started researching far-from-equilibrium optimal protocols in wanting to discover a fundamental, non-approximate structure in nonequilibrium thermodynamics, beyond the Sivak and Crooks thermodynamic geometry framework (which had always been known as an approximate framework). And yet, after all these years, it was exactly the Sivak and Crooks friction tensor that turned out to encode this fundamental geometric structure in the end! 

\section{Chapter introduction}

A consequence of the Second Law of Thermodynamics is that finite-time processes require work to be irretrievably lost as dissipation. Recent studies in stochastic thermodynamics have aimed to characterize minimal-work protocols, which have applications for nanoscopic engineering \cite{diana2013finite, zulkowski2014optimal, proesmans2020finite, proesmans2020optimal, whitelam2023train, schmiedl2007efficiency, martinez2017colloidal, martin2018extracting, abiuso2020optimal, brandner2020thermodynamic, frim2022optimal, frim2022geometric} and for understanding biophysical systems \cite{geiger2010optimum, dellago2013computing, sivak2016thermodynamic, lucero2019optimal,  blaber2022efficient, davis2024active}. In this chapter we unify disparate geometric approaches and arrive at a novel framework for obtaining and better understanding thermodynamically optimal protocols.

The problem statement is: Given a configuration space $x \in \mathbb{R}^d$, inverse temperature $\beta$, and potential energy function $U_\lambda(x)$ parameterized by $\lambda \in \mathcal{M}$, what is the optimal protocol $\lambda^*(t)$ connecting the parameter values $\lambda_i$ and $\lambda_f$ in a finite time $\tau$ that minimizes the work 
\begin{equation}
  W[\lambda(t)] = \int_0^\tau \frac{\dee \lambda^\mu}{\dt} \bigg\langle \frac{\partial U_{\lambda}}{\partial \lambda^\mu } \bigg\rangle \dt \label{eq:work-definition} \, ?
\end{equation}
Here $\mathcal{M} \subseteq \mathbb{R}^n$ is an orientable $m$-dimensional manifold, locally resembling $\mathbb{R}^m$ everywhere with $m \leq n$. We use Greek indices to denote local coordinates of $\lambda \in \mathcal{M}$, and the Einstein summation convention (i.e., repeated Greek indices within a term are implicitly summed). The ensemble average $\langle \cdot \rangle$ is over trajectories $X(t)|_{t \in [0, \tau]}$ that start in equilibrium with $\lambda_i$ and evolve via some specified Langevin dynamics under $U_{\lambda(t)}|_{t \in [0, \tau]}$. 

Schmiedl and Seifert \cite{schmiedl2007optimal} showed that optimal protocols minimizing Eq.~\eqref{eq:work-definition} have intriguing \textit{discontinuous jumps} at the beginning and end times, which have proven to be ubiquitous~\cite{schmiedl2007optimal, 
then2008computing, bonancca2018minimal, naze2022optimal, blaber2021steps, zhong2022limited, whitelam2023train, rolandi2023optimal, engel2023optimal, esposito2010finite}.\footnote{The discontinuities in the protocol are also taken into account in the integral Eq.~\eqref{eq:work-definition}}\footnote{Gomez-Marin et al found that optimal protocols for under-damped dyanmics have even worse regularity \cite{gomez2008optimal}; at starting and end times, they have Dirac impulses!} Furthermore, optimal protocols can even be \textit{non-monotonic} in time \cite{zhong2022limited, whitelam2023train}. 

Sivak and Crooks demonstrated through linear response \cite{zwanzig2001nonequilibrium} that in the slow-driving limit ($\tau \gg \tau_\mathrm{R}$, an appropriate relaxation time-scale), optimal protocols are \textit{geodesics} of a symmetric positive-definite\footnote{We assume the parameterization of $U_\lambda$ is such that $\rho_{\lambda}^\mathrm{eq}$ are different for different $\lambda$; otherwise, the tensor could be semi-positive definite.} 
friction tensor defined in terms of equilibrium time-correlation functions. Treating the friction tensor as a Riemannian metric induces a geometric structure on the space of control parameters, known as ``Thermodynamic geometry.'' This approach is computationally tractable, as the friction tensor can be obtained through measurement, and geodesics on $\mathcal{M}$ can be determined by solving an ordinary differential equation. Geodesic protocols have been studied for a variety of systems including the Ising model \cite{rotskoff2015optimal, gingrich2016near, rotskoff2017geometric, louwerse2022multidimensional}, barrier crossing \cite{sivak2016thermodynamic, lucero2019optimal, blaber2022efficient}, bit-erasure \cite{zulkowski2014optimal, scandi2022minimally}, and nanoscopic heat engines (after allowing temperature to be controlled) \cite{abiuso2020optimal, brandner2020thermodynamic, frim2022optimal, frim2022geometric, chennakesavalu2023unified}, but unfortunately their performance can degrade past the slow-driving regime \cite{zhong2022limited}. 

Alternatively, when the ensemble of trajectories is additionally constrained to end in equilibrium with $\lambda_f$, finding the work-minimizing protocol for overdamped dynamics is equivalent to the Benamou-Brenier formulation of the $L^2$ optimal transport problem~\cite{aurell2011optimal, ito2024geometric} finding the dynamical mapping between the two distributions that has minimal integrated squared distance \cite{benamou2000computational} which itself yields a Riemannian-geometric structure \cite{otto2001geometry, kniefacz2017otto, villani2021topics}. The Benamou-Brenier solution is a time-dependent distribution and time-dependent velocity field that solves a continuity equation, which in this chapter we explicitly identify as a desired probability density evolution and the additional counterdiabatic forcing needed to effectuate its faster-than-quasistatic time evolution (as studied in so-called Engineered Swift Equilibriation \cite{martinez2016engineered, baldassarri2020engineered, frim2021engineered}, Counterdiabatic Driving \cite{iram2021controlling}, and Shortcuts to Adiabaticity \cite{torrontegui2013shortcuts, jarzynski2013generating, campbell2017trade, guery2019shortcuts, plata2021taming, guery2023driving, patra2021semiclassical, li2017shortcuts, li2022geodesic}). Remarkably, optimal protocols obtained in this manner are exact for arbitrary protocol durations $\tau$~\cite{aurell2011optimal}. Unfortunately, this approach involves solving coupled PDEs in configuration space ($\mathbb{R}^d$), which is typically infeasible for dimension $d \gtrsim 5$. Furthermore, the control space $\mathcal{M}$ must be sufficiently expressive in order to implement the optimal transport solution, which is often overly-restrictive \cite{klinger2024universal}. 

For overdamped dynamics in arbitrary dimension, \cite{chennakesavalu2023unified} showed that the friction tensor may be obtained via a perturbative expansion of the Benamou-Brenier objective function. Here we derive an even stronger result, that thermodynamic geometry is in fact \textit{equivalent} to optimal transport geometry, in the sense that the friction tensor and the Benamou-Brenier problem restricted to equilibrium distributions parameterized by $\lambda$ have identical geodesics and geodesic distances. Surprisingly, we find that a counterdiabatic component may be calculated using the Fisher information metric from information geometry \cite{fisher1922mathematical, rao1992information, amari2016information}. We demonstrate that protocols obtained by adding this counterdiabatic term to thermodynamic geometry geodesics are analytically exact for parametric harmonic oscillators, reproduce recently discovered non-monotonic behavior in certain optimal protocols \cite{zhong2022limited}, and satisfyingly explain the origin of jumps at beginning and end times.

\begin{figure} 
\centering
\includegraphics[width=10cm]
{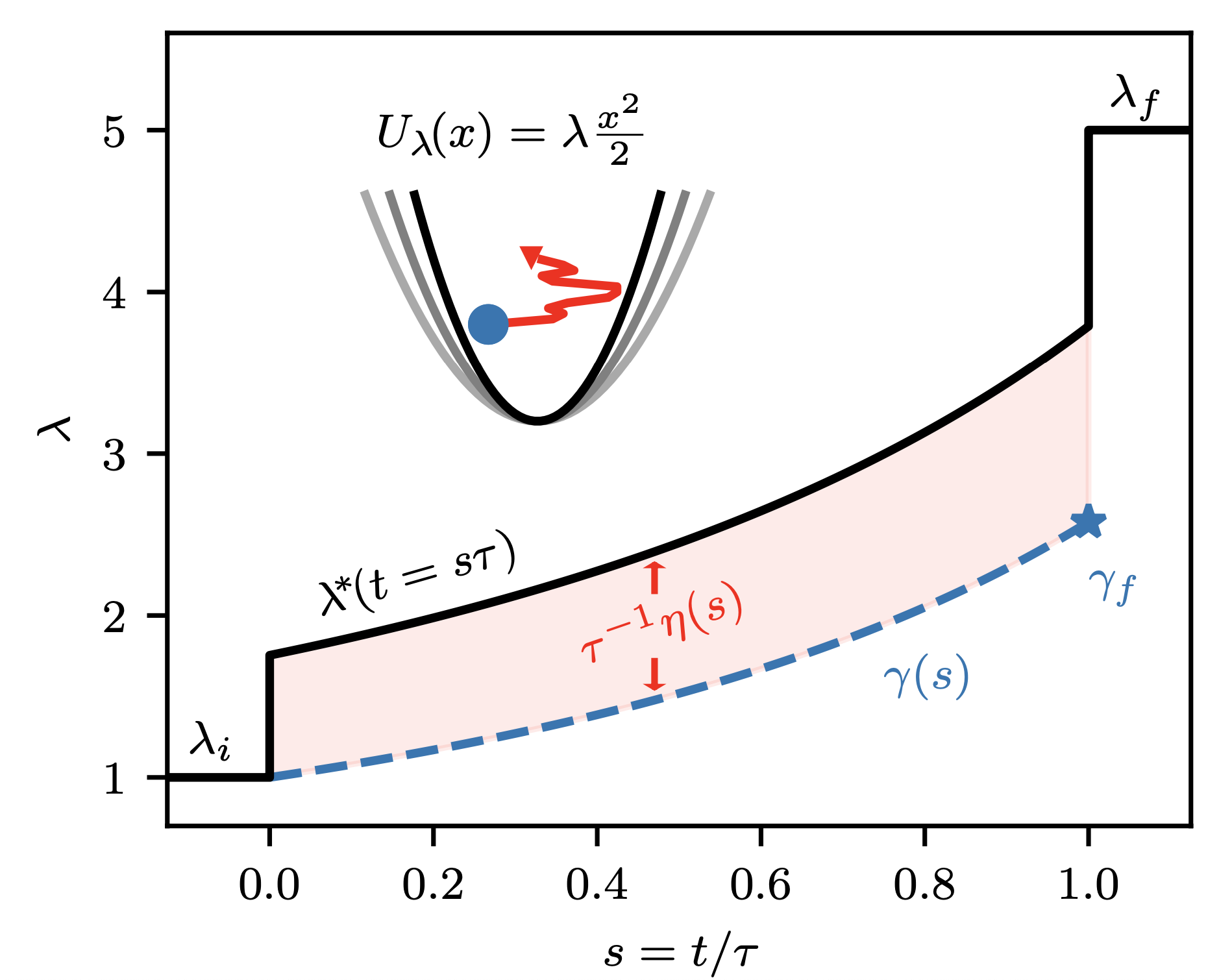}
\caption[Revisiting discontinuous optimal protocols, and decomposing in terms of geodesic and counterdiabatic components.]{The exact optimal protocol (black) for the variable-stiffness harmonic oscillator (depicted here, $\lambda_i = 1$, $\lambda_f = 5$, and $\tau = 0.5$) found in \cite{schmiedl2007optimal} can be explained as the sum of geodesic (dashed blue) and counterdiabatic (shaded red) components. The geodesic $\gamma(s)$ connects $\gamma(0) = \lambda_i$ to $\gamma(1) = \gamma_f$ (blue star) solving Eq.~\eqref{eq:gamma-f}. 
} 
\label{fig:fig1}
\end{figure}

\section{Preliminaries}

For each $\lambda \in \mathcal{M}$ there is a corresponding equilibrium distribution 
\begin{equation}
    \rho^\mathrm{eq}_\lambda(x) = \exp \{ - \beta[U_\lambda(x) - F(\lambda)] \}, \label{eq:equilibrium-dist}
\end{equation}
where $F(\lambda) = -\beta^{-1} \ln \int \exp[-\beta U_\lambda(x')] \dx'$ is the equilibrium free energy of the potential energy $U_\lambda(\cdot)$. For ease of notation we will denote $\rho_i^\mathrm{eq} = \rho^\mathrm{eq}_{\lambda_i}$, $\rho_f^\mathrm{eq} = \rho^\mathrm{eq}_{\lambda_f}$, and $\Delta F = F(\lambda_f) - F(\lambda_i)$. 

We consider overdamped Langevin equations, such that
trajectories $X(t) \in \mathbb{R}^d$ follow the stochastic ODE
\begin{gather}
  \dee X(t) = - \nabla U_{\lambda(t)}(X(t)) \dt  + \sqrt{2 \beta^{-1}} \dee B(t), \label{eq:beyond-lin-response-langevin-eq}
\end{gather}
where $B(t) \in \mathbb{R}^d$ is an instantiation of standard Brownian motion \cite{oksendal2013stochastic}. Here we will consider only isothermal protocols, so WLOG we set $\beta = 1$. 

The probability density $\rho(x, t)$ corresponding to Eq.~\eqref{eq:beyond-lin-response-langevin-eq} undergoes a time-evolution expressible either as a Fokker-Planck equation or a continuity equation of a gradient field 
\begin{equation}
    \frac{\partial \rho }{\partial t} = \mathcal{L}_{\lambda(t)} \rho \quad\mathrm{or}\quad \frac{\partial \rho }{\partial t} = \nabla \cdot (\rho \nabla \phi )  \label{eq:fokker-planck-eq}, 
\end{equation}
where $\mathcal{L}_\lambda$ is the Fokker-Planck operator~\cite{risken1996fokker}
\begin{equation}
    \mathcal{L}_\lambda \rho = \nabla^2 \rho + \nabla \cdot (\rho \, \nabla U_\lambda), \label{eq:fokker-planck-operator}
\end{equation}
while $\phi$ is a scalar field that depends on both $\rho$ and $\lambda$
\begin{equation}
  \phi(x, t) = \ln \rho(x, t) + U_{\lambda(t)}(x). \label{eq:effective-velocity}
\end{equation}
The adjoint operator $\mathcal{L}_\lambda^\dagger$ acts on a scalar field $\psi(x)$ via\footnote{The operator $\mathcal{L}_\lambda^\dagger$ is also known as the Backward Kolmogorov operator or the ``generator'' of Eq.~\eqref{eq:beyond-lin-response-langevin-eq}. By substituting in Eq.~\eqref{eq:equilibrium-dist} into Eq.~\eqref{eq:BK-1}, one readily recovers the perhaps more familiar expression $\mathcal{L}^\dagger_\lambda \psi = \nabla^2 \psi - \nabla U_\lambda \cdot \nabla \psi$ \cite{risken1996fokker}.} 
\begin{align}
    [\mathcal{L}^\dagger_\lambda \psi] (x) &= \rho_\lambda^\mathrm{eq}(x)^{-1} \nabla \cdot [ \rho_\lambda^\mathrm{eq} (x)  \nabla \psi(x) ].  \label{eq:BK-1} 
\end{align}
Finally, $f_\mu(x) := - \partial U_\lambda(x) / \partial \lambda^\mu$ is the conjugate force to $\lambda^\mu$. The excess conjugate force is then 
\begin{equation}
    \delta f_\mu(x) = - \bigg[ \frac{\partial U_\lambda(x)}{\partial \lambda^\mu} - \bigg\langle \frac{\partial U_\lambda}{\partial \lambda^\mu} \bigg\rangle_\lambda^\mathrm{eq} \bigg]  = \frac{\partial \ln \rho_\lambda^\mathrm{eq}(x)}{\partial \lambda^\mu}. \label{eq:excess-conj-force}
\end{equation}

\section{Thermodynamic geometry}

In the slow-driving limit, the excess work, defined as the work (Eq.~\eqref{eq:work-definition}) minus the equilibrium free energy difference $W_\mathrm{ex} = W - \Delta F$, is~\cite{sivak2012thermodynamic}  
\begin{equation}
  W_\mathrm{ex}[\lambda(t)] \approx \int_0^\tau \frac{\dee \lambda ^\mu}{\dt}  \frac{\dee \lambda^\nu }{\dt}  g_{\mu \nu} (\lambda(t)) \dt, \label{eq:SC-cost}
\end{equation}
where
\begin{equation}
  g_{\mu \nu}(\lambda) = \int_0^\infty \big\langle \delta f_\mu(X(t')) \, \delta f_\nu(X(0))  \big\rangle^\mathrm{eq}_\lambda \dt' \label{eq:friction-SC}
\end{equation}
is the symmetric positive-definite 
friction tensor. Here, $\langle \cdot \rangle^\mathrm{eq}_\lambda$ denotes an equilibrium average (i.e., $X(0) \sim \rho_\lambda^\mathrm{eq}$, and trajectories undergo Langevin dynamics (Eq.~\eqref{eq:beyond-lin-response-langevin-eq}) with constant $\lambda$).

Remarkably, the friction tensor induces a Riemannian geometry on control space $(\mathcal{M}, g)$ known as ``Thermodynamic geometry,'' with squared thermodynamic length between $\lambda_A, \lambda_B \in \mathcal{M}$ given by minimizing the path action
\begin{gather}
    \mathcal{T}^2 (\lambda_A, \lambda_B) = \min_{\lambda(s) |_{s \in [0, 1]}} \bigg\{ \int_0^1  \frac{\dee \lambda ^\mu}{\ds}  \frac{\dee \lambda^\nu }{\ds}  g_{\mu \nu} (\lambda(s)) \ds \ \bigg| \nonumber \\
    \mathrm{satisfying \ } \lambda(0) = \lambda_A, \lambda(1) = \lambda_B \bigg\} . \label{eq:thermodynamic-distance-SC}
\end{gather}
In the slow-driving limit, optimal protocols $\lambda^*(t)$ connecting $\lambda_i$ and $\lambda_f$ in time $\tau$ are time-rescaled versions of geodesics of Eq.~\eqref{eq:thermodynamic-distance-SC}, and the optimal excess work scales inversely with protocol time $W_\mathrm{ex}^* \approx \mathcal{T}^2 (\lambda_i, \lambda_f) / \tau$ \cite{sivak2012thermodynamic, zulkowski2012geometry}. 

While this geometric framework is both mathematically elegant and computationally tractable, geodesic protocols are fundamentally approximate; their performance often degrades for sufficiently small protocol times, in some cases performing even worse than a linear interpolation protocol \cite{zhong2022limited}.

\section{Optimal transport geometry}

Optimal transport is traditionally formulated as finding the transport map sending a distribution $\rho_A$ to another $\rho_B$ that minimizes an integrated ($L^2$) squared distance. This minimal integrated squared distance defines 
the squared $L^2$-Wasserstein metric distance between probability distributions, which was shown in \cite{benamou2000computational} to also be the minimum of a path action 
\begin{gather}
   \mathcal{W}_2^2[\rho_A, \rho_B] = \min_{\rho_s, \phi_s |_{s \in [0,1]}} \bigg\{ \int_0^{1} \int \rho_s(x) | \nabla \phi_s (x)|^2 \dx \ds \ \bigg| \nonumber \\ \quad\quad\quad\quad\quad\mathrm{satisfying \ } \frac{\partial \rho_s}{\partial s} = \nabla \cdot (\rho_s \nabla \phi_s), \, \rho_0 = \rho_A, \, \rho_1 = \rho_B \bigg\}. \label{eq:BB-cost-function}
\end{gather}
Here $\rho_s(\cdot)|_{s \in [0,1]}$ is a trajectory of configuration space probability densities $\mathcal{P}(\mathbb{R}^d)$,\footnote{For technical reasons, each probability distribution must have finite second moment \cite{villani2021topics}.} 
and $\phi_s(\cdot)|_{s \in [0, 1]}$ is a trajectory of scalar fields that yield gradient velocity fields $v_s = -\nabla \phi_s$ satisfying the continuity equation $\partial_s \rho_s = - \nabla \cdot (\rho_s v_s)$.\footnote{In the original formulation the optimization is over trajectories of effective vector fields $v_s|_{s \in [0, 1]}$ as opposed to scalar fields $\phi_s|_{s \in [0, 1]}$ \cite{benamou2000computational}; however, it is straightfoward to prove that under optimality, each $v_s$ \textit{must} be a gradient field \cite{benamou2000computational, ito2024geometric}.} 
This so-called Benamou-Brenier formulation of optimal transport (Eq.~\eqref{eq:BB-cost-function}) reveals a Riemmanian structure on the space of probability distributions known as Otto calculus \cite{otto2001geometry, kniefacz2017otto, villani2021topics}: on this manifold of probability distributions $M := \mathcal{P}(\mathbb{R}^d)$, a ``point'' is a probability distribution $\rho \in M$, a ``tangent space vector'' is a gradient velocity field identifiable (up to a constant offset) by a scalar field $\phi \in T_\rho(M)$, and geodesics are the argmin of Eq.~\eqref{eq:BB-cost-function}.

For overdamped dynamics, the work-minimizing protocol satisfying boundary conditions $\rho(\cdot, 0) = \rho_i^\mathrm{eq}$ and $\rho(\cdot, \tau) = \rho_f^\mathrm{eq}$ is a time-scaled solution of Eq.~\eqref{eq:BB-cost-function} for $\rho_A = \rho_i^\mathrm{eq}, \rho_B = \rho_f^\mathrm{eq}$, assuming sufficiently expressive control {(described in the following paragraph)} \cite{aurell2011optimal}. (See the SM for a concise derivation \cite{zhong2024beyond}.) From the continuity equation form of $\partial \rho(\cdot, t) / \partial t$ (Eqs.~\eqref{eq:fokker-planck-eq} and~\eqref{eq:effective-velocity}), the optimal protocol $\lambda^*(t)$ for finite time $\tau$ can be expressed in terms of $\rho_s^*$ and $\phi_s^*$ that solve Eq.~\eqref{eq:BB-cost-function}, as satisfying (up to a constant offset)
\begin{equation}
  U_{\lambda^*(t)}(x) = -\ln \rho_{t/\tau}^*(x) + \tau^{-1} \phi_{t/\tau}^*(x) .\label{eq:OT-solution}
\end{equation}
The first term corresponds to the Benamou-Brenier geodesic $\rho_{s}^*$, and the second one with $\phi_{s}^*$ is a counterdiabatic term that drives the probability distribution solving Eq.~\eqref{eq:fokker-planck-eq} to match the geodesic $\rho(\cdot, t) = \rho^*_{t / \tau}.$\footnote{Note that the two terms $-\ln \rho^*_s$ and $\phi^*_s$ corresponding exactly to time-reversal-symmetric $U_0(\cdot, t)$ and time-reversal-asymmetric $U_1(\cdot, t)$ found in  Eqs.~(6)-(8) in \cite{li2017shortcuts} and Eq.~(8) in \cite{zhong2023time}; also, see Eq.~(6) in \cite{proesmans2020optimal} and Eq.~(2) in the SI of \cite{martinez2016engineered} for dimension $d = 1$.} 

Remarkably, this solution is \textit{exact} for any finite $\tau$, and it provides a geometric interpretation for these work-minimizing protocols as optimal transport geodesics connecting $\rho_i^\mathrm{eq}$ to $\rho_f^\mathrm{eq}$. Through the time-scaling $t = \tau s$, it follows that $W_\mathrm{ex}^* = \mathcal{W}_2^2[\rho_i^\mathrm{eq}, \rho_f^\mathrm{eq}] / \tau$ is a \textit{tight} lower bound for excess dissipation in this additionally-constrained setting \cite{dechant2019thermodynamic, nakazato2021geometrical}. However, there are two important caveats to this approach: first, solving Eq.~\eqref{eq:BB-cost-function} involves PDEs on configuration space, which generally for dimension $d \gtrsim 5$ is computationally intractable (although, see \cite{beck2021solving, han2018solving, chen2021solving, albergo2023stochastic} for sophisticated modern machine learning methods, as well as \cite{klinger2024universal}). Second, the control parameters must be sufficiently expressive in the sense that for all $t \in (0, \tau)$ there \text{has to be} a $\lambda \in \mathcal{M}$ that satisfies Eq.~\eqref{eq:OT-solution}. Worse yet, there might not be \textit{any} admissible protocols that can satisfy the terminal constraint $\rho(\cdot, \tau) = \rho_f^\mathrm{eq}$~\cite{klinger2024universal}. 

Without the terminal condition, this problem is no longer over-constrained. The optimal excess work can be expressed as a minimum over $\rho_f = \rho(\cdot, \tau)$ (see the SM \cite{zhong2024beyond}), 
\begin{equation}
    W_\mathrm{ex}^* = \min_{\rho_f} \ \mathcal{W}_2^2[\rho_i^\mathrm{eq}, \rho_f]  /  \tau + D_\mathrm{KL}(\rho_f \, | \, \rho^\mathrm{eq}_f) , \label{eq:free-endpoint}
\end{equation}
where the additional KL-divergence cost
\begin{equation}
    D_\mathrm{KL}(\rho_A \, | \, \rho_B) := \int \rho_A (x)  \ln  \frac{\rho_A(x)}{\rho_B(x)}  \dx , \label{eq:KL-divergence}
\end{equation}
is the dissipation from the equilibration $\rho_f \rightarrow \rho^\mathrm{eq}_f$ that occurs for $t > \tau$.\footnote{Interestingly, this cost function is equivalent to the celebrated JKO scheme used to study convergence properties of the Fokker-Planck equation \cite{jordan1998variational}, as observed in \cite{chen2019stochastic}, with effective time-step $h = \tau/2$. We study the time-scaling of these unconstrained-end-state optimal protocols in a separate upcoming work \cite{unpublished}.} 
Optimal protocols $\lambda^*(t)$ are also obtained via Eqs.~\eqref{eq:BB-cost-function} and~\eqref{eq:OT-solution}, but now with $\rho_A = \rho_i^\mathrm{eq}$ and $\rho_B = \rho_f^*$ that minimizes Eq.~\eqref{eq:free-endpoint}. Without the restrictive terminal constraint, protocols that approximate Eq.~\eqref{eq:OT-solution} are allowed (in the case of limited expressivity), and may be near-optimal in performance \cite{gingrich2016near, engel2023optimal}.

\section{Demonstrating equivalence of geometries}

We start by expressing Eq.~\eqref{eq:friction-SC} with the time-propagator (e.g., see Ch. 4.2 of \cite{risken1996fokker}):
\begin{align}
 g_{\mu \nu}(\lambda)\ &= \int_0^\infty \int \rho_\lambda^\mathrm{eq}(x) \, \delta f_\mu(x) \, e^{\mathcal{L}_\lambda^\dagger t'} \, \delta f_\nu(x)  \dx \dt' \nonumber \\
 &=  -\int \rho_\lambda^\mathrm{eq}(x)  \, \delta f_\mu(x) \, \big\{\mathcal{L}_\lambda^\dagger \big\}^{-1}\, [\delta f_\nu ]  (x) \dx.  \label{eq:friction-tensor-3}
\end{align}
The second line comes from taking the time-integral, where the inverse operator $\big\{\mathcal{L}_\lambda^\dagger \big\}^{-1}$ is defined in terms of a properly constructed Green's function (Eq.~(40) in \cite{wadia2022solution}). This expression is the lowest order tensor found in a perturbative expansion of the Fokker-Planck equation~\cite{wadia2022solution}.

By formally defining $\phi_\mu = \big\{\mathcal{L}_\lambda^\dagger \big\}^{-1} \delta f_\mu$ as (up to a constant offset) the scalar field solving $\mathcal{L}_\lambda^\dagger \phi_\mu = \delta f_\mu$, it is straightfoward to show with Eqs.~\eqref{eq:BK-1} and~\eqref{eq:excess-conj-force} that, for any protocol $\lambda(s)|_{s \in [0, 1]}$,
\begin{equation}
    \frac{\partial \rho_{\lambda(s)}^\mathrm{eq} }{\partial s} = \nabla \cdot ( \rho_{\lambda(s)}^\mathrm{eq} \nabla \phi_s), \  \ \mathrm{where} \ \ \phi_s(x) = \frac{\dee \lambda^\mu}{\ds} \phi_\mu(x).  \label{eq:friction-phi}
\end{equation}
Applying $\delta f_\mu = \mathcal{L}^\dagger_\lambda \phi_\mu$ and Eq.~\eqref{eq:BK-1} to Eq.~\eqref{eq:friction-tensor-3} shows that the thermodynamic distance (Eq.~\eqref{eq:thermodynamic-distance-SC}) may be expressed as
\begin{gather}
    \mathcal{T}^2(\lambda_A, \lambda_B) = \min_{\lambda(s), \phi_s|_{s \in [0, 1]}} \bigg\{ \int_0^{1} \int \rho_{\lambda(s)}^\mathrm{eq}(x) |\nabla \phi_s(x)|^2 \dx \ds \ \bigg| \nonumber \\ \quad\quad  \mathrm{satisfying \ } \frac{\partial \rho_{\lambda(s)}^\mathrm{eq}}{\partial s} = \nabla \cdot (\rho_{\lambda(s)}^\mathrm{eq} \nabla \phi_s), \, \lambda(0) = \lambda_A, \lambda(1) = \lambda_B \bigg\}.
    \label{eq:squared-thermodynamic-length-2}
\end{gather} 
This is our first major result in this chapter: this expression is equivalent to the squared $L^2$-Wasserstein distance (Eq.~\eqref{eq:BB-cost-function}) with the constraint that $\rho_s|_{s \in [0, 1]}$ is a trajectory of equilibrium distributions $\rho^\mathrm{eq}_{\lambda(s)}|_{s \in [0, 1]}$. In other words, thermodynamic geometry induced by the friction tensor (Eq.~\eqref{eq:friction-SC}) on $\mathcal{M}$ is \textit{equivalent} to optimal transport geometry restricted to the equilibrium distributions $\mathcal{P}^\mathrm{eq}_\mathcal{M}(\mathbb{R}^d)$ corresponding to $\mathcal{M}$,\footnote{This is entirely analogous to how both the unit ball $B^2 = \{ (x, y, z) \, | \, x^2 + y^2 + z^2 \leq 1 \}$ and unit sphere $S^2 = \{ (x, y, z)  \, | \, x^2 + y^2 + z^2 = 1 \}$ can be seen as submanifolds of Euclidean space $\mathbb{R}^3$; while geodesics connecting two points in the unit ball are equivalent to straight-line Euclidean geodesics, on the unit sphere they differ.} 
and thus share the same geodesics and geodesic distances.

Up till now, thermodynamic geometry has prescribed optimal protocols as friction tensor geodesics joining $\lambda_i$ and $\lambda_f$, which are approximate for finite $\tau$. Optimal transport solutions require solving PDEs, but yield exact optimal protocols containing both geodesic and counterdiabatic components (Eq.~\eqref{eq:OT-solution}). Our unification of geometries suggests that thermodynamic geometry protocols may be made exact by including a counterdiabatic term.

\section{Geodesic-counterdiabatic optimal protocols}

From here we consider the control-affine parameterization\footnote{The term $U_\mathrm{offset}(\lambda)$ is constant in $x$ so that $\nabla U_\mathrm{offset} = 0$, but is allowed to be non-linear in $\lambda$. It is necessary to express certain potentials, e.g., the variable center harmonic trap $U_\lambda(x) = (x - \lambda)^2 / 2 = x^2 / 2 + \lambda^2 / 2 - \lambda x$.} 
\begin{equation}
  U_\lambda(x) = U_\mathrm{fixed}(x) + U_\mathrm{offset}(\lambda) + \lambda^\mu U_\mu (x), \label{eq:affine-U}
\end{equation}
and control space $\lambda \in \mathcal{M} = \mathbb{R}^m$. 
It follows from the equivalence of thermodynamic and optimal transport geometries that the optimal protocol should have the form
\begin{equation}
    \lambda^*(t) = \gamma(t / \tau) + \tau^{-1} \eta(t / \tau), \label{eq:counterdiabatic-protocol}
\end{equation}
namely the sum of a geodesic term and a counterdiabatic term that correspond to the two terms in  Eq.~\eqref{eq:OT-solution}, where $\rho_s^*|_{s\in[0,1]}$ and $\phi^*_s|_{s\in[0,1]}$ solve Eq.~\eqref{eq:BB-cost-function} with $\rho_A = \rho_i^\mathrm{eq}$ and $\rho_B = \rho_f^*$ from Eq.~\eqref{eq:free-endpoint}. Here, $\gamma(s)$ will be a geodesic of $g(\lambda)$ joining $\gamma(0) = \lambda_i$ to $\gamma(1) = \gamma_f$, where
\begin{equation}
  \gamma_f = \arg \min_{\lambda} \ \mathcal{T}^2(\lambda_i, \lambda) / \tau + D_\mathrm{KL}(\rho_\lambda^\mathrm{eq} | \rho_f^\mathrm{eq} ).
  \label{eq:gamma-f}
\end{equation}
We show in Appendix A that the counterdiabatic term is
\begin{equation} \label{eq:eta-solution}
    \eta(s) = h^{-1}(\gamma(s)) \, g(\gamma(s))  \bigg[ \frac{\dee \gamma(s)}{\ds} \bigg], 
\end{equation}
where, intriguingly, $h$ is the Fisher information metric\footnote{Traditionally the Fisher information metric is denoted as $h_{\mu \nu}(\lambda) = \int \rho_\lambda^\mathrm{eq} \, (\partial \ln \rho_\lambda^\mathrm{eq} / \partial \lambda^\mu) \, (\partial \ln \rho_\lambda^\mathrm{eq} / \partial \lambda^\nu) \dx$ (which can be seen as equivalent via Eq.~\eqref{eq:excess-conj-force}).} 
\begin{equation}
  h_{\mu \nu}(\lambda) = \int \rho_\lambda^\mathrm{eq}(x) \, 
  \delta f_\mu(x) \, \delta f_\nu(x) \dx, \label{eq:fisher-information-metric}
\end{equation}
which also induces a Riemannian geometry on the space of parametric equilibrium probability distributions $(\mathcal{M}, h)$ known as ``Information geometry'' \cite{fisher1922mathematical, rao1992information, amari2016information}. Eq.~\eqref{eq:eta-solution} is exact in cases of sufficient expressivity (i.e., when Eq.~\eqref{eq:OT-solution} can be satisfied); otherwise, $\eta(s)$ is the full solution projected onto $\mathcal{M}$. 

This is our second major result in this chapter: the equivalence between thermodynamic and optimal transport geometries implies that optimal protocols beyond linear response require counterdiabatic forcing, and can be obtained {for control-affine potentials (Eq.}~\eqref{eq:affine-U}\footnote{In the event that the potential energy $U_\lambda$ may not be expressed in control-affine form (Eq.~\eqref{eq:affine-U}, the construction of the counterdiabatic term is still valid, granted that for all $s \in [0, 1]$ the potential energy is locally linearizable around $\lambda = \gamma(s)$, and the magnitude of the resulting $\eta(s)$ is small enough so that the local linearization is a good approximation.} via:
\begin{algorithm}[H] \label{alg:algorithm}
  \caption{Geodesic-counterdiabatic optimal protocols {for control-affine potentials (Eq.}~\eqref{eq:affine-U}{).} \\ 
  \textbf{Input:} $\lambda_i, \lambda_f$, protocol time $\tau$, metrics $g_{\mu \nu}(\lambda)$, $h_{\mu \nu}(\lambda)$ (Eqs.~\eqref{eq:friction-SC}, \eqref{eq:fisher-information-metric}), KL divergence $D_\mathrm{KL}(\cdot \, |\,  \rho_f^\mathrm{eq})$ (Eq.~\eqref{eq:KL-divergence}).}
   \begin{algorithmic}[1]
   \State Solve geodesic $\gamma(s)|_{s\in[0,1]}$ connecting $\gamma(0) = \lambda_i$ and $\gamma(1) = \gamma_f$ (obtained from Eq.~\eqref{eq:gamma-f}) under $g_{\mu \nu}$.
   \State Calculate counterdiabatic term $\eta(s) = h^{-1} g \, [\dee \gamma / \dee s]$
   \State Return optimal protocol $\lambda^*(t) = \gamma(t / \tau) + \tau^{-1} \eta(t / \tau)$
   \end{algorithmic}
\end{algorithm}
We emphasize that this procedure does not require solving any configuration-space PDEs. Moreover, in the limit $\tau \rightarrow \infty$, the counterdiabatic component in Eq.~\eqref{eq:counterdiabatic-protocol} vanishes and Eq.~\eqref{eq:gamma-f} is solved by $\gamma_f = \lambda_f$, and thus geodesic protocols from thermodynamic geometry are reproduced.

\section{Examples}

We show in Appendix B that Alg.~1. reproduces exact optimal protocols solved in \cite{schmiedl2007optimal} for controlling a parameteric harmonic potential. Fig. 1. illustrates an optimal protocol for $U_\lambda(x) = \lambda x^2 / 2$. Notice that at $t = 0$ the counterdiabatic term is suddenly turned on, while at $t = \tau$ the geodesic ends at $\gamma_f \neq \lambda_f$ and the counterdiabatic term is suddenly turned off. Seen in this light, the discontinuous jumps in optimal protocols $\lambda^*(t)$ arise from the sudden turning on and off of counterdiabatic forcing, and the discontinuity of the geodesic at $t = \tau$. We note that starting in equilibrium at $t = 0$ and suddenly ending control at $t = \tau$ are both unnatural in biological settings; these generic discontinuous jumps can be seen as artifacts of the imposed boundary conditions. 

\begin{figure}[h]
\centering
\includegraphics[width=0.6\textwidth]{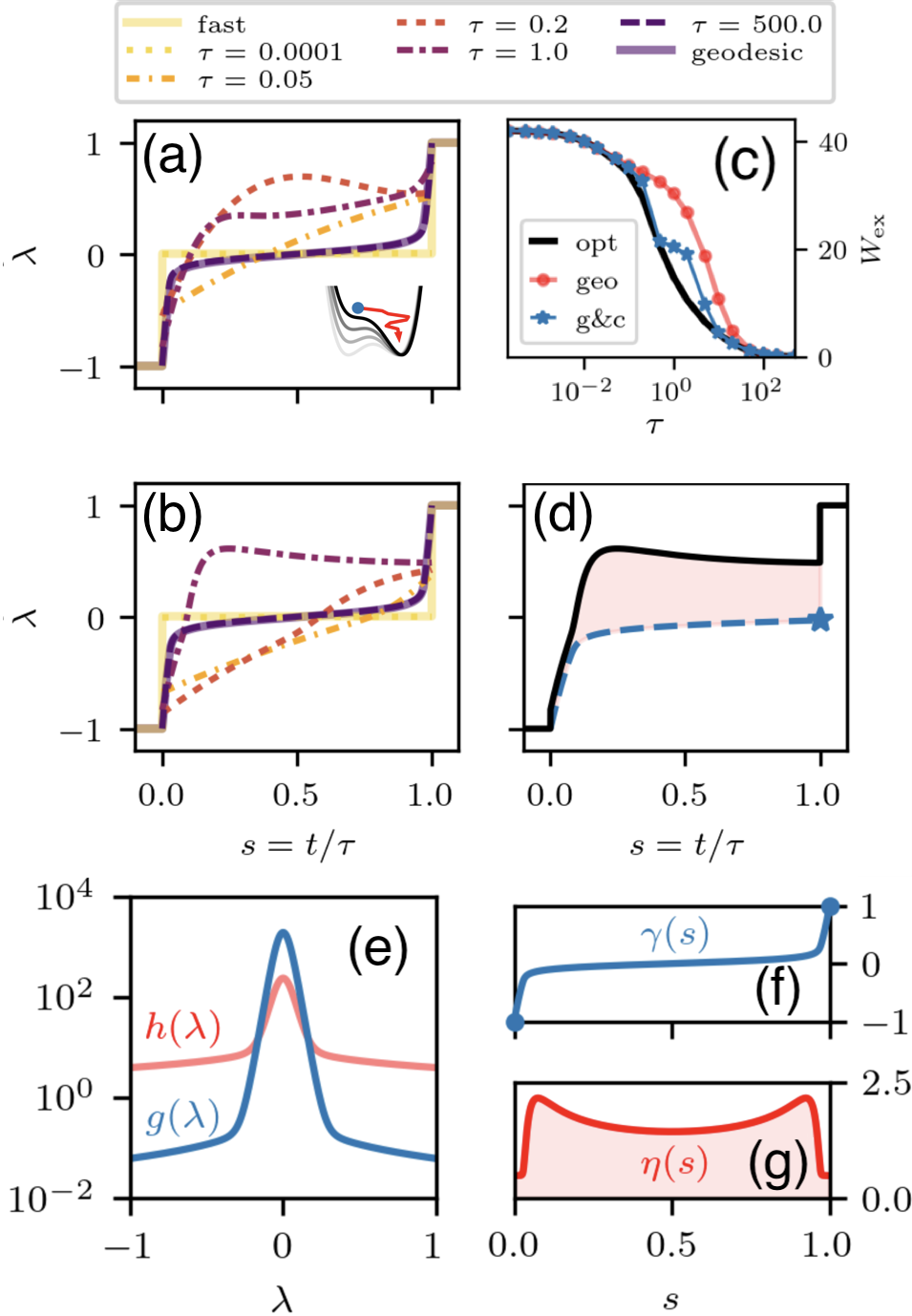}
\caption[Applying the geodesic-counterdiabatic perspective of optimal protocols for the linearly biased double well. ]{(a) Exact optimal protocols obtained in \cite{zhong2022limited} from solving PDEs, for the linearly-biased double well (Eq.~\eqref{eq:double-well-potential}; $E_0 = 16$) for different protocol durations $\tau$ including the fast protocol $\tau \rightarrow 0$ \cite{blaber2021steps} (solid yellow) and the friction tensor geodesic protocol (dark purple). (b) Geodesic-counterdiabatic protocols numerically obtained from Alg.~1. (c) Geodesic-counterdiabatic protocols (blue stars) outperform the geodesic protocol (red circles) for all $\tau$; cf. performance of exact optimal protocols (black). We examine the reduction in performance at $\tau \sim 2$ in the SM \cite{zhong2024beyond}. (d) The $\tau = 1$ protocol numerically obtained via Alg.~1 (here $\gamma_f = 0.0291$), same coloring as Fig.~1. (e) The friction and Fisher information tensors yield (f) the geodesic $\gamma(s)$ (here $\lambda_A = -1$, $\lambda_B = 1$) and (g) the non-monotonic
counterdiabatic forcing $\eta(s)$.}
\label{fig:fig2}
\end{figure}

Surprisingly, \emph{non-monotonic} optimal protocols have been found for the linearly-biased double well ~\cite{zhong2022limited} 
\begin{equation}
  U_\lambda(x) = E_0[(x^2 - 1)^2 / 4 - \lambda x],   \label{eq:double-well-potential}
\end{equation}
for certain values of $E_0$ and $\tau$ (e.g., $\tau = 0.2$ in Fig.~2(a) for $E_0 = 16$). Fig.~2(b) illustrates protocols numerically obtained from Alg.~1 (details given in the SM in web version of \cite{zhong2024beyond}). Due to the limited expressivity of the controls, these protocols are not identical to the exact optimal protocols obtained by solving PDEs~\cite{zhong2022limited} (Fig.~2(a)). However, they reproduce significant properties (e.g., discontinuous jumps and non-monotonicity, becoming exact in $\tau \rightarrow 0$ and $\tau \rightarrow \infty$), and lead to improved performance over geodesic protocols (Fig.~2(c)). Fig.~2(d) illustrates the non-monotonic $\tau = 1$ protocol as a sum of geodesic and counterdiabatic terms. The tensors $g$ and $h$ (Fig.~2(e)) yield necessarily monotonic geodesics (Fig.~2(f)), and non-monotonic counterdiabatic forcing (Fig.~2(g)) that leads to non-monotonic optimal protocols. 

\section{Discussion}

We have demonstrated the equivalence between overdamped thermodynamic geometry on $\mathcal{M}$---previously seen as an approximate framework---and $L^2$ optimal transport geometry on equilibrium distributions $\mathcal{P}^\mathrm{eq}_\mathcal{M}(\mathbb{R}^d) \subset \mathcal{P}(\mathbb{R}^d)$. The resulting geodesic-counterdiabatic optimal protocols from Alg.~1 are exact for parameteric harmonic traps, and explain both the ubiquitous discontinuous jumps and the non-monotonic behavior observed in optimal protocols. 

We note that \cite{li2022geodesic} presents a geodesic-counterdiabatic PDEs approach for underdamped dynamics. Additionally, underdamped optimal control has recently been related to a modified optimal transport problem \cite{sabbagh2023wasserstein, sanders2024optimal}. We expect that the metric tensor in \cite{li2022geodesic}, the friction tensor (Eq.~\eqref{eq:friction-SC}) for underdamped dynamics \cite{zulkowski2012geometry}, and the optimal transport specified in \cite{sabbagh2023wasserstein, sanders2024optimal} may also be geometrically unified through methods similar to ours. 

An interesting future direction will be to apply our findings to heat engines \cite{abiuso2020optimal, brandner2020thermodynamic, frim2022optimal, frim2022geometric, chennakesavalu2023unified} and active matter systems \cite{zulkowski2013optimal, davis2024active}, which have been studied with approximate geodesic protocols. We hope that the insight that minimal-work protocols require both geodesic and counterdiabatic components will prove to be useful in understanding the cyclic and fundamentally non-equilibrium processes that occur in biological systems. 


This work greatly benefited from conversations with Adam Frim.  A.Z. was supported by the Department of Defense (DoD) through the National Defense Science \& Engineering Graduate (NDSEG) Fellowship Program.
MRD thanks Steve Strong and Fenrir LLC for supporting this project. This work was supported in part by the U.S. Army Research Laboratory and the U.S. Army Research Office under Contract No. W911NF-20-1-0151.


\begin{subappendices}
\renewcommand{\thesection}{\thechapter.\alph{section}} %

\section{Counterdiabatic driving expression}

In this Appendix we derive our expression for the counterdiabatic term (Eq.~\eqref{eq:eta-solution}). Per definition, the counterdiabatic term $\phi_s(x) = \eta^\mu(s) U_\mu(x)$ is constructed to solve the continuity equation 
\begin{equation}
  \frac{\partial \rho_{\gamma(s)}^\mathrm{eq} }{\partial s}(x) = \nabla \cdot ( \rho_{\gamma(s)}^\mathrm{eq}(x) \nabla \phi_s(x)) .\label{eq:counterdiabatic-PDE}
\end{equation}
We can divide by $\rho^\mathrm{eq}_\lambda(x)$ and plug in $\partial_s \ln \rho_{\gamma(s)}^\mathrm{eq}  = [ \dee \gamma(s) / \ds] \, \delta f_\mu(x) $  to obtain
\begin{equation}
    \frac{\dee \gamma^\mu}{\ds} \delta f_\mu(x) = - \eta^\nu(s) [\mathcal{L}_{\gamma(s)}^\dagger \, \delta f_\nu ](x), \label{eq:counterdiabatic-1}
\end{equation}
where we have used the fact that $U_\nu(x) = -\delta f_\nu + \mathrm{const}$, and that the adjoint operator (Eq.~\eqref{eq:BK-1}) satisfies $\mathcal{L}^\dagger_\lambda [\psi + c] = \mathcal{L}^\dagger_\lambda [\psi]$ for any scalar field $\psi(x)$ and constant $c \in \mathbb{R}$.

Due to the limited expressivity of available controls in a given problem, it might not be possible to satisfy Eq.~\eqref{eq:counterdiabatic-1}. However, this potential insolubility is resolved by applying a projection operator to both sides
\begin{gather}
    \frac{\dee \gamma^\mu}{\ds}  \int \bigg[ - \rho^\mathrm{eq}_{\gamma(s)}(x) \, \delta f_\alpha (x) \, \big\{ \mathcal{L}_{\gamma(s)}^\dagger \big\}^{-1} \bigg] \delta f_\mu(x) \dx \nonumber \\ = - \eta^\nu(s) \int \bigg[ - \rho^\mathrm{eq}_{\gamma(s)}(x) \, \delta f_\alpha (x) \, \big\{ \mathcal{L}_{\gamma(s)}^\dagger \big\}^{-1} \bigg] \big[\mathcal{L}_{\gamma(s)}^\dagger \, \delta f_\nu \big](x) \dx, 
\end{gather}
which yields 
\begin{equation}
    g_{\alpha \mu}(\gamma(s)) \bigg[ \frac{\dee \gamma^\mu}{\ds} \bigg] = h_{\alpha \nu} (\gamma(s)) \, \eta^\nu(s), \label{eq:counterdiabatic-2}
\end{equation}
using the friction tensor and Fisher information metric expressions Eq.~\eqref{eq:friction-tensor-3} and Eq.~\eqref{eq:fisher-information-metric}. Optimal transport geometry measures ``horizontal'' displacement while information geometry measures ``vertical'' displacement \cite{khan2022optimal}, so $g$ and $h$ can be seen to give the conversion between the left and right hand sides of\footnote{One may choose other projection operators to act on Eq.~\eqref{eq:counterdiabatic-1}, but then non-conventional tensors would have to be defined. Furthermore, the particular interpretation for Eq.~\eqref{eq:counterdiabatic-2}, that the geometries associated with $g$ and $h$ convert between horizontal and vertical displacements, will be lost.}. 

Because the Fisher information metric is symmetric and positive-definite, 
we can apply its inverse to Eq.~\eqref{eq:counterdiabatic-2}, thus reproducing Eq.~\eqref{eq:eta-solution}. 

\section{Analytic reproduction of harmonic oscillator optimal protocols}

In this Appendix we show that Alg. 1 exactly reproduces the optimal protocols for harmonic potentials first found in \cite{schmiedl2007optimal}. 

We first consider a variable-center harmonic potential in one dimension $U_\lambda(x) = (x - \lambda)^2/2$, both the friction and Fisher information tensors are spatially constant $g(\lambda) = f(\lambda) = 1$ due to translation symmetry. For any $\lambda_f$, the KL-divergence given by $D_\mathrm{KL}(\rho_\lambda^\mathrm{eq} | \rho_f^\mathrm{eq} ) = \int \rho_\lambda^\mathrm{eq}(x) \ln \,  [ \rho_\lambda^\mathrm{eq}(x) / \rho_f^\mathrm{eq}(x) ] \dx = (\lambda - \lambda_f)^2 / 2$, while the squared thermodynamic length is given by $\mathcal{T}^2(\lambda_i, \lambda) = (\lambda - \lambda_i)^2$. Without loss of generality we fix $\lambda_i = 0$. Following Alg. 1, we obtain $\gamma_f = (1 + 2 / \tau)^{-1} \lambda_f$ with geodesic $\gamma(s) = s \, \gamma_f$, and $\eta(s) = \gamma_f$. Ultimately, this yields the optimal protocol
\begin{equation}
    \lambda^*(t) = \underbrace{[ \lambda_f / (2 + \tau)] \, t}_\text{geodesic} \hspace{5pt}  + \underbrace{ 1 / (2 + \tau)}_\text{counterdiabatic}\hspace{-5pt},  \quad\mathrm{for}\quad t \in (0, \tau), 
\end{equation}
which yields the original analytic solution reported by Schmiedl and Seifert $\lambda^*(t) = \lambda_f(1 + t) / (2 + \tau)$, Eq.~(9) in \cite{schmiedl2007optimal} (note that they use $t$ to denote protocol duration and $\tau$ to denote time, which is swapped with respect to our notation). However, we now have a refined interpretation of this optimal protocol: as consisting of a geodesic component that connects $\lambda_i$ to $\gamma_f \neq \lambda_f$, and a counterdiabatic component necessary to achieve the geodesic trajectory for finite protocol durations $\tau$.

We now solve for the optimal protocol for a variable-stiffness harmonic trap $U_\lambda(x) = \lambda x^2 / 2$. We will defer solving for $\gamma_f$ until after obtaining the analytic form for $\lambda^*(t)$. The friction tensor for this potential has previously been shown to be $g(\lambda) = 1/4 \lambda^3$ \cite{sivak2012thermodynamic}, and the Fisher information metric can be calculated to be $h(\lambda) = \big\langle x^4 / 4 \big\rangle_\lambda^\mathrm{eq} - \big( \langle x^2 / 2 \rangle_\lambda^\mathrm{eq} \big)^2  = 1/2 \lambda^2$. As shown in \cite{sivak2012thermodynamic}, by switching to standard deviation coordinates $\sigma = \lambda^{-1/2}$, the friction tensor is constant $\tilde{g}(\sigma) = 1$, and thus geodesics are $\sigma(s) = (1 - s) \, \sigma_A + s \, \sigma_B$ with thermodynamic length $\mathcal{\tilde{T}}^2(\sigma_A, \sigma_B) = (\sigma_B - \sigma_A)^2$, where $\sigma_A = \lambda_i^{-1/2}$ and $\sigma_B = \gamma_f^{-1/2}$. This yields the geodesic
\begin{equation}
    \gamma(s) = [(1 - s) \sigma_A + s \, \sigma_B ]^{-2},
\end{equation}
with the corresponding counterdiabatic term given by
\begin{equation}
  \eta(s) = (\sigma_A - \sigma_B) / [(1 - s) \sigma_A + s \, \sigma_B ].
\end{equation}
Plugging these into $\lambda^*(t) = \gamma(t / \tau) + \tau^{-1} \eta (t / \tau)$ leads to the interpretable optimal protocol expression
\begin{equation}
  \lambda^*(t) = \underbrace{\big[ 1 / \sigma^*(t) \big]^{2} }_\text{geodesic} + \underbrace{\tau^{-1} \big[ \Delta \sigma / \sigma^*(t) \big] }_\text{counterdiabatic}, \label{eq:solution-HO-stiffening}
\end{equation}
where $\sigma^*(t) = \sigma_i + (t / \tau) \Delta \sigma $ is the linear interpolation between the endpoints $\sigma_i = \lambda_i^{-1/2}$ and $\sigma_i + \Delta \sigma = \gamma_f^{-1/2}$. 

Finally we solve for $\gamma_f$ via
\begin{equation}
    \gamma_f = \arg \min_{\lambda} \ \mathcal{T}^2(\lambda_i, \lambda) / \tau + D_\mathrm{KL}(\rho_\lambda^\mathrm{eq} | \rho_f^\mathrm{eq} ), \label{eq:gamma_f_def}
\end{equation}
using $\mathcal{T}^2(\lambda_i, \lambda) = (\lambda_i^{-1/2} - \lambda^{-1/2})^2$ (recall that $\lambda^{-1/2} = \sigma$ is the standard deviation) and $D_\mathrm{KL}(\rho_\lambda^\mathrm{eq} | \rho_f^\mathrm{eq} ) = (1/2)[(\lambda_f/\lambda - 1) + \ln (\lambda / \lambda_f)]$ \cite{7440}. This yields
\begin{equation}
    \gamma_f =  \big( \sqrt{1 + 2 \lambda_i \tau + \lambda_i \lambda_f \tau^2 } - 1 \big)^2 / \lambda_i \tau^2,
\end{equation}
or when expressed in $\sigma$ coordinates
\begin{equation}
    \Delta \sigma = \sigma_i  \bigg( \frac{ 1 + \lambda_f \tau - \sqrt{1 + \lambda_k \tau + \lambda_i \lambda_f \tau^2 }}{ 2 + \lambda_f \tau } \bigg) . \label{eq:delta-sigma}
\end{equation}
Substituting this expression into Eq.~\eqref{eq:solution-HO-stiffening} reproduces the exact optimal protocol, Eqs.~(18) and~(19) in \cite{schmiedl2007optimal}. (Note that they use $t$ to denote protocol duration and $\tau$ to denote time, which is swapped with respect to our notation.)

Fig.~1 illustrates the obtained exact optimal protocol for this problem (Eqs.~\eqref{eq:solution-HO-stiffening} and~\eqref{eq:delta-sigma}) as a sum of geodesic and counterdiabatic components. This clarifies the origin of the jumps in optimal protocols: at $t = 0$, the counterdiabatic component is suddenly turned on, and at $t = \tau$ it is abruptly turned off.

\end{subappendices}

\chapter{Conclusion}

Thank you dear reader, for making it this far into reading this thesis! 

Rewinding back to 2019 when I first began my PhD research, all I knew about optimal protocols was that in the large-$\tau$ limit, the excess work of a finite-time protocol could be approximated as
\begin{equation}
    W_\mathrm{ex} = \int_0^\tau \dot{\lambda}^\mu \bigg[ \bigg\langle \frac{\partial U_\lambda}{ \partial \lambda^\mu }\bigg\rangle - \bigg\langle \frac{\partial U_\lambda}{ \partial \lambda^\mu }\bigg\rangle^\mathrm{eq}_\lambda \, \bigg] \dt \approx \int_0^\tau |\dot{\lambda} |^2_\mathrm{thermo} \dt \label{eq:work-thermo-geo}
\end{equation}
via the Sivak and Crooks thermodynamic geometry framework \cite{sivak2012thermodynamic}. Although a very useful framework---optimal protocols in the large-$\tau$ regime may be solved as finding friction-tensor geodesics---I was not particularly satisfied by the approximation sign ($\approx$), and spent the entirety of my PhD trying to understand whether there was an exact geometric structure behind optimal non-equilibrium protocols.

In Chapters 2 and 3, we developed numerical methods to solve for optimal protocols beyond the large-$\tau$ regime, through (respectively) adapting optimal control theory to the Fokker-Planck equation, and applying importance sampling of the Langevin equation to tractably optimize protocols on-the-fly.\footnote{While Ch. 3 solved for optimal protocols minimizing the \emph{time-asymmetric} work, the objective function Eq.~\eqref{eq:objective-function} could be constructed just with the first term $\langle W\rangle_F$ to solve for traditional optimal protocols. We have found that this reproduces all the optimal protocols in Ch. 2 \cite{unpublished}.} The obtained optimal protocols reproduced the discontinuous jumps at $t = 0$ and $t = \tau$, and revealed new \emph{non-monotonic} optimal protocols. 

The final theoretical insights were demonstrated in Chapter 5: the excess thermodynamic work to enact a finite-time protocol may be rewritten in terms of the thermodynamic state trajectory
\begin{equation}
    W_\mathrm{ex} = \mu^{-1} \int_0^\tau \| \dot{\rho}_t\|^2_\mathrm{transport} \dt + D_\mathrm{KL}( \rho_\tau | \rhoeq_{\lambda_f}). \label{eq:work-geo-again}
\end{equation}
The first term is purely geometric, and surprisingly can also be expressed equivalently with the Sivak and Crooks friction tensor $\int_0^\tau \| \dot{\rho}_t\|^2_\mathrm{transport} \dt = \mu \int_0^\tau \dot{\gamma}^\mu \dot{\gamma}^\nu g_{\mu \nu}\big( \gamma(t) \big) \dt$ in the case that the thermodynamic is steered to follow $\rho_t = \rhoeq_{\gamma(t)}$ for a path $\gamma(t) \in \mathcal{M}$. This yields the \emph{exact} expression 
\begin{equation}
    W_\mathrm{ex} = \int_0^\tau |\dot{\gamma} |^2_\mathrm{thermo} \dt + D_\mathrm{KL}( \rhoeq_{\gamma(\tau)} | \rhoeq_{\lambda_f}), \label{eq:conclusion-exact-expression}
\end{equation}
which notably does \emph{not} have the approximation sign, c.f., Eq.~\eqref{eq:work-thermo-geo})!\footnote{How funny is it that the first term in the \emph{exact} expression involves the same friction tensor $g_{\mu \nu}$ that was first developed in this approximate framework! This really was a full-circle moment for me.} Solving for exact optimal protocols now involves finding a geodesic $\gamma(t)|_{t \in [0,\tau]} \in \mathcal{M}$ (minimizing Eq.~\eqref{eq:conclusion-exact-expression} for $\gamma(0) = \lambda_i$ and $\gamma(\tau)$ unconstrained), and then solving for the protocol $\lambda^*(t)$ that can steer the thermodynamic state to follow this trajectory $\rho_t = \rhoeq_{\gamma(t)}$ through including an additional counterdiabatic driving. Together, the boundary term in Eq.~\eqref{eq:conclusion-exact-expression} and the counterdiabatic driving needed to steer the thermodynamic state explain both the generic discontinuities at $t = 0$ and $t = \tau$  (Fig.~\ref{fig:jumps}) and observed non-monotonic behavior (Fig.~\ref{fig:double-well}.b for $t_f = 0.2$) in globally optimal protocols. 

We now finally return to the first sentence in this thesis:\\[0.25\baselineskip]
What is the \emph{laziest way possible} to do something, subject to the fundamental laws of physics?\\[0.25\baselineskip]
The answer, at the very least in the setting of overdamped stochastic thermodynamics, is geometric: Steer the thermodynamic state to follow a geodesic under the friction tensor!

\subsection{The Big Remaining Question}

The big remaining question that has not been addressed in my completed PhD work is: how does this exact geometric expression relate to the \emph{in vivo} behavior of proteins in biology that operate nontrivially far from equilibrium? See the following subsection ``Optimal active engines and the Lorentz force law'' for an initial attempt. We hope to have inspired the reader to consider this as an interesting research problem to tackle!

\section{Next steps}

Before finishing the thesis, upon the insistence of my thesis committee, we will sketch out projects that are \emph{so close} to being finished---coming soon!

\subsection{Higher-order approximations for optimal protocols in $\lambda(t)$}

Here, working with PhD student Sam D'Ambrosia, we work with the original Sivak and Crooks framework to \emph{rigorously} extend
\begin{equation}
    W_\mathrm{ex} = \int_0^\tau |\dot{\lambda}|^2_\mathrm{thermo} \dt + o(\tau^{-3})
\end{equation}
to include higher-order terms 
\begin{equation}
    W_\mathrm{ex} = \int_0^\tau  \bigg\{ \sum_{k = 2}^K  \dot{\lambda}^{\mu_1} ... \dot{\lambda}^{\mu_k} g^{(k)}_{\mu_1 ... \mu_k} \big(\lambda(t)\big) \bigg\} \dt +  o(\tau^{-{K+1}}),
\end{equation}
where $g^{(k)}_{\mu_1 ... \mu_k} (\lambda)$ are generalized thermodynamic tensors also obtained via time-correlation expressions similar to Eq.~\eqref{eq:friction-SC}. 

We have found that unfortunately, including each additional term makes the optimization more and more difficult, and the cost function still remains \emph{approximate}; instead, we argue that, once again, via the \emph{exact} expression Eq.~\eqref{eq:conclusion-exact-expression}, optimal protocols can be efficiently obtained by optimizing a path for the thermodynamic state $\gamma^*(t)$ as opposed to optimizing a path for the controls $\lambda^*(t)$.

\subsection{Optimal transport, information geometry, and the Onsager coefficient matrix}

Looking back at Chapter 4, the Sivak and Crooks thermodynamic metric may be written as 
\begin{equation}
    g_{\mu \nu}(\lambda) = -\int \frac{\partial \ln \rhoeq_{\lambda}}{\partial \lambda^\mu}  \mathcal{L}_{\lambda}^{-1} \bigg[ \frac{\partial \ln \rhoeq_{\lambda}}{\partial \lambda^\nu} \rhoeq_\lambda \bigg] \dx \label{eq:g-new}
\end{equation}
(see Eq.~\eqref{eq:friction-tensor-3}), while the Fisher Information Metric (Eq.~\eqref{eq:fisher-information-metric}) is 
\begin{equation}
    h_{\mu \nu}(\lambda) = \int \frac{\partial \ln \rhoeq_{\lambda}}{\partial \lambda^\mu} \frac{\partial \ln \rhoeq_{\lambda}}{\partial \lambda^\nu} \rhoeq_\lambda  \dx,  \label{eq:h-new}
\end{equation}
see also \cite{amari2016information}.

What is nice about metric tensors is that locally, as symmetric positive bilinear forms, they may be simultaneously diagonalized \cite{eremenko2018simultaneous}\footnote{Thank you David Theurel for pointing this out to me!}, and the way to do this for $g_{\mu \nu}(\lambda)$ and $h_{\mu \nu}(\lambda)$ is to consider the \emph{left} eigenvectors $\{ \phi^{(i)}_\lambda (x) \}_{i = 0, 1, 2...}$ of the Fokker-Planck operator, satisfying
\begin{equation}
    \mathcal{L}^\dagger_\mathrm{\lambda} [\phi^{(i)}_\lambda]  = \alpha_i \phi^{(i)}_\lambda
\end{equation}
with $\alpha_0 = 0$ and (negative) $\alpha_{i+1} < \alpha_i$. These orthogonal eigenvectors may be normalized to be orthonormal, as to satisfy 
\begin{equation}
    \int \phi^{(i)}_\lambda(x) \, \phi^{(j)}_\lambda(x) \, \rhoeq_\lambda(x)  \dx = \delta_{ij}. 
\end{equation}

We have then 
\begin{equation}
    -\int \phi^{(i)}_\lambda \, \mathcal{L}_\lambda^{-1} \{ \phi^{(j)}_\lambda  \rhoeq_\lambda \}  \dx = -\alpha_j^{-1} \delta_{ij}. 
\end{equation}
(See also \cite{wadia2022solution}, and more recently, \cite{sawchuk2024dynamical}.)

Comparing these two equations to Eqs.~ \eqref{eq:g-new} and \eqref{eq:h-new} suggest that: if at a given $\lambda$ we specify the local coordinates for $\{\lambda^i\}_{i =  1,2 ...}$,\footnote{Here we use $i$ instead of $\mu$ to emphasize the specific choice of coordinates.} so that $\partial \ln \rhoeq_\lambda/ \partial \lambda^i = \phi^{(i)}_\lambda$, then the friction tensor becomes 
\begin{equation}
  g_{ij}(\lambda) = -\alpha_j^{-1} \delta_{ij}, 
\end{equation}
and the Fisher information metric is
\begin{equation}
    h_{ij}(\lambda) = \delta_{ij}. 
\end{equation}
In other words, using the eigenmodes of the Fokker-Planck equation as the coordinate basis simultaneously diagonalizes both tensors!

\subsubsection{Onsager Transport Coefficient Matrix}

In the early 1930s, Lars Onsager introduced what is today known as the Onsager transfer coefficient matrix \cite{onsager1931reciprocal}: given a thermodynamic system in equilibrium with a potential energy function $U(x)$ and the scalar functions (``observables'') $O_A(x)$ and $O_B(x)$, the transfer coefficient matrix is defined as 
\begin{equation}
    L_{AB } = \lim_{\epsilon \rightarrow 0} \, \frac{\mathrm{change \ in \ } \langle O_A \rangle^\mathrm{eq}_{U \rightarrow U - \epsilon O_B} }{ \epsilon 
    } . 
\end{equation}
If a group of observables $\{O_\mu(x)\}_{\mu = 1, 2, ...}$ are chosen to be the conjugate forces with regards to $\{\lambda^\mu\}_{\mu = 1, 2, ...}$
\begin{equation}
    O_\mu(x) = -\frac{\partial U_\lambda}{\partial \lambda^\mu}(x), 
\end{equation}
then we have that 
\begin{equation}
    L_{\mu \nu}(\lambda) = -\int \frac{\partial U_\lambda}{\partial \lambda^\mu} \mathcal{L}_\mu \bigg[ \frac{\partial U_\lambda}{\partial \lambda^\nu} \rhoeq_\lambda \bigg] \dx. 
\end{equation}

The final observation we made is that this may be also written as 
\begin{equation}
    L_{\mu \nu}(\lambda) = -\int \frac{\partial \ln \rhoeq_\lambda}{\partial \lambda^\mu} \mathcal{L}_\mu \bigg[ \frac{\partial \ln \rhoeq_\lambda}{\partial \lambda^\nu} \rhoeq_\lambda \bigg] \dx,  \label{eq:L-new} 
\end{equation}
which allows directly comparison to Eqs.~\eqref{eq:g-new} and \eqref{eq:h-new}! 

In this way, the Onsager transport coefficient matrix can also be viewed as another Riemmanian metric tensor, defining another thermodynamic geometry on $\lambda \in \mathcal{M}$!

In particular, using the same basis $\partial \ln \rhoeq_\lambda/ \partial \lambda^i = \phi^{(i)}_\lambda$, the Onsager transport coefficient matrix is diagonalized
\begin{equation}
    L_{ij} (\lambda) = -\alpha_i \delta_{ij}. 
\end{equation}
We are left with the trio of identities: 
\begin{equation}
    g_{ij}(\lambda) = -\alpha_j^{-1} \delta_{ij}, \quad\quad h_{ij}(\lambda) = \delta_{ij}, \quad\mathrm{and}\quad L_{ij} = -\alpha_i \delta_{ij}. 
\end{equation}
In other words, the eigenmodes of $\mathcal{L}_\lambda$ simultaneously diagonalizes these three tensors!

\subsection{Optimal active engines and the Lorentz force law}

It is known that Newton's equations could be derived from of the principle of least action, e.g., considering all possible paths $x(t) \in \mathbb{R}^3$ subject to boundary conditions $x(0) = x_A$ and $x(\tau) = x_B$, the one that minimizes the path action functional
\begin{equation}
    S[x(t)] = \int_0^\tau \frac{m |\dot{x}(t)|^2}{2} - V\big( x(t)\big) \dt
\end{equation}
must satisfy $m\ddot{x}(t) = -\nabla V\big(x(t)\big)$.\footnote{This equation $-\nabla V\big(x(t)\big) =m\ddot{x}(t)$ is the famous expression $F = ma$!} 

Also established (although not taught as commonly in the undergraduate physics curriculum) is that for the path action functional
\begin{equation}
    S^\mathrm{EM}[x(t)] = \int_0^\tau  \frac{m |\dot{x}(t)|^2}{2} + q\dot{x}(t) \cdot \vec{A}\big(x(t) \big)  \dt \label{eq:lorentz-force}
\end{equation}
paths of least action satisfy the differential equation $m \ddot{x}= q\dot{x} \times \vec{B}$, where $\vec{B} = \vec{\nabla} \times \vec{A}$---this is the Lorentz force law for a particle in a magnetic field!

In an ongoing project with Adam Frim that is very close to completion, we show that the input thermodynamic work
\begin{equation}
    W_\mathrm{in} = \int_0^\tau \dot{\lambda}^\mu \bigg\langle \frac{\partial U_\lambda}{\partial \lambda^\mu} \bigg\rangle \dt 
\end{equation}
for the \emph{non-conservative} overdamped Langevin equation 
\begin{equation}
    \dot{X}_t = - \mu \nabla U_{\lambda(t)}(X_t) + \mu F_\mathrm{nc}(X_t) + \sqrt{2 \mu \kBT} \, \xi_t,
\end{equation}
where $F_\mathrm{nc}(x)$ is a constant \emph{non-conservative} vector field satisfying $\nabla \cdot F_\mathrm{nc} = 0$, can be written in terms of the thermodynamic state trajectory $\rho_t|_{t\in[0,\tau]}$ as 
\begin{equation}
    W_\mathrm{in} = \int_0^\tau \mu^{-1} \| \dot{\rho}_t\|^2_\mathrm{transport} +  (A\cdot\dot{\rho}_t) _\mathrm{transport} \dt + F^\mathrm{neq}[ \rho_\tau | U_{\lambda_f}].
\end{equation}
The integral has the exact form as the Lorentz force path action Eq.~\eqref{eq:lorentz-force}!\footnote{For the thermdynamic state trajectory solving $\partial_t \rho_t = \nabla \cdot (\rho_t \nabla \phi_t),$, we have $\| \dot{\rho}_t\|^2_\mathrm{transport} = \int \rho_t |\nabla \phi_t|^2 \dx$, and $( A, \dot{\rho}_t)_\mathrm{transport} = \int  \rho_t F_\mathrm{nc} \cdot \nabla \phi_t \dx$.} In particular time-periodic protocols can obtain \emph{positive} output work $W_\mathrm{out} = -W_\mathrm{in}$ as $\int_0^\tau (A \cdot \dot{\rho}_t)_\mathrm{transport} \dt$ can be \emph{negative}, due to $F_\mathrm{nc}$. Thus, optimal protocols maximizing work extraction from active systems steer the thermodynamic state to satisfy the Lorentz force law defined on the space of thermodynamic states / probability distributions! We also construct optimal \emph{active engines} by considering time-periodic protocols yielding time-periodic steady states. 

What is particularly exciting about this result is the possibility to finally connect to actual biological motors like ATP-synthase found in the mitochondria\footnote{also known as the powerhouse of the cell.}, that may be modeled with non-conservatively driven Langevin equations \cite{dimroth1999energy}, and compare to experimental results measuring the thermodynamic efficiencies of optimal protocols controlling these sorts of systems \cite{mishima2025efficiently}!

\section{\emph{Fin}}

Dear reader, thank you so, so much for reading my thesis! Hopefully you have now an understanding and appreciation for the cover art on page 0 of this thesis. If you have any questions or comments, feel free to get in touch through email at \url{adrizhong@berkeley.edu}. Until next time!

\subsection{The End ... for now!  }



\newcommand{\boldtheta}{\theta} 
\newcommand{\Ndata}{N_\mathrm{data}}
\newcommand{\Nesctheta}{N^\mathrm{esc}_\boldtheta}
\newcommand{\Vb}{V_\mathrm{b}}
\newcommand{\ntheta}{n^\mathrm{ }_\theta}
\newcommand{\phitheta}{\phi^\mathrm{ }_\theta}

\appendix

\chapter{Physical Derivation of the Langevin Equation} \label{chapter:physical-derivation-langevin}

This appendix is based off of a derivation I learned from Phill Geissler's course in nonequilibrium statistical mechanics. 

Here we derive the overdamped Langevin equation  (Eq.~\eqref{eq:langevin-eq}), reproduced here for a fixed potential energy $U(x)$:
\begin{equation}
    \dot{X}_t = -\mu \nabla U(X_t) + \sqrt{2 \mu \kBT} \, \boldxi_t , \label{eq:langevin-rerwritten} 
\end{equation} 
from purely deterministic, reversible Hamiltonian equations of the system linearly coupled to a large ensemble of harmonic oscillators. The derivation is actually a bit cursed, in the sense that there are a number of seemingly contrived steps, and that many of these steps are quite difficult to perform rigorously. Following the tradition of physics, the sketch we present here forgoes this rigor. 

In order to give a physical derivation, we will first derive the underdamped Langevin equation, which is 
\begin{equation}
    M\ddot{X}_t = -\gamma \dot{X}_t  - \nabla U(X_t) + \sqrt{2 \gamma \kBT \,} \xi_t \label{eq:langevin-underdamped}
\end{equation}
(here $M$ is the system's mass and $\gamma = \mu^{-1}$ is the drag coefficient) from a system linearly coupled to a bath of harmonic oscillators  $\{Y_i(t)\}$ with individual mass and spring constants $\{m_i, k_i\}$: 
\begin{align}
    M\ddot{X}_t &= -\nabla U(X_t) +\underbrace{ \sum_i c_i Y_i(t)}_\mathrm{from \ coupling \ to \ bath} \label{eq:deterministic-with-bath} \\ 
    m_i\ddot{Y}_i(t) &= - k_i Y_i(t) \quad -\underbrace{c_i X_t}_\mathrm{frm \ Newton's \ 2nd \ Law}. \label{eq:HO}
\end{align}
Of course, this assumes a particular kind of bath (that of a collection of harmonic oscillators) and a linear coupling interaction, but hey, we have to start somewhere.\footnote{Note that here we use a subscript $X_t$ to denote the ``stochastic'' appearance of $X_t$, while the for the bath variables we write $Y_i(t)$ to emphasize that they really are deterministic underlying variables.} For simplicity, we choose our coordinates for $x$ to have the initial condition $X_0 = 0$.

What is so nice about Eq.~\eqref{eq:HO} is that as a linear differential equation, there is an exact solution: 
\begin{equation}
    Y_i(t) = Y^\mathrm{h}_i(t) + \frac{c_i}{m_i } \int^t_{0} G_i(t -t') {X}_{t'} \dt', \label{eq:HO-solution}
\end{equation}
where $Y^\mathrm{h}_i(t) = Y_i(0) \cos (\omega_i t) + (\dot{Y}_i(0) / \omega_i) \sin (\omega_i t)$ is the homogeneous solution that is completely determined by the initial conditions of the bath, and $G_i(t - t') = \omega_i^{-1} \sin\big(\omega_i(t - t')\big)$ is the Green's function kernel for a undamped harmonic oscillator. Here $\omega_i = \sqrt{k_i/m_i}$ is the fundamental frequency of each of the bath's harmonic oscillators. We can consider $Y_i(0)$ and $\dot{Y}_i(0)$ as random variables, as the exact initial conditions of the harmonic oscillator bath are unknown. 

We can plug Eq.~\eqref{eq:HO-solution} into Eq.~\eqref{eq:deterministic-with-bath} to get the expression
\begin{equation}
    M\ddot{X}_t = - \nabla U(X_t) + \bigg[ \sum_i c_i Y^\mathrm{h}_i(t) \bigg]  + \int_0^t \bigg[ \sum_i \frac{c_i}{m_i \omega_i} \,\sin \big( \omega_i (t - t') \big)  \bigg]  X_{t'} \dt',
\end{equation}
which is beginning to resemble Eq.~\eqref{eq:langevin-underdamped}. 

The next step is to do an integration-by-parts in $t'$ for the third term.\footnote{This is probably the most contrived steps in this derivation, and to be honest I am not quite sure how this step was devised in the first place!} We have 
\begin{equation}
    \int_0^{t} \sin (\omega_i (t - t')) X_{t'} \dt' = \cancelto{0}{\bigg[ \frac{1 - \cos\big(\omega_i (t - t')\big)}{\omega_i}    X_{t'} \bigg]\bigg|_{t' = 0}^t}  - \int_0^{t}  \frac{1 - \cos (\omega_i (t - t'))}{\omega_i} \,\dot{X}_{t'} \dt'  ,  
\end{equation}
whose latter term cancels to zero as we have chosen coordinates with the initial condition $X_0 = 0$.

This may all be substituted back to get 
\begin{equation}
    M\ddot{X}_t = - \nabla U(X_t) + \underbrace{\bigg[ \sum_i c_i Y^\mathrm{h}_i(t) \bigg]}_{ C \, \xi_t}  - \int_0^t  \underbrace{\bigg[ \sum_i \frac{c_i [1 - \cos\big( \omega_i(t - t') \big) ]}{m_i \omega_i^2}  \bigg]}_{K(t - t')}  \dot{X}_{t'} \dt',
\end{equation}
or, with some rearranging, 
\begin{equation}
    M\ddot{X}_t = -\int_0^t K(t - t') \dot{X}_{t'} \dt' - \nabla U(X_t)  +  C \, \xi_t. 
\end{equation}
Here, $K(t - t')$ is known as the \emph{memory kernel}, $C$ is a a chosen constant dependent on the couplings $\{c_i\}$, and $\xi_t$ is an effectively random time series whose stochasticity comes from the initial bath conditions $\{ Y_i(0), \dot{Y}_i(0)\}$ that we do not know.\footnote{This is why we switched back to the subscript to denote the time-dependence for $\xi_t$.}

Finally, in one last step of contrivance, we define define the variables $\gamma$ and $\kBT$ from the harmonic oscillator constants $\{k_i, m_i\}$ and couplings $c_i$ so that $C = \sqrt{2 \gamma \kBT }$, the memory kernel integrates to $\int_{-\infty}^\infty K(s)  \ds = 2 \gamma$, and the $\xi_t$ has the statistics $\langle \xi_t\rangle = 0$ and $ \langle \xi_t \,\xi_{t'}^\mathsf{T} \rangle = \boldsymbol{I}_d \,\delta(t - t') $. We can further approximate $\int_0^t K(t - t') \dot{X}_{t'} \dt' \approx  \dot{X}_t \int_{-\infty}^0 K(s) \ds = \gamma \dot{X}_t$, as long as $\dot{X}_{t'}$ varies much more slowly than the $K(\cdot)$. Now we are ultimately left with
\begin{equation}
    M\ddot{X}_t = - \gamma \dot{X}_t - \nabla U(X_t) + \sqrt{2 \gamma \kBT} \, \xi_t,
\end{equation}
which is the underdamped Langevin equation!\footnote{To ensure that $\xi_t$ is a memory-less Gaussian white noise, the  choice of the distribution of couplings $c_i$, bath oscillator constants $(k_i, m_i)$, and initial conditions $Y_i(0), \dot{Y}_i(0)$ must be done carefully. It may be seen through $\xi_t = \sum_i  \big\{ (c_i Y_i(0) / C) \cos(\omega_i t) + (c_i \dot{Y}_i(0) / C\omega_i) \cos(\omega_i t)\big\} $ that $c_i Y_i(0) / C$ and $c_i \dot{Y_i}(0) / C \omega_i$ are Fourier coefficients for $\xi_t$! }

The last step to recover the overdamped Langevin equation is through crossing out
\begin{equation}
    \cancelto{0}{M\ddot{X}_t} = - \gamma \dot{X}_t - \nabla U(X_t) + \sqrt{2 \gamma \kBT} \, \xi_t,
\end{equation}
which anecdotally is performed to the verbalized words ``by taking the overdamped limit, we can ignore the `inertia' of the system''.\footnote{Of course, to do this more rigorously, we must consider the underdamped Fokker-Planck equation for $\rho_t^\mathrm{ud} (x, v)$ corresponding to the underdamped Langevin equation written as $\dot{X}_t = V_t, \ \dot{V_t} = -M^{-1}\nabla U(X_t) -  \gamma  M^{-1}V_t  + \sqrt{2 \gamma M^{-1} \kBT} \,\xi_t$---here $V_t = \dot{X}_t$ is the velocity variable---and expand $\rho^\mathrm{ud}_t$ in terms of $\gamma^{-1}$. In the overdamped, small-$\gamma^{-1}$ limit, the lowest order nontrivial term is the overdamped Fokker-Planck equation $\partial_t \rho_t(x) = \gamma^{-1} \nabla \cdot (\rho_t \nabla (U + \kBT \ln \rho_t))$, see pg. 71 ``Integrating over Velocity'' in \cite{tong2012kinetic}.} Defining $\mu = \gamma^{-1}$ and rearranging finally gets us the overdamped Langevin equation Eq.~\eqref{eq:langevin-rerwritten}.

And thus, in three short pages we have sketched the derivation of the stochastic overdamped Langevin equation from considering the deterministic equations of the system coupled to a large bath of harmonic oscillators with unknown initial conditions! In this way, the unknown variables of the large environment give rise to the stochasticity of dynamics, and thus we can get irreversible dynamics: starting at a known $\rho_0(x) = \delta(x - x_0)$, the thermodynamic state $\rho_t$ relaxes to the equilibrium distribution following the arrow of time.

\chapter{Counterdiabatic Driving for Arbitrary Geometries}

Counterdiabatic driving, referred to here as engineered swift equilibration (ESE) is a class of driving protocols that enforce an instantaneous equilibrium distribution with respect to external control parameters at all times during rapid state transformation of open, classical non-equilibrium systems. ESE protocols have previously been derived and experimentally realized for Brownian particles in simple, one-dimensional, time-varying trapping potentials; one recent study considered ESE in two-dimensional Euclidean configuration space. Here we extend the ESE framework to generic, overdamped Brownian systems in arbitrary curved configuration spaces and illustrate our results with specific examples not amenable to previous techniques. Our approach may be used to impose the necessary dynamics to control the full configurational distribution in a wide variety of experimentally realizable settings.

\subsection{Note}

This appendix is based off of: Adam G. Frim, Adrianne Zhong, Shi-Fan Chen, Dibyendu Mandal, and Michael R. DeWeese. ``Engineered swift equilibration for arbitrary geometries.'' \emph{Physical Review E} 103, no. 3 (2021): L030102. This paper is the first paper that I had my name on as a PhD student. I owe very much gratitude to Adam G. Frim for completing this project, as during that time I was experiencing health issues. 

This work builds off of the observation that a for a specified thermodynamic state trajectory $\rho_{\gamma([0,\tau])}$, a \textit{typically non-gradient} vector field solution $\boldv_{[0,\tau]}$ to the continuity equation 
\begin{equation}
    \partial_t \rho_t(x) = -\boldnabla \cdot \big( \rho_t(x) \,  \boldv_t(x) \big) \nonumber
\end{equation}
may be obtained via considering the so-called probability current as a gradient field $J_t(x) = \rho_t (x) \,v_t (x)= -\boldnabla A_t(x)$, which then transforms the continuity equation into the Poisson's equation
\begin{equation}
    \partial_t \rho_t =  \Delta A_t \nonumber
\end{equation}
where $\Delta$ is the scalar Laplacian. This has the formal solution $A_t  = \Delta^{-1} [\partial_t \rho_t]$, and so this vector field solution may be recovered as 
\begin{equation}
    v_t(x) = - \rho_t^{-1} \nabla  \big( \Delta^{-1} [\partial_t \rho_t] \big) \nonumber. 
\end{equation}
This is a correct expression for a possible counterdiabatic forcing $F_t(x) = \mu^{-1} v_t(x)$ added to the Langevin equation 
\begin{equation}
    \dot{X}_t = - \mu \boldnabla U_{\gamma(t)} (X_t) + \sqrt{2\mu \kBT} \, \xi_t + \mu F_t(X_t) \nonumber
\end{equation}
to ensure the desired thermodynamic state trajectory is $\rho_t = \rhoeq_{\gamma(t)}$. 

Of course, the main difficulty is to evaluate $A_t = \Delta^{-1} \big[ \partial_t \rho_t \big]$ is given by the Green's function expression 
\begin{equation}
    A_t(x) = \int G(x, x') \big( \partial_t \rho_t(x')\big) \dx', \nonumber
\end{equation}
where $G(x, x')$ is the Green's function corresponding to the Laplacian operator. This may be high dimensional integral which cannot be easily computed. Still, this formal expression $v_t = -\rho_t^{-1} \nabla \{ \Delta^{-1} [\partial_t \rho_t ] \}$ is correct and can be practically useful for low-to-medium dimensions. (See, for instance \cite{pfau2020integrable}). 

This work extends to Langevin equations where the configuration space $X_t \in \mathcal{X}$ is not necessarily Euclidean space (i.e., $\mathcal{X} \neq \mathbb{R}^d$), through considering the calculus of differential forms. 

\indent \section{Chapter introduction}
In any process that transforms 
a system from one state to another, 
there exists some intrinsic relaxation time for the final state to be reached. Recently, a number of studies have attempted to manipulate or, in principle, eliminate altogether this relaxation time by means of alternative driving protocols. 
Such strategies are generally known as
shortcuts to adiabaticity (STA), in which one attempts to rapidly transform from a specified initial state or distribution at time $t_0$ to a target state at a specified final time $t_f$ \cite{2000_PRL_Emmanouilidou,2009_JPhysA_Berry,2013_Torrontegui,2013_PRL_del_Campo,jarzynski2013generating,2014_PRX_Jarzynski,2014_PRX_Jarzynski,2017_NJP_Patra}. 
These types of protocols
have seen success in a variety of settings, including both open and closed quantum and classical systems. A more specific class of strategies 
goes by the name of 
Engineered Swift Equilibration (ESE) and focuses on enforcing internal equilibrium in an open classical system during the entire transformation \cite{martinez2016engineered,2016_NC_an,2018_PRE_Chupeau,guery2019shortcuts}. This is typically done by means of an added driving force: in essence, by following a specified driving protocol, a rapidly-transforming system assumes the trajectory of a quasistatic transformation.

To make this more concrete,
consider a physical system described by some time-dependent Hamiltonian $H(\lambda_i(t))$, where $t$ is the time, and all time dependence is prescribed by parameters $\lambda_i(t)$.  
Following standard Boltzmann statistics \cite{Kadanoff}, the equilibrium probability distribution at a given time is 
\begin{equation} \label{eq:equilibrium}
    \rho_\mathrm{eq}(\mathbf{x}; \lambda_i(t)) = \frac{\exp(-\beta H_0(\mathbf{x}; \lambda_i(t)))}{Z(\lambda_i(t))},
\end{equation}
where $\beta = (k_BT)^{-1}$ is the inverse temperature,
$k_B$ is Boltzmann's constant,
and $Z(\lambda_i(t))$ is the partition function, explicitly dependent on parameters $\lambda_i(t)$. If the $\{\lambda_i(t)\}$ are changed adiabatically, the system will be well-described by Boltzmann statistics at all times. However, if $\lambda_i(t)$ changes sufficiently rapidly, the system deviates from its equilibrium distribution specified by Eq.~(\ref{eq:equilibrium}). The purpose of ESE is to introduce a modified Hamiltonian $H(\mathbf{x},t) = H_0(\mathbf{x},\lambda_i(t))+H_1(\mathbf{x},t)$ such that under the full dynamics of $H$, the system assumes 
the
internal equilibrium distribution of $H_0$ alone [Eq.~(\ref{eq:equilibrium})], at all times.

To date, ESE protocols have 
been successfully derived 
for a Brownian particle trapped in a variety of simple one-dimensional potentials \cite{li2017shortcuts, 2018_NJP_Chupeau} and realized experimentally for a Brownian particle in a 1D harmonic trap \cite{martinez2016engineered}. The framework was also recently analytically solved for the Brownian gyrator \cite{2020_PRE_Baldassarri}, a two-dimensional (Euclidean) system in contact with two heat baths that admits non-equilibrium steady states. In this chapter, we extend the ESE framework to generic overdamped Brownian systems, including those with arbitrarily high dimensional, non-Euclidean configuration spaces. We demonstrate the utility of our framework by numerically finding the ESE forcing for previously intractable systems.



\indent \section{Theory}
We 
first
consider a particle undergoing Brownian motion in the overdamped limit, whose dynamics are governed by the Langevin equation:
\begin{equation}
\label{eq:Langevin}
    \gamma \frac{d\mathbf{x}(t)}{dt} = - \mathbf{\nabla} V(\mathbf{x}(t); \lambda_i(t)) 
    + \bm{\eta}(t) + \mathbf{F}_{\mathrm{ext}}(t),
\end{equation} %
\noindent
where $\mathbf{x}$ is the position of the particle, $\gamma$ is the viscosity, $V(\mathbf{x}; \lambda_i(t))$ is the potential acting on the particle parameterized by control parameters $\lambda_i(t)$, $\bm{\eta}$ is 
Gaussian
noise with delta function autocorrelation $\langle \eta_i(t) \eta_j(t') \rangle = (2\gamma/\beta)\delta_{ij}\delta(t - t')$, where $i,j$ index Euclidean coordinates and $\mathbf{F}_{\mathrm{ext}}(t)$ is an external force on the particle.

Following standard procedures \cite{Kadanoff}, this leads to a Fokker-Planck equation of the form:
\begin{equation} \label{eq:FP}
    \partial_t \rho(\mathbf{x}, t) = \mathbf{\nabla} \cdot \bigg[ \bigg(\frac{\mathbf{\nabla} V(\mathbf{x}; \lambda_i(t)) - \mathbf{F}_\mathrm{ext}}{\gamma} + \frac{\mathbf{\nabla} }{\beta \gamma} \bigg) \rho(\mathbf{x}, t) \bigg],
\end{equation}
where $\rho(\mathbf{x}, t)$ is the configuration space probability distribution at a given time. In the absence of an external force, the steady state solution is found by taking the LHS 
of Eq.~\eqref{eq:FP}
to zero, yielding the usual Boltzmann distribution, $\rho_\mathrm{eq}$.

Now suppose that the control parameters $\lambda_i(t)$ are time-dependent, and varied quickly enough so that we cannot assume a quasistatic transition. We desire to find a $\mathbf{F}_\mathrm{ext}(t)$ such that  $\rho(\mathbf{x}(t), t; \lambda_i(t)) =\rho_\text{eq}(\mathbf{x}; \lambda_i(t))$ for all times $ t$. Clearly, this can only be satisfied if all explicitly time-dependent terms in Eq.~(\ref{eq:FP}) independently cancel. Defining $\mathbf{P} \equiv \rho_\mathrm{eq}\mathbf{F}_\mathrm{ext}$, this constraint 
can be written
\begin{equation} \label{eq:ESE}
\nabla \cdot \mathbf{P} = -\gamma \partial_t \rho_\mathrm{eq} .
\end{equation}%
We now generalize to systems whose configuration space is an arbitrary compact, orientable Riemannian manifold $M$ with metric $g$, and write the vector $\mathbf{P}$ as a differential one-form $P = P_i \mathrm{d} x^{i}$. Equation~(\ref{eq:ESE}) then generalizes to
\begin{equation} \label{eq:ESEdiff}
  \mathrm{d}^{\dagger} P = -\gamma \partial_t \rho_\text{eq} ,
\end{equation}
where $\mathrm{d}$ is the exterior derivative and $\mathrm{d}^\dagger$ is its Hodge dual: $\mathrm{d}^\dagger = \star \mathrm{d}\star$, where $\star$ is the Hodge star operator. We now invoke the Hodge Decomposition, which states that, for any k-form $P_k$ on $M$, there exists a unique decomposition \cite{Nakahara}:
\begin{equation} \label{eq:hodgeD}
    P_k = \mathrm{d} A_{k-1} + \mathrm{d}^{\dagger}B_{k+1} + C_k ,
\end{equation}
where $A_{k-1}$ and $B_{k+1}$ are $(k-1)$- and $(k+1)$-forms, respectively, and $C_k$ is a harmonic $k$-form; {\em i.e.} $\Delta C_k = 0$, and $\Delta = \mathrm{d}\mathrm{d}^{\dagger} + \mathrm{d}^{\dagger} \mathrm{d}$ is the generalized Laplace operator. Therefore, we may write Eq.~(\ref{eq:ESE}) as
\begin{equation} \label{eq:ESEHodge}
  \mathrm{d}^{\dagger} P = \mathrm{d}^{\dagger}(\mathrm{d} A + \mathrm{d}^{\dagger}B + C)= \mathrm{d}^{\dagger} \mathrm{d} A + \mathrm{d}^{\dagger} C = \Delta A + \mathrm{d}^{\dagger} C ,
\end{equation}
where $A$ is a 0-form, $B$ is a 2-form, and $C$ is a harmonic 1-form. For simplicity, we choose $C=0$ to yield a generalized Poisson's equation:
\begin{equation} \label{eq:ESEManifold}
  \Delta A = -\gamma \partial_t \rho_{\mathrm{eq}} .
\end{equation}
Using this result, coupled with Eq.~(\ref{eq:hodgeD}), we arrive at
our main result:
\begin{equation} \label{eq:ESEManifoldForce}
  \mathbf{F}_{\text{ext}} = \frac{\mathrm{d} A}{\rho_{\mathrm{eq}}}.
\end{equation}
To demonstrate the utility of this expression, we now apply it to 
multiple physically realizable example systems.

\indent \section{Higher Dimensional Euclidean Configuraton Space}
In a $d$-dimensional Euclidean space, the Hodge decomposition trivializes to the Helmholtz decomposition \cite{Nakahara}. Our results still hold, though for certain cases the decomposition may no longer be unique. Noting this caveat, we continue with our analysis. For this space, the generalized Laplace operator is simply the standard Laplace operator. Note that our choice $C = 0$ amounts to a no-curl gauge:
$\nabla \times \mathbf{P} = 0$. We may thus define a scalar potential $A$ 
that satisfies
$\nabla A = \mathbf{P}$. This is analogous to standard electrostatics, where $A$ and $\mathbf{P}$ play of the roles the electric potential and field, respectively \cite{Jackson}.


Given the general solution to the Laplace operator for Euclidean space, we have, for $d\neq 2$,
\begin{equation}
A(\mathbf{x}) = -\frac{\Gamma(\frac{d-2}{2})}{4\pi^\frac{d}{2}}\int d^d\mathbf{x}'  \ |\mathbf{x}-\mathbf{x}'|^{2-d}(-\gamma \partial_t\rho_{\text{eq}}(\mathbf{x}'))
\end{equation}
and $\mathbf{F}_{\text{ext}}(\mathbf{x}) =\rho^{-1}_{\text{eq}} \mathbf{\nabla} A$. Note that this solution reproduces Eq.~(12) of  \cite{li2017shortcuts} for 
1D systems. 
For the special case of $d=2$, the potential is given by
\begin{equation}
A(\mathbf{x}) = \frac{1}{2\pi}\int d^2\mathbf{x}'  \ \log |\mathbf{x}-\mathbf{x}'|(-\gamma \partial_t\rho_{\text{eq}}(\mathbf{x}')).
\end{equation}

\begin{figure*}[t]
\label{fig:ESEfig}
    \centering
    \includegraphics[width=\textwidth]{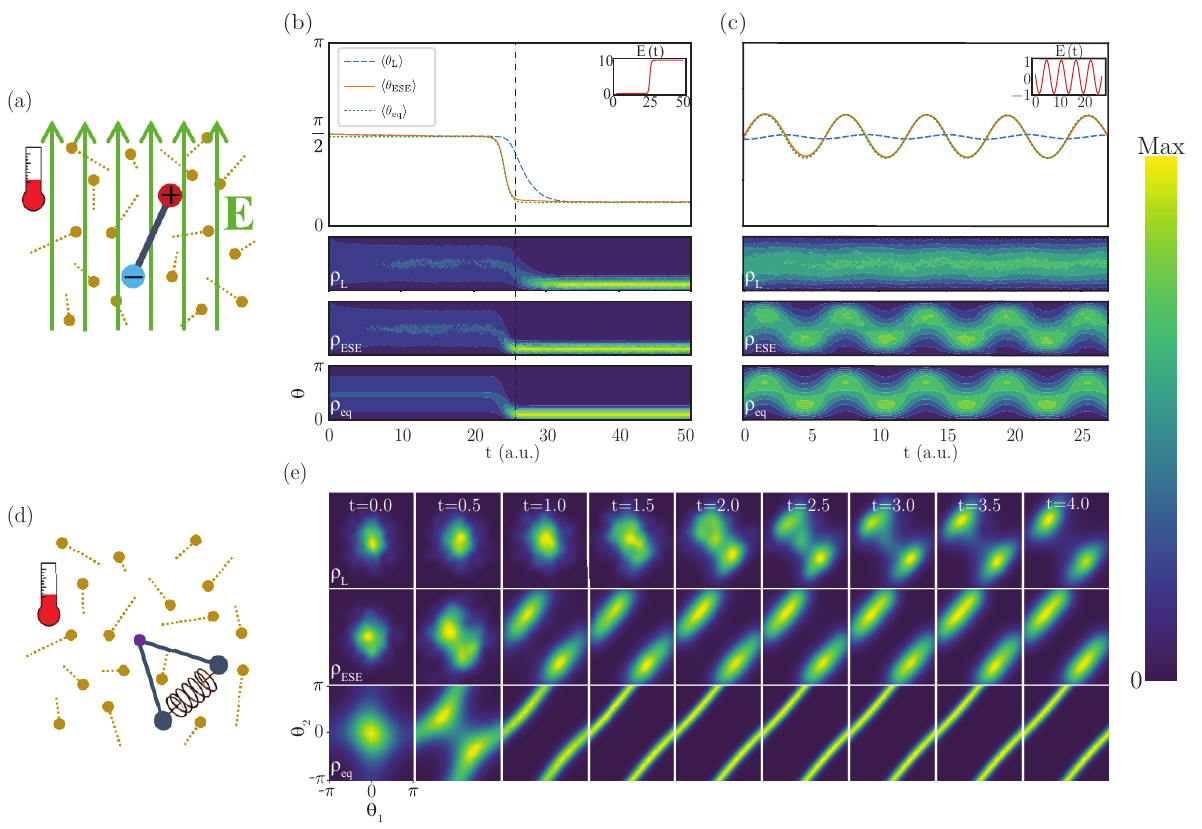}
    \caption[Efficacy of counterdiabatic protocols for two non-Euclidean topologies: dipole in a magnetic field, and toroidal configuration space]{Simulations demonstrate that ESE protocols produce densities closely tracking the equilibrium distribution in configuration space corresponding to the control parameters at each moment in time.
    \textbf{(a)} Schematic diagram of the ensemble considered in subplots \textbf{(b)} and \textbf{(c)}. An ensemble of electric dipoles are placed in a uniform, time-varying electric field. Due to these constraints,
    the state of a dipole is specified by a polar coordinate $\theta$ and an azimuthal angle $\phi$ such that the system's configuration space is a (2D) sphere. \textbf{(b)} The electric field is varied sigmoidally in time, as shown in the inset. In the top panel, we show the average values of $\theta$ over the entire ensemble for a system undergoing Langevin dynamics (dashed blue), ESE dynamics (solid orange), and for a Boltzmann distibution for the given electric field $E(t)$, denoted $\rho_{\text{eq}}$ (dotted green). The three lower panels are the corresponding full distributions for $\theta$ at all times. \textbf{(c)} Same as panel \text{b} for the sinusoidally varying electric field shown in the inset. \textbf{(d)} Schematic diagram of the ensemble considered in subplot \textbf{(e)}. An ensemble of two coupled pendula with time-varying coupling constant. The state of a coupled pendulum is specified by the two angular coordinates of the pendula, $\theta_1$, $\theta_2 \in [-\pi,\pi)$, such that the full system's configuration space is a torus. \textbf{(e)} The probability distributions of an ensemble of coupled pairs of pendula undergoing Langevin dynamics (top) and ESE dynamics (middle) plotted against the instantaneous Boltzmann distribution (bottom) for a coupling constant that changes from zero at small times to a negative value. See Supplementary Material for a movie of this process \cite{frim2021engineered}.
    }
\end{figure*}

\indent \section{Spherical Configuration Space}
We now consider topologically nontrivial configuration spaces. As a first example, consider an electric dipole with dipole moment $\mathbf{p} = p\hat{\mathbf{p}}$ placed in a time-varying electric field pointed in the $z$-direction, $\mathbf{E} = E(t) \hat{\mathbf{z}}$, as shown in Fig.~1(a). The potential energy for this system is 
\begin{equation}
\label{eq:dipole}
    V(\theta, \phi, t)  = -p E(t) \cos (\theta),
\end{equation}
where $\cos\theta = \hat{\mathbf{p}}\cdot \hat{\mathbf{z}}$ and $\phi$ measures the azimuthal angle about to the 
$z$-axis.
The configuration space of this system is the 2-sphere, $M = S^2$, for which the Laplace operator is
\begin{equation}
    \Delta_{S^2}  = \frac{1}{\sin \theta} \frac{\partial}{\partial \theta} \bigg(\sin \theta \frac{\partial }{\partial \theta} \bigg) + \frac{\partial^2 }{\partial \phi^2}.
\end{equation}
To calculate the required ESE force, we find $\rho_{\text{eq}}(t)$, the instantaneous Boltzmann distribution for this system: 
\begin{equation}
\rho_{\text{eq}}(t) = \left[\frac{4\pi \sinh(\beta pE(t))}{\beta p E(t)}\right]^{-1} \exp{\beta pE(t)\cos \theta},
\end{equation}
such that the governing equation is given by $\Delta_{S^2} A = -\gamma \partial_t \rho_\text{eq}$. In the high temperature limit, one may find an analytic expression for $A$ (see Supplemental Materials Section SM.1 for details \cite{frim2021engineered}); however, finding a closed-form expression is, in general, intractable. Instead, we employ a series expansion in the spherical harmonics, $Y_{\ell}^m(\theta,\phi)$. The spherical harmonics are the eigenfunctions of the spherical Laplace operator, {\em i.e.} $\Delta_{S^2}Y_{\ell}^m(\theta,\phi) = -\ell(\ell+1)Y_{\ell}^m(\theta,\phi) $, such that if we write $\rho_{\text{eq}} = \sum_{\ell,m}c_{\ell,m}(t)Y_\ell^m(\theta,\phi)$, then by the orthogonality and completeness of the spherical harmonics, we have
\begin{equation}
\label{eq:SphericalHarms}
A = \gamma\sum_{\ell=0}^\infty \sum_{m=-\ell}^\ell\frac{\partial_t c_{\ell,m}(t)}{\ell(\ell+1)}Y_\ell^m(\theta,\phi).
\end{equation}
Note that $c_{\ell,m}$ may likewise be  computed:
\begin{equation}
\label{eq:SphericalHarmsCoeffs}
c_{\ell,m}(t) = \int_\Omega \rho_{\text{eq}}(\theta,\phi,t) Y_\ell^{m*}(\theta,\phi)d\Omega.
\end{equation}
Due to the azimuthal symmetry of $\rho_{\text{eq}}(\theta,\phi,t)$, only $m=0$ terms will be nonzero, simplifying our analysis. Finally, 
$\mathbf{F}_\text{ext}$ is 
found by taking $\mathbf{P} = \mathrm{d} A \implies \mathbf{F}_\text{ext} = \rho_\text{eq}^{-1} (\mathrm{\nabla}_{S^2}) A$.

We now simulate the system for specified functions $E(t)$. The dynamics of this system are governed by a set of Langevin equations:
\begin{align}
&m(\ddot{\theta}-\sin\theta\cos\theta \dot{\phi}^2)=-\gamma \dot{\theta} -pE\cos\theta+\eta_\theta+F_{\text{ext},\theta} , \label{eq:sphericaldynamics1}\\
&m(\sin\theta \ddot{\phi}+2\cos\theta\dot{\theta}\dot{\phi})=-\gamma \sin\theta \dot{\phi}+\eta_\phi+F_{\text{ext},\phi}. \label{eq:sphericaldynamics2}
\end{align}
Note that, though we do not explicitly enforce the overdamped limit in Eqs.~(\ref{eq:sphericaldynamics1}) and (\ref{eq:sphericaldynamics2}), we will effectively do so by means of parameter choices in our simulations. For a specified $E(t)$, we numerically solve Eqs.~(\ref{eq:SphericalHarms}) and (\ref{eq:SphericalHarmsCoeffs}) to find the ESE force (truncating above $\ell=5$) and then simulate the Langevin dynamics in both the presence and absence of this force for an ``ensemble" of $10^4$ dipoles. In units of $\beta = m = p = 1$, we simulate for $\gamma = 20$. Due to the noise terms in Eqs.~(\ref{eq:sphericaldynamics1}) and (\ref{eq:sphericaldynamics2}), these are stochastic differential equations, which we simulate by means of a first-order Euler-Maruyama algorithm~\cite{Kloeden}, with step sizes of $dt = 0.01$ time units. 
Given the promotion of configuration space to a non-Euclidean manifold, the relation for the noise term is 
modified~\cite{2009_EPL_Kumar,2017_PRE_apaza}: $\langle \eta_i(t) \eta_j(t') \rangle =(2\gamma/\beta)g^{ij}\delta(t-t') $, where $g^{ij}$ is the inverse metric of the manifold in question. For the (unit) sphere, the inverse metric is $g^{\theta\theta} = 1$, $g^{\phi\phi} = 1/\sin\theta$, and all other entries are zero. Therefore, following the standard Euler-Maruyama treatment, we take $\eta_\theta dt = \sqrt{2\gamma dt/\beta}\mathcal{N}(0,1)$ and $\eta_\phi dt = \sqrt{2\gamma dt \csc\theta/\beta}\mathcal{N}(0,1)$, where $\mathcal{N}(0,1)$ is the 
Normal distribution with zero mean and unit variance. To deal with the spherical-polar coordinate singularities at $\theta = 0$ and $\theta = \pi$, we temporarily rotate to a different local coordinate system where the poles are shifted away and numerically integrate a single time-step, whenever $0<\theta<\pi/10$ or $9\pi/10<\theta<\pi$; see~\cite{frim2021engineered} for further details.

In Fig.~1 we plot both the mean value and 
the full 
probability distribution of $\theta(t)$ for both the standard Langevin and ESE dynamics for two representative temporally varying electric fields, as described below. Due to the azimuthal symmetry of the problem, the azimuthal angle $\phi$ is not affected by any temporal change in $E(t)$.

In Fig.~1(b), we consider a sigmoidally varying electric field, $E(t) \sim E_0 + (\Delta E) S(t - t_c)$, where 
\begin{equation} \label{eq:logistic}
S(t) = (1 + e^{-t})^{-1}
\end{equation}
is the logistic function, $\Delta E = 10$ is the amplitude of the change of the electric field,
and 
$t_c$ is the transition time. For $E_0$ small, 
$\theta$ is primarily distributed about $\pi/2$, which corresponds to the equator. We note that for $pE_0 \ll k_B T$, the dipoles have no preferred direction, so they should be uniformly distributed throughout configuration space. However, the distribution appears non-uniform as a function of $\theta$. This is an artifact of our coordinate system: there is simply more phase space area at $\theta = \pi/2$ (the equator) than elsewhere, such that the probability as a function of $\theta$ should be non-uniform. For $t\gg t_c$, we find that, on average, $\theta<\pi/2$ . This is physically sensible: the electric field is strong and directed in the $\hat{z}$ direction such that the tendency of the dipoles will be to align with it. However, again as an artifact our coordinate system, there is no probability density at $\theta = 0$ as there is no phase space area at this pole. For $t\sim t_c$, in the absence of the ESE force, the system remains out of equilibrium for a finite period of time before eventually relaxing to the new equilibrium. However, when the ESE force is introduced, the system remains 
close to
the equilibrium distribution at all times.

In Fig 1(c), we consider a sinusoidal electric field, $E(t)\sim E\sin(t)$. In this case, we see that the constantly changing field never allows the standard Langevin system to fully equilibrate; instead, the system oscillates with an approximate $\pi/2$ phase-shift at a significantly smaller amplitude. The ESE dynamics, on the other hand, converge 
close 
to the equilibrium distribution at all points in time.

\indent \section{Toroidal configuration space}
Next, we consider toroidal configuration space, as exemplified by a system of two pendula, each of mass $m$ and unit length suspended vertically, and coupled to each other with a time-varying coupling constant $\kappa(t)$, as illustrated in Fig.~1(d). The potential may be modeled as
 \begin{equation}
    V(\theta_1, \theta_2, t) \simeq -(mg\cos\theta_1+mg\cos\theta_2-\kappa(t) \cos(\theta_1-\theta_2)),
\end{equation}
where $\theta_1$ and $\theta_2$ are angles of the respective pendula with respect to the $z$-axis. Given the periodicity in $\theta_i$, the configuration space of this system is the 2-torus, $M = T^2 = [0,2\pi]\times [0,2\pi]$. The Laplace operator for this manifold is
\begin{equation}
    \Delta_{T^2}  =  \frac{\partial^2}{\partial\theta_1^2}+\frac{\partial^2}{\partial\theta_2^2} , 
   \end{equation}
where one must recall the periodicity of the coordinates: $\theta_i \sim \theta_i+2\pi n$ for $n\in \mathbb{Z}$. We again employ a series expansion to solve Eq.~(\ref{eq:ESE}). In this case, we carry out a 2D Fourier series. Considering that $\Delta_{T^2} \exp{i(m_1\theta_1 + m_2\theta_2)} = -(m_1^2+m_2^2)\exp{i(m_1\theta_1 + m_2\theta_2)}$, we may write $\rho_\text{eq} = \sum_{m_1,m_2} c_{m_1,m_2}(t) \exp{i(m_1\theta_1 + m_2\theta_2)}$ and easily deduce that
\begin{equation}
	A = \gamma \sum_{m_1,m_2} \frac{\partial_tc_{m_1,m_2}(t)}{m_1^2+m_2^2} e^{i(m_1\theta_1 + m_2\theta_2)}.
 \end{equation}
We again simulate this dynamical system for a given $\kappa(t)$. The governing Langevin equations are now 
\begin{equation}
m\ddot{\theta}_{1/2}= -\gamma \dot{\theta}_{1/2}-mg\sin\theta_{1/2}  + \kappa(t)\sin(\theta_{1/2}-\theta_{2/1})+\eta_{\theta_{1/2}}+F_{\text{ext},\theta_{1/2}} . \label{eq:Toroiddynamics}
\end{equation}
As with the last example, our framework is not limited to
the overdamped limit. For a specified $\kappa(t)$, we numerically solve the ESE force 
$F_{ext}(t)$
by means of the Fourier series expansion (truncating above $m_1=m_2=10$) and then simulate the Langevin dynamics in both the presence and absence of this force. In units of $\beta = m = g = 1$, we simulate for $\gamma = 20$. We again employ an Euler-Maruyama algorithm with time step $dt = 0.01$. Due to the locally Euclidean metric of the torus, we need not deal with any geometric contributions to the noise terms and have $\eta_{\theta_1}dt = \eta_{\theta_2} dt = \sqrt{2\gamma dt/\beta}\mathcal{N}(0,1)$. For our simulations, we choose $\kappa(t) \sim -\kappa_0 S(t)$, where $S(t)$ is defined by (\ref{eq:logistic}) 
so that the pendula are initially uncoupled but after some critical time $t_c$ they are anti-coupled (we choose anti-coupling for clarity in the resulting figures). In Fig.~1(e), we display the resulting probability distribution for the coordinates $(\theta_1,\theta_2)$ at 
several times near 
$t_c$. In the absence of the ESE force (top row), equilibration happens over a finite amount of time as the system relaxes to its new anti-coupled distribution. However, when the ESE force is added (second row), the resulting distribution 
agrees well with the calculated equilibrium distribution $\rho_\text{eq}(t)$ corresponding to the control parameter values at each moment in time (bottom row).

\indent \section{Discussion}
In previous studies, the notions of optimality and control often refer to specific protocols designed to minimize excess work or some other performance index when changing between two equilibria or non-equilibrium steady states
in finite time
\cite{schmiedl2007optimal, schmiedl2007efficiency, zulkowski2012geometry, 2013_plos_zulkowski_optimal,2015_PRE_zulkowski,plata2019optimal}. 
The ESE framework we develop here may also be considered a control strategy, but
ESE seeks only to minimize time to equilibration throughout the protocol without any constraints or penalties on the work required to do this.
In principle, provided a smooth trajectory of control parameters, ESE should allow for arbitrarily rapid equilibrium switching of ensemble distributions. However, realizing the required forces in a laboratory setting would presumably preclude such a situation. Further, the theory itself breaks down in such a limit due to higher-order effects ignored in a basic Langevin treatment, such as a finite characteristic timescale of the noise correlations. Therefore, for the range of timescales for which Eq.~(\ref{eq:Langevin}) applies, ESE yields a method to achieve
controlled, swift equilibration. Furthermore, previous studies have analyzed the relation between 
the duration $\tau$ of a protocol and the energy dissipated in carrying out a drive, concluding that for a variety of model systems, the energy dissipated is proportional to $1/\tau$  \cite{esposito2010finite,2014_erasure_PRE_Zulkowski,li2017shortcuts,2017_PRL_Campbell}. 

ESE protocols ensure a high degree of control throughout the drive. Not only do we enforce the mean, or the mean and variance (or any finite combination of moments) of the probability distribution, we place demands on the \emph{entire} probability distribution at all times during the protocol. In fact,
we can
make an even stronger claim. Following \cite{Nakahara}, if a scalar field integrates to zero over a full manifold (without boundary), it may be written as the divergence of a vector field. Importantly, by integrating the LHS of Eq.~(\ref{eq:ESE}) over any phase space manifold $M$, we see that
\begin{equation}
    \int_M (-\gamma \partial_t \rho_\text{eq}) dV = -\gamma \frac{d}{dt} \int_M \rho_\text{eq} dV = 0.
\end{equation}
following conservation of probability. We conclude that for any arbitrary time-dependent potential described by some smooth set of coordinates, there will always be a corresponding ESE force that can enforce
swift equilibration.

Finally, we note a further degree of freedom in the ESE condition defined by Eq.~(\ref{eq:ESEHodge}): The differential form $d^\dagger P = \Delta A+d^\dagger C$ allows for an alternative, arbitrary choice of a harmonic one-form $C$ if we make the replacement $\gamma \partial_t \rho \rightarrow \gamma \partial_t \rho+d^\dagger C$ in Eq.~(\ref{eq:ESEManifold}). We also may include an arbitrary 2-form $B$ by using Eq.~(\ref{eq:ESEHodge}), which leads to a class of inequivalent, though perhaps nonconservative, driving forces that can each enforce equilibration. These choices may allow for greater flexibility in solving Eq.~(\ref{eq:ESEManifold}) and finding appropriate forces for practical laboratory applications.

\indent \section{Conclusion}
In this Appendix, we have successfully extended the ESE protocol to systems with nontrivial phase space topology. 
Our results could be useful for designing optimal strategies for manipulating a thermalized system of multiple canonical position variables swiftly through controlled parameter changes. Our methods can be used to calculate the necessary auxiliary forces to impose internal equilibrium dynamics in experimental settings. 
The derivation we present here is only valid for the overdamped limit. In future work, it will be interesting and useful to generalize our framework to include underdamped systems.

\printbibliography
\end{document}

%% file: abstract.tex

\begin{abstract}

What are the fundamental limitations placed by the laws of thermodynamics on the energy expenditure needed to carry out a given task in a nonequilibrium environment in finite time? In this thesis we investigate ``optimal nonequilibrium processes'': how nonequilibrium state changes in a thermodynamic system may be performed most efficiently, in the sense of requiring the least amount of thermodynamic work. Surprisingly, there is a hidden, fundamental geometric structure in this optimization problem that is related to the mathematics of optimal transport theory: how to optimally send, e.g., baguettes from bakeries to cafés, given supply and demand constraints, that requires the least amount of total distance traveled by the baguettes. After giving a brief overview on the mathematical framework of stochastic thermodynamics for the overdamped Langevin equation, we present a trio of works: (1) applying optimal control theory to the Fokker-Planck equation to calculate exact optimal protocols for low-dimensional systems, which reproduces the intriguing previously-discovered discontinuities in globally optimal protocols and reveals new \emph{non-monotonic} optimal protocols for a certain system; (2) exploiting the importance sampling of Langevin trajectories under different protocols, to adaptively optimize protocols by controlling the thermodynamic state trajectory, which is useful for efficiently calculating free energy differences between different Hamiltonians; and finally, (3) deriving an exact geometric description of optimal nonequilibrium processes and a geodesic-counterdiabatic decomposition for the optimal protocols that enact them, which satisfyingly explains the highly non-intuitive properties of discontinuities and possible non-monotonicity in globally optimal protocols. 

\end{abstract}